\documentclass[11pt]{article}

\newcommand{\veps}{\varepsilon}

\newcommand{\argmin}{{\rm argmin}}

\makeatletter
\newcommand*{\rom}[1]{\expandafter\@slowromancap\romannumeral #1@}
\makeatother

\def\##1\#{\begin{align}#1\end{align}}
\def\$#1\${\begin{align*}#1\end{align*}}

\def\vmu{{\mathbf{\upmu}}}
\def\vtheta{{\mathbf{\uptheta}}}

\def\vdelta{{\mathbf{\updelta}}}

\def\vc{{\mathbf{c}}}
\def\vd{{\mathbf{d}}}
\def\ve{{\mathbf{e}}}
\def\vf{{\mathbf{f}}}

\def\vp{{\mathbf{p}}}
\def\vq{{\mathbf{q}}}

\def\vs{{\mathbf{s}}}

\def\vx{{\mathbf{x}}}
\def\vy{{\mathbf{y}}}

\def\mH{{\mathbf{H}}}
\def\mI{{\mathbf{I}}}

\def\mX{{\mathbf{X}}}
\def\mY{{\mathbf{Y}}}

\def\mSigma{{\mathbf{\Upsigma}}}

\def\gA{{\mathcal{A}}}

\def\gI{{\mathcal{I}}}

\newcommand{\R}{\mathbb{R}}

\renewcommand{\tilde}{\widetilde}

\usepackage[left=1in,right=1in,top=1.1in,bottom=1.2in]{geometry}
\usepackage{amsmath,amssymb,amsfonts,amsthm,bbm,bm}
\usepackage{times}
\RequirePackage[colorlinks,citecolor=blue,urlcolor=blue,linkcolor=blue,linktocpage=true]{hyperref}
\usepackage{url}
\usepackage{graphicx}
\usepackage{array}
\usepackage{hhline}
\usepackage{wrapfig,lipsum}
\usepackage{soul}
\usepackage{booktabs}
\usepackage[numbers]{natbib}
\usepackage{verbatim}
\usepackage{setspace}
\usepackage{graphicx}
\usepackage{enumitem}
\usepackage{color}
\usepackage[dvipsnames,table]{xcolor}
\usepackage{bbm}
\usepackage{multirow}
\usepackage{caption}
\usepackage{subcaption}
\usepackage{booktabs}
\usepackage{float}
\usepackage{placeins}

\usepackage[Symbol]{upgreek}

\usepackage{xr}

\newcommand{\vF}{\mathbf{F}}
\renewcommand{\thefootnote}{\arabic{footnote}}

\newtheorem{thm}{Theorem}

\newcommand{\parallelsum}{\mathbin{\!/\mkern-5mu/\!}}

\makeatletter
\newcommand{\myfnsymbol}[1]{%
  \expandafter\@myfnsymbol\csname c@#1\endcsname
}
\newcommand{\@myfnsymbol}[1]{%
  \ifcase #1
  \or 1
  \or 2
  \or 3
  \or \TextOrMath{\textasteriskcentered}{*}
  \fi
}
\newcommand{\affiliationA}{\@myfnsymbol{1}}
\newcommand{\affiliationB}{\@myfnsymbol{2}}
\newcommand{\affiliationC}{\@myfnsymbol{3}}
\newcommand{\correspondingA}{\@myfnsymbol{4}}
\makeatother

\title{Assessing and improving reliability of neighbor embedding methods: a map-continuity perspective}
\author{Zhexuan Liu\thanks{Department of Statistics, University of Wisconsin--Madison, Madison, WI, 53706, USA.}, Rong Ma\thanks{Department of Biostatistics, T.H.~Chan School of Public Health, Harvard University, Boston, MA 02115, USA. Email: \texttt{rongma@hsph.harvard.edu}}, and Yiqiao Zhong}

\author{
  Zhexuan Liu\textsuperscript{\affiliationA},
  Rong Ma\textsuperscript{\affiliationB,\affiliationC},
  Yiqiao Zhong\textsuperscript{\affiliationA,\correspondingA}
}
\date{\today}

\begin{document}

\renewcommand{\thefootnote}{\myfnsymbol{footnote}}
\maketitle
\footnotetext[1]{Department of Statistics, University of Wisconsin--Madison, Madison, WI, 53706, USA.}%
\footnotetext[2]{Department of Biostatistics, T.H.~Chan School of Public Health, Harvard University, Boston, MA 02115, USA.}%
\footnotetext[3]{Department of Data Science, Dana-Farber Cancer Institute, Boston, MA 02115, USA.}%
\footnotetext[4]{Email: \texttt{yiqiao.zhong@wisc.edu}}%

\setcounter{footnote}{0}
\renewcommand{\thefootnote}{\fnsymbol{footnote}}

\abstract{

Visualizing high-dimensional data is essential for understanding biomedical data and deep learning models. Neighbor embedding methods, such as t-SNE and UMAP, are widely used but can introduce misleading visual artifacts. We find that the manifold learning interpretations from many prior works are inaccurate and that the misuse stems from a lack of data-independent notions of embedding maps, which project high-dimensional data into a lower-dimensional space. Leveraging the leave-one-out principle, we introduce LOO-map, a framework that extends embedding maps beyond discrete points to the entire input space. We identify two forms of map discontinuity that distort visualizations: one exaggerates cluster separation and the other creates spurious local structures. As a remedy, we develop two types of point-wise diagnostic scores to detect unreliable embedding points and improve hyperparameter selection, which are validated on datasets from computer vision and single-cell omics.
}



\section*{Introduction}\label{sec:intro}

Data visualization plays a crucial role in modern data science, as it offers essential and intuitive insights into high-dimensional datasets by providing low-dimensional embeddings of the data. For visualizing high-dimensional data, the last two decades have witnessed the rising popularity of 
t-SNE \cite{HintonTSNE2008} and UMAP \cite{mcinnesUMAP2018}, which are extensively used in, e.g., single-cell analysis \cite{kobak2019art,linderman2019fast,luecken2019current} and feature interpretations for deep learning models \cite{JING2022LayerUMAP,Islam2023DNNManifoldLearning}. 

The neighbor embedding methods \cite{probabilistic_view_of_NE, BoydMDE2021} are a family of visualization methods, which include t-SNE, UMAP, and LargeVis \cite{largevis} as popular examples, that determine embedding points directly by solving a complicated optimization algorithm to minimize the discrepancy between similarities of input points and those of the corresponding low-dimensional points.  Given input data $\vx_1,\ldots,\vx_n$, a neighbor embedding algorithm $\gA$ computes the points $(\vy_1, \ldots, \vy_n) = \gA(\vx_1,\ldots,\vx_n)$ in the 2D plane, aiming to preserve the essential structures of $\vx_1,\ldots,\vx_n$. Due to algorithmic complexity, $\gA$ is often used as a black-box visualization tool.

These visualization methods are often interpreted as manifold learning algorithms, which extract and represent latent low-dimensional manifolds in 2D and 3D spaces \cite{ghojogh2023elements, wei2025diffusive, kim2024inductive}. However, unlike classical dimension reduction methods such as PCA \cite{kppca1901}, where a parametric mapping $\vf_{\vtheta}$ is determined and any input point $\vx$ is embedded through $\vy = \vf_{\vtheta}(\vx)$,
there is no globally defined embedding map for neighbor embedding methods as the ``embedding points'' $\vy_1, \ldots, \vy_n$ are determined in a discrete manner.

A key conceptual difficulty is the lack of sample-independent notion of embedding maps, since the embedding points $\vy_1,\ldots, \vy_n$ depend on $n$ input points $\vx_1,\ldots, \vx_n$ collectively, which makes it challenging to understand the correspondence between an input point $\vx_i$ and an embedding point $\vy_i$. Thus, it is unclear what structures the embedding points inherit from the input points, even in ideal settings where inputs are drawn from known distributions or simple manifolds. The lack of continuous-space embedding maps leads to recent recognition that neighbor embedding methods often produce misleading results by creating severe distortion through the embedding maps and introduce spurious clusters in low-dimensional visualization \cite{WangPacMap2021, ChariSpeciousArt2023}. Moreover, neighbor embedding methods are sensitive to the choice of optimization algorithms \cite{YangOptimizationTSNE2015}, initialization schemes \cite{KobakInitializationImportantForTSNEUMAP2019},
and hyperparameters \cite{KobakInitializationImportantForTSNEUMAP2019,Ma2022theoreticalfoundationoft-SNE}, leading to inconsistent interpretations \cite{AllOfUs2024, marx2024seeing}.  

Some progress has been made to improve the reliability of these visualization methods, including insights on
embedding stages \cite{Ma2022theoreticalfoundationoft-SNE,arora2018analysistsnealgorithmdata,linderman2019Clusteringwitht-SNEProvably}, force-based interpretations \cite{steinerberger2022t-SNEforcefulcolorings}, visualization quality \cite{arora2018analysistsnealgorithmdata,linderman2019Clusteringwitht-SNEProvably,shaham2017stochasticneighborembeddingseparates}, initialization schemes, and hyperparameter selection \cite{kobak2019art,ChariSpeciousArt2023, wattenberg2016how,  BelkinaAutoOptParaTSNE2019, scDEED2023}. To enhance the faithfulness of neighbor embedding methods, multiple diagnostic approaches have been proposed \cite{kobak2019art,ChariSpeciousArt2023, wattenberg2016how,  BelkinaAutoOptParaTSNE2019, EMBEDR, dynamicviz, heiter2024pattern}. 
However, most existing diagnostic methods offer only partial solutions and rely on ad hoc fixes, sometimes even introducing new artifacts.

In this work, we show that the manifold learning interpretation, which implicitly assumes a continuous mapping, is inaccurate. Our analyses reveal intrinsic discontinuity points in the embeddings that result in severe distortions. Our results imply that t-SNE and UMAP---which can induce topological changes to visualization---are fundamentally different from PCA and other parametric embedding methods.

We address the conceptual difficulty by proposing a notion of embedding map---which we call LOO-map---induced by a given neighbor embedding method $\gA$. LOO-map is a mapping in the classical sense and approximates the properties of $\gA$ around each embedding point. It is based on a well-established strategy from statistics known as the leave-one-out (LOO) method, which posits that adding, deleting, or changing a single input point has negligible effects on the overall inferential results. Using LOO, we can decouple the pairwise interaction in the algorithm $\gA$: we add a new input point $\vx$ to $\vx_1,\ldots, \vx_{n}$ and freeze $\vy_1,\ldots, \vy_{n}$ in the optimization problem, allowing only one free variable $\vy$. We call the resulting minimizer $\vf(\vx)$ the LOO-map, which satisfies the approximation $(\vy_1,\ldots, \vy_{n}, \vf(\vx)) \approx \gA(\vx_1,\ldots, \vx_{n}, \vx)$. By design, the LOO-map $\vf$ not only satisfies $\vf(\vx_i) \approx \vy_i$ for all $i$'s, but also reveals the  embedding point $\vf(\vx)$ of a potential new input point $\vx$. As such, LOO-map extends the mapping defined over the discrete input set $\{\vx_1, \ldots, \vx_n\}$ to the entire input space.

LOO-map offers a unified framework for understanding known issues like distance distortion \cite{ChariSpeciousArt2023,globaltsne}, low stability \cite{dynamicviz}, and poor neighborhood preservation \cite{ChariSpeciousArt2023, Cooley689851}, while also revealing new insights into embedding discontinuity. In our view, discontinuities of $\vf(\vx)$ represent an extreme form of distortion that accompanies topological changes in the embedding space, e.g., connected clusters become separated and a uniform shape is fractured into pieces. In contrast, classical dimension reduction methods such as PCA do not suffer from map discontinuity since a continuous parametric map $\vf_{\theta}(\vx)$ is constructed explicitly. In this regard, embedding discontinuity is an innate issue of the family of neighbor embedding methods.

Using LOO-map, we identify two types of observed distortion patterns, one affecting global properties of the embedding map and the other affecting local relationships. Both types of distortion are a consequence of discontinuities in $\vf(\vx)$ and can cause topological changes in the embedding structures.

\begin{itemize}
    \item \textbf{Overconfidence-inducing (OI) discontinuity}. Overlapping clusters or data mixtures in the input space are embedded into well-separated clusters, which creates a misleading visual impression of over-confidence that there is less uncertainty in the datasets.
    This biased perception of uncertainty can, in turn, lead to overly confident scientific conclusions. 
    \item \textbf{Fracture-inducing (FI) discontinuity}. Small spurious and artificial clusters form in the embedding space, even for non-clustered data. 
    Unlike OI discontinuity, such spurious clusters are small, localized, and formed in arbitrary locations. 
\end{itemize}
%
%
We propose two types of point-wise diagnostic scores, namely \textit{perturbation scores} and \textit{singularity scores}, to quantify the severity of the two types of map discontinuity at each embedding point. Our approach is flexible and works as a wrapper around many neighbor embedding algorithms (Supplementary File Section~\ref{sec: appendix: derivation of sscore}) without any label information.

In this work, we demonstrate the utility of our method through two use cases: detecting out-of-distribution data (or distribution shifts) in computer vision using the perturbation score, and selecting hyperparameters in single-cell data analysis using the singularity score. We evaluate our method on multiple simulated and real-world datasets (Supplementary Table~\ref{tab: datasetoverview}, Methods). Comparisons with existing approaches show that our method achieves superior performance in detecting topological changes in embedding and hyperparameter selection. The R package implementing our method, along with a tutorial, is publicly available on GitHub:
\begin{center}\url{https://github.com/zhexuandliu/MapContinuity-NE-Reliability}. \end{center}

\section*{Results}

\subsection*{Overview of methods}\label{sec:overview}

\begin{figure}[htbp]
    \centering
    \includegraphics[scale = 0.9]{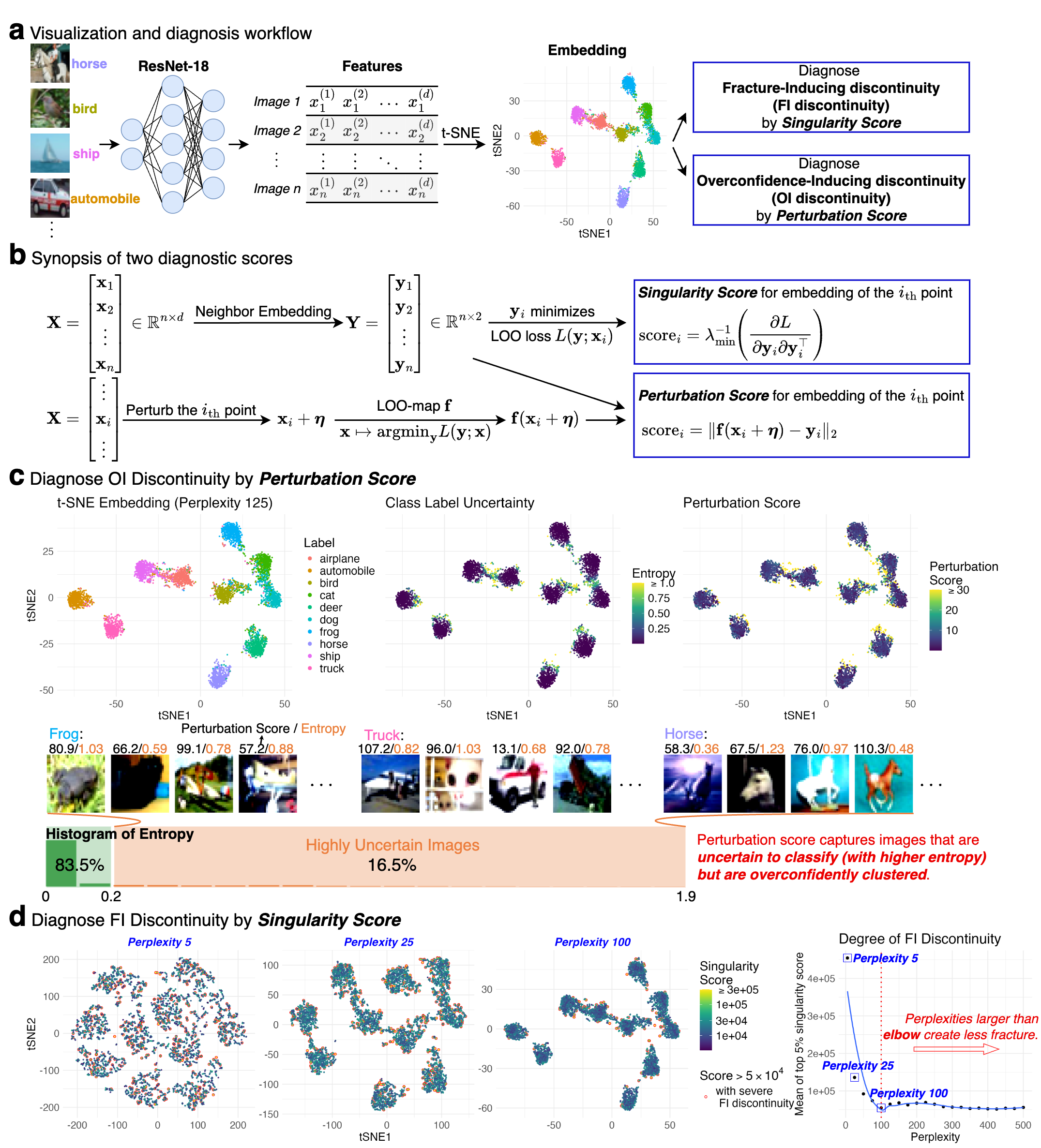}
    \caption{\textbf{Overview: assessment of embeddings generated by neighbor embedding methods, illustrated with image data. }\textbf{a} We use a standard pre-trained convolutional neural network (CNN) to obtain features of image samples from the CIFAR10 dataset, and then 
    visualize the features using a neighbor embedding method, specifically t-SNE. \textbf{b} Basic ideas of singularity scores and perturbation scores. 
    \textbf{c} 
    t-SNE tends to embed image features into separated clusters even for images with ambiguous semantic meanings (as quantified by higher entropies of predicted class probabilities by the CNN). Perturbation scores identify the embedding points that have ambiguous class membership but less visual uncertainty. 
    \textbf{d} An incorrect choice of perplexity leads to visual fractures (FI discontinuity), which is more severe with a smaller perplexity.
    We recommend choosing the perplexity no smaller than the elbow point.}
    \label{fig: overview}
\end{figure}

We provide an overview of LOO-map and demonstrate the proposed two diagnostic scores (Fig.~\ref{fig: overview}).


First, we introduce a general strategy to discern and analyze discontinuities in neighbor embedding methods (e.g., t-SNE, UMAP). Given input points $\vx_1, \ldots, \vx_n$ in a potentially high-dimensional space, e.g., attribute vectors or feature vectors, an embedding algorithm $\gA$ maps them to 2D points $\vy_1, \ldots, \vy_n$ by solving an optimization problem involving $O(n^2)$ pairwise interaction terms. The LOO strategy assumes no dominant interaction term so that perturbing any single input point has negligible effects on the overall embedding. We extensively verify this assumption on simulated and real datasets (Table~\ref{tab: LOO epsilon (small)}, Supplementary Table~\ref{tab: LOO Epsilon}, Methods). By adding a new input $\vx$ and optimizing its corresponding $\vy$ while freezing $(\vy_j)_{j\le n}$, LOO-map reduces the optimization problem to only $O(n)$ effective interaction terms. We identify the discontinuity points of $\vf(\vx)$ as the source of the observed distortions and artifacts.





Then, we devise two label-free point-wise diagnostic scores to quantitatively assess embedding quality (Fig.~\ref{fig: overview}a). The first quantity, namely the perturbation score, quantifies how much an embedding point $\vy_i$ moves when the input $\vx_i$ is moderately perturbed, which helps to probe the discontinuity of $\vf (\vx)$ from the input space. The second quantity, namely the singularity score, measures how sensitive an embedding point is to an infinitesimal input perturbation, thus providing insights into $\vf (\vx)$ at each specific location $\vx=\vx_i$. The two scores, as we will show below, are motivated by different considerations and reveal qualitatively distinct features of the visualizations (Fig.~\ref{fig: overview}b-d).



Finally, we demonstrate how our scores can improve the reliability of neighbor embedding methods. Following the workflow in Fig.~\ref{fig: overview}a, we extract high-dimensional features of image data using a deep learning model (e.g., ResNet-18 \cite{resnet}) and apply t-SNE for the 2D embedding. We observe that some inputs with ambiguous (mixed) class membership are misleadingly embedded into well-separated clusters (Fig.~\ref{fig: overview}c), creating overconfidence in the cluster structure. Ground-truth labels and label-informed entropy scores confirm that the visualization under-represents the uncertainty for mixed points, making them appear more distinct than they should be (Fig.~\ref{fig: overview}c). Further examination of image examples confirms such an artifact of reduced uncertainty in the embedding space. As a diagnosis, we find that embedding points with high perturbation scores correlate well with such observed (OI) discontinuity. 



Our second diagnostic score can help with hyperparameter selection. A practical challenge of interpreting t-SNE embeddings is that the results may be sensitive to tuning parameters. In fact, we find that a small perplexity tends to induce small spurious structures similar to fractures visually speaking, suggesting the presence of local (FI) discontinuity in the LOO-map $\vf$ (Fig.~\ref{fig: overview}d). Our singularity score captures such FI discontinuity as more high-scoring points emerge under smaller perplexities. With this diagnosis, we recommend choosing a perplexity no smaller than the elbow point of the FI discontinuity curve.

\subsection*{Leave-one-out as a general diagnosis technique}\label{sec:loo}
We start with a generic setup for neighbor embedding methods that encompasses SNE \cite{HintonSNE2002}, t-SNE \cite{HintonTSNE2008}, UMAP \cite{mcinnesUMAP2018}, LargeVis \cite{largevis}, PaCMAP \cite{WangPacMap2021}, among others. First, we introduce basic mathematical concepts and their interpretations.
\begin{itemize}
    \item Input data matrix $\mX = [\vx_1,\ldots,\vx_n]^\top\in\R^{n\times d}$: the input data to be visualized. Dimension $d$ may be large (e.g., thousands).
    \item Embedding matrix $\mY = [\vy_1,\ldots,\vy_n]^\top\in\R^{n\times p}$: the embedding points we aim to determine for visualization, where $p$ can be 2 or 3.
    \item (Pairwise) similarity scores $(v_{i,j})_{i<j}$: a measure of how close two input points are in the input space, often calculated based on a Gaussian kernel.
    \item (Pairwise) embedding similarity scores $(w_{i,j})_{i<j}$: a measure of how close two embedding points are, which takes the form of a heavy-tailed kernel (e.g., t-distribution). The computation often requires a normalization step.
    \item (Pairwise) loss function $\mathcal{L}$: a measure of discrepancy between $v_{i,j}$ and $w_{i,j}$. An NE method minimizes this loss function over embedding points to preserve local neighborhood structures. The algorithms of NE methods aim to find the embedding $\mY$ by minimizing the total loss composed of the sum of the divergences between $v_{i,j}$ and $w_{i,j}$ of all pairs of points and a normalization factor $Z(\mY)$.
\end{itemize}
For convenience, We introduce a generic optimization problem that neighbor embedding methods aim to solve as follows:
\begin{equation}
\min_{\vy_1, \ldots, \vy_n \in \R^2} \sum_{1\le i<j\le n} \underbrace{\mathcal{L}(w(\vy_i, \vy_j); v_{i,j}(\mX))}_{\text{unnormalized pairwise loss}} + \underbrace{Z(\mY)}_{\text{normalization factor}}.
\label{opt:neighbor}
\end{equation}
Particularly, in the t-SNE algorithm (see Supplementary Methods~\ref{sec: method: derivation of sscore} for other algorithms), we have, 
\begin{align}
& \mathcal{L}(w_{i,j}; v_{i,j}) = -2 v_{i,j} \log(w_{i,j}), \notag \\
& w_{i,j} = w(\vy_i, \vy_j) = (1+\|\vy_{i}-\vy_{j}\|^2)^{-1}, \quad
Z(\mY) = \log\Big(\sum_{k\neq l}w(\vy_k, \vy_l)\Big). \label{def:tsne-w}
\end{align}
A fundamental challenge of assessing the embeddings is that we only know how \textit{discrete points}---not the \textit{input space}---are mapped since the optimization problem is solved numerically by a complicated algorithm. Consequently, it is unclear if underlying structures (e.g., clusters, low-dimensional manifolds) in the input space are faithfully preserved in the embedding space.


Consider adding a new point $\vx$ to existing data points. We may wish to fix $\vx_1,\ldots,\vx_n$ and analyze how embedding points $\gA(\vx_1,\ldots, \vx_{n}, \vx)$ change as we vary $\vx$, thereby quantifying the mapping of $\vx$ under $\gA$. However, the embedding points would depend on all $n+1$ input points, and require re-running the neighbor embedding algorithm for each new $\vx$. 

\begin{figure}[t]
    \centering
    \includegraphics[scale = 0.9]{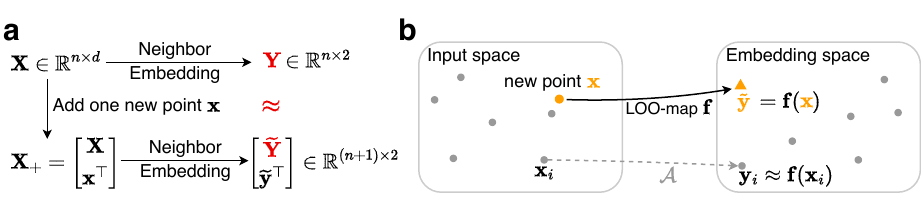}
    \caption{\textbf{Diagrams showing the idea of Leave-one-out (LOO) and LOO-map.} \textbf{a} Idea of LOO. Adding one input point does not significantly change the overall positions of embedding points. The assumption allows us to analyze the properties of the embedding map over the entire input space via an approximated loss which we call LOO loss. \textbf{b} We introduce a global embedding map (LOO-map) $\vf(\vx) = \argmin_{\vy}L(\vy;\vx)$ defined in the entire input space as an approximation to the neighbor embedding method $\mathcal{A}$.}
    \label{LOO diagram}
\end{figure}

\paragraph{LOO loss function and LOO-map.} LOO is a generic decoupling technique that allows us to isolate the changes of one embedding point versus the others \cite{jacknife1, crossvalidation, crossvalidation2, sample-reuse-method, Wahba1978,Wahba1979}. We introduce the LOO assumption, the LOO loss function, and LOO-map as follows.
\begin{itemize}
    \item LOO assumption: adding (or deleting/modifying) a single input point does not change embedding points significantly (Figure~\ref{LOO diagram}a).
    \item LOO loss function $L(\vy; \vx)$: it consists of $n$ pairwise loss terms relevant to the newly added point $\vx$. We aim to determine the embedding $\vy$ for $\vx$ (Fig.~\ref{LOO diagram}b).
    \item LOO-map $\vf$: it is defined as $\vf: \vx \mapsto \argmin_{\vy} L(\vy; \vx)$ for all possible input $\vx$ (Fig.~\ref{LOO diagram}b). This definition allows us to examine the map property in the entire region. 
\end{itemize}
Rooted in the \textit{stability} idea \cite{leobreiman1996,leobreiman2001,stability}, LOO assumes that adding (or deleting/modifying) a single input point does not change embedding points significantly (Fig.~\ref{LOO diagram}a). This assumption allows us to study the map $\vx \mapsto \gA(\vx_1,\ldots, \vx_{n}, \vx)$ approximately. Consider the optimization problem in Equation~\ref{opt:neighbor} with $n+1$ inputs points $\vx_1,\ldots,\vx_n, \vx$. Under the LOO assumption, when adding the new $(n+1)$-th input point $\vx$, we can freeze the embedding matrix $\mY = [\vy_1,\ldots,\vy_n]^\top$ and allow only one free variable $\vy$ in the optimization problem. More precisely, the mathematical formulation of LOO loss function is given by
\begin{equation}\label{def:loo}
L(\vy; \vx) = \sum_{1\le i \le n} \mathcal{L}\Big(w(\vy_i, \vy); v_{i,n+1}\big(\begin{bmatrix}\mX \\  \vx^\top \end{bmatrix}\big)\Big) 
+ Z\Big(\begin{bmatrix}\mY \\ \vy^\top \end{bmatrix}\Big)  \; .
\end{equation}
The LOO loss is motivated by the following observation: suppose $\begin{bmatrix}\widetilde{\mY} \\ \tilde\vy^\top \end{bmatrix}$ is the embedding of $\mX_+ = \begin{bmatrix}\mX \\  \vx^\top \end{bmatrix}$, i.e., it reaches the minimum of the original loss, then $\vy = \tilde\vy$ is necessarily the minimizer of a partial loss involving the embedding point of the point $\vx$:
\begin{equation*}
    \begin{aligned}
        \tilde\vy 
        &= \argmin_{\vy \in \R^2} \sum_{1\le i \le n} \mathcal{L}\Big(w(\tilde\vy_i, \vy); v_{i,n+1}\big(\begin{bmatrix}\mX \\  \vx^\top \end{bmatrix}\big)\Big) 
+ Z\Big(\begin{bmatrix}\widetilde\mY \\ \vy^\top \end{bmatrix}\Big) \\
&\approx \argmin_{\vy \in \R^2} \sum_{1\le i \le n} \mathcal{L}\Big(w(\vy_i, \vy); v_{i,n+1}\big(\begin{bmatrix}\mX \\  \vx^\top \end{bmatrix}\big)\Big) 
+ Z\Big(\begin{bmatrix}\mY \\ \vy^\top \end{bmatrix}\Big)
    \end{aligned}
\end{equation*}
where the approximation is based on the LOO assumption $\widetilde{\mY} \approx \mY$. This approximation allows us to decouple the dependence of $\tilde \vy_i$ on $\vx$. We then define the LOO-map as $\vf: \vx \mapsto \argmin_{\vy} L(\vy; \vx)$.

\paragraph{Empirical validation of the LOO assumption.} We empirically validate the LOO assumption by showing that $\mY$ and $\widetilde \mY$ are very close for a large sample size $n$. Define the normalized error between embeddings before and after deleting a data point by
\begin{equation}\label{def:error}
    \epsilon_n = \frac{1}{\|\mY\|_F}\|\mY-\widetilde \mY\|_F,
\end{equation}
where $\| \cdot \|_F$ means the Frobenius norm of a matrix.
A sufficiently small $\epsilon_n$ will support the approximation in our derivation of the LOO-map.

We calculate this error extensively on both simulated and real datasets (Methods). The results support our LOO assumption (Table~\ref{tab: LOO epsilon (small)}, Supplementary Table~\ref{tab: LOO Epsilon}).
We observe that the approximation errors are small and generally decreasing in $n$, which validates our LOO assumption. 




\renewcommand{\arraystretch}{1.2}

\definecolor{row1}{RGB}{222,220,204}
\definecolor{row2}{RGB}{252,251,247}
\definecolor{row3}{RGB}{241,239,230}

\begin{table}[ht]
\centering
\small
\begin{tabular}{p{3.6cm}p{3.6cm}p{3.6cm}p{2.25cm}}
\hline
\rowcolor{row1} \textbf{Number of points} & $\boldsymbol{n=1000}$          & $\boldsymbol{n=3000}$          & $\boldsymbol{n=5000}$          \\ \hline
\rowcolor{row2} \textbf{2-GMM}            & $0.068\ (0.0017)$ & $0.044\ (0.0018)$ & $0.034\ (0.0007)$ \\ \hline
\rowcolor{row3} \textbf{Swissroll}        & $0.074\ (0.0110)$ & $0.043\ (0.0041)$ & $0.033\ (0.0014)$ \\ \hline
\rowcolor{row2} \textbf{CIFAR10}          & $0.042\ (0.0081)$ & $0.044\ (0.0013)$ & $0.039\ (0.0006)$ \\ \hline
\rowcolor{row3} \textbf{IFNB}             & $0.069\ (0.0022)$ & $0.049\ (0.0019)$ & $0.044\ (0.0010)$ \\ \hline
\end{tabular}
\caption{\textbf{Empirical validation of LOO on both simulated datasets and real datasets.} We measured the approximation error $\epsilon_n$ defined in Equation~\ref{def:error} across 20 independent trials and reported the average (and the standard deviation) of $\epsilon_n$. We find that the errors are small and generally decreasing in $n$, which supports our LOO assumption.
}
\label{tab: LOO epsilon (small)}
\end{table}

\subsection*{LOO-map reveals intrinsic map discontinuities} \label{sec:instability}


\begin{figure}[htbp]
    \centering 
    \includegraphics[scale = 0.9]{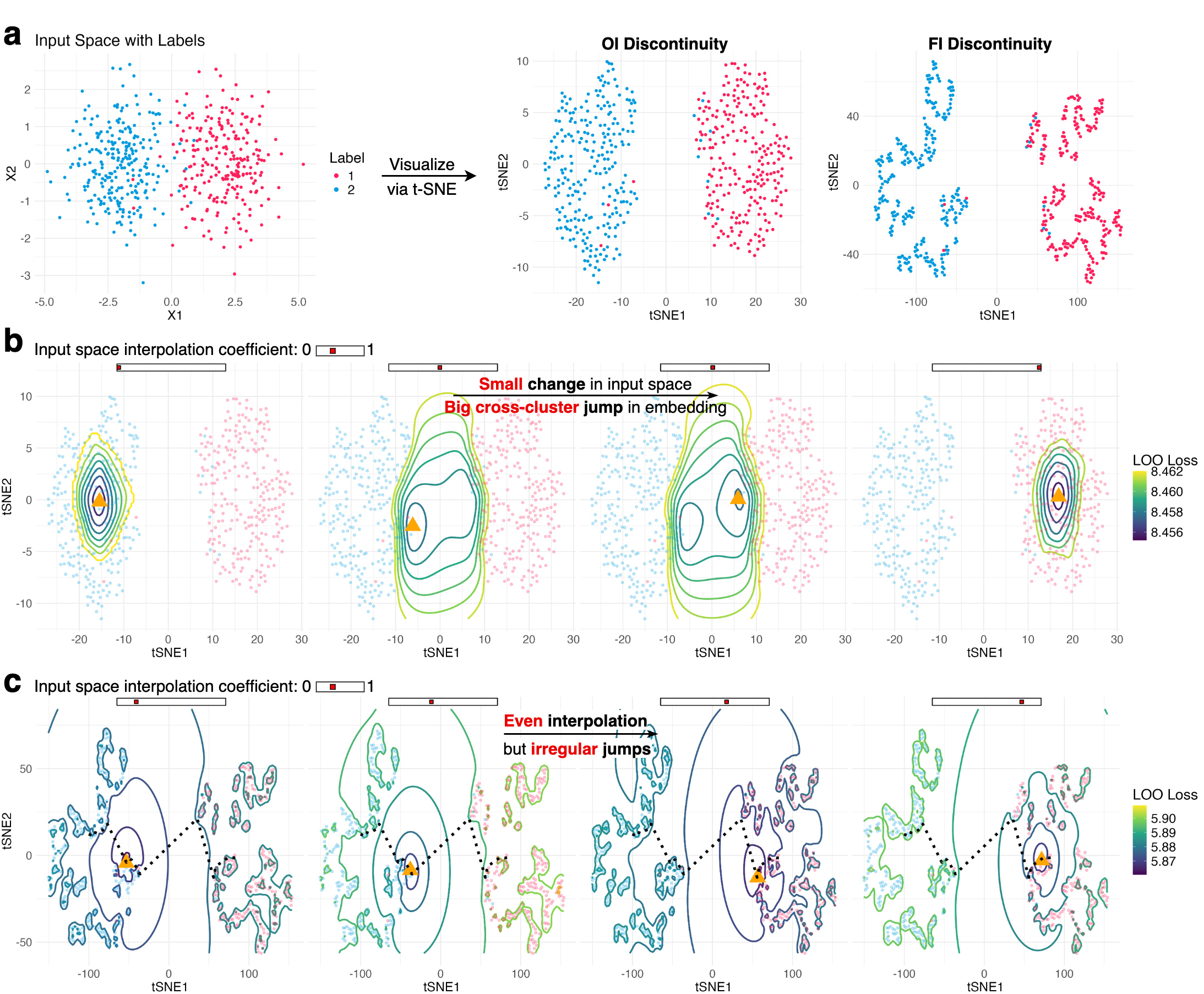}
    \caption{
    \textbf{LOO loss landscape reveals the origins of two distortion patterns.} \textbf{a} We illustrate two discontinuity patterns on simulated Gaussian mixture data. \textit{OI discontinuity:} t-SNE embeds points into well-separated clusters and creates visual overconfidence. \textit{FI discontinuity:} t-SNE with an inappropriate perplexity creates many artificial fractures.
    \textbf{b} Origin of OI discontinuity: LOO loss contour plot shows distantly separated minima. 
    We add a new input point $\vx$ at one of the $4$ interpolated locations $\vx = t\vc_1 + (1-t)\vc_2$ where $t\in\{0,0.47,0.48,1\}$ and then visualize the landscape of the LOO loss $L(\vy; \vx)$ using contour plots in the space of $\vy$.
    The middle two plots exhibit two well-separated minima (orange triangle), which cause a huge jump of the embedding point (as the minimizer of the LOO loss) under a small perturbation of $\vx$.
    \textbf{c} Origin of FI discontinuity: We show LOO loss contour plots with interpolation coefficient $t\in\{0.2,0.4,0.6,0.8\}$. The plots show many local minima and irregular jumps. Under an inappropriate perplexity, the loss landscape is consistently fractured.
    Numerous local minima cause an uneven trajectory of embedding points (dashed line) when adding $\vx$ at evenly interpolated locations.
    }
    \label{fig: discontinuity examples}
\end{figure}




By analyzing the LOO loss, we identify the two observed discontinuity patterns as a result of the map discontinuities of $\vf(\vx)$. We use t-SNE as an example to illustrate the main results. 


We generate mixture data by sampling 500 points from two overlapping 2D Gaussian distributions and run t-SNE with two representative choices of perplexity, 5 and 50. The resulting visualizations confirm the two discontinuity patterns (Fig.~\ref{fig: discontinuity examples}a). OI discontinuity pushes mixed points to cluster boundaries, creating overly tight structures, while FI discontinuity fragments embeddings into small pieces, leading to many sub-clusters. Similar discontinuity patterns are also common among other neighbor embedding methods (Supplementary Fig.~\ref{fig: appendix: umapdiscontinuity}).

We trace the origins of the observed discontinuity patterns by the LOO loss function. To this end, we add a single point $\vx$ at varying locations to the input data and track how $\vx$ is mapped. By visualizing the landscape of the LOO loss $L(\vy; \vx)$ at four different inputs $\vx$, we provide snapshots of the LOO-map $\vx \mapsto \argmin_{\vy}L(\vy; \vx)$. More specifically, we choose the centers $\vc_1, \vc_2$ of the two Gaussian distributions and consider the interpolated input $\vx(t) = t \vc_1 + (1-t)\vc_2$, $t \in [0,1]$. Since $\vx(t)$ is mapped to the LOO loss minimizer, tracking the loss minima reveals the trajectory of the corresponding embedding point $\vy(t)$ under varying $t$.


We find that the observed OI discontinuity is caused by a discontinuity point of $\vf(\vx)$ in the midpoint of two mixtures. To demonstrate this, we visualize the LOO loss landscape and the embedding of the added point $\vx(t)$ at four interpolated locations where $t \in \{0, 0.47, 0.48, 1\}$. There are two clearly well-separated minima in the LOO loss landscape when $t \approx 0.5$ (Fig.~\ref{fig: discontinuity examples}b). As a result, the embedding point $\vy(t)$ jumps abruptly between local minima with a slight change in $t$. A further gradient field analysis shows a hyperbolic geometry around the discontinuity point of $\vf(\vx)$ (Fig.~\ref{fig: appendix: 2_10_gradient_field}).


We also find that the FI discontinuity is caused by numerous irregular local minima of $L(\vy;\vx)$ under an inappropriate choice of perplexity. This conclusion is supported by the observation that the loss landscape of $L(\vy; \vx)$ is consistently irregular and contains many local valleys under a small perplexity (Fig.~\ref{fig: discontinuity examples}c). Moreover, varying the interpolation coefficient $t$ from 0 to 1 at a constant speed results in an uneven trajectory of the embedding point $\vy(t)$. Because of many irregularities, the embedding points tend to get stuck at these local minima, thus forming spurious sub-clusters. In addition, we find that larger perplexity typically lessens FI discontinuity (Supplementary Fig.~\ref{fig: appendix: 2_3}, \ref{fig: appendix: use case 2}).
\subsection*{LOO-map motivates diagnostic scores for capturing topological changes}



OI discontinuity and FI discontinuity reflect the properties of $\vf(\vx)$ at different levels: OI discontinuity is relatively global, while FI discontinuity is relatively local. To quantify their severity respectively, we introduce two point-wise scores (Methods): (i) perturbation scores for OI discontinuity and (ii) singularity scores for FI discontinuity. For computational efficiency, both scores are based on modifying individual input points instead of adding a new point so that we maintain $n$ data points in total. This approach is justified by the LOO assumption, allowing using the partial loss as LOO loss.

Briefly speaking, we define the perturbation score as the amount of change of an embedding point $\vy_i$ under the perturbation of an input point $\vx_i$ of a moderate length. As the data distribution is not known a priori, we search the perturbation directions using the top principal directions of the data (Methods). 

We define the singularity score as the inverse of the smallest eigenvalue of a Hessian matrix that represents the sensitivity of the embedding point $\vy_i$ under infinitesimal perturbations. Our derivation (Supplementary Methods~\ref{sec: appendix: derivation of sscore}) reveals that small eigenvalues can produce substantial local discontinuities, whereas a singular Hessian matrix leads to the most severe discontinuity. We find that infinitesimal perturbations are particularly effective for capturing the local characteristics of FI discontinuities. Detailed expressions for the singularity scores of t-SNE, UMAP and LargeVis are provided in Supplementary Methods~\ref{sec: method: derivation of sscore}.

Generally, we recommend using the perturbation score to diagnose the trustworthiness of cluster structures, and the singular score to detect spurious local structures.

\subsection*{Simulation studies}

\begin{figure}[htbp]
    \centering
    \includegraphics[scale = 0.9]{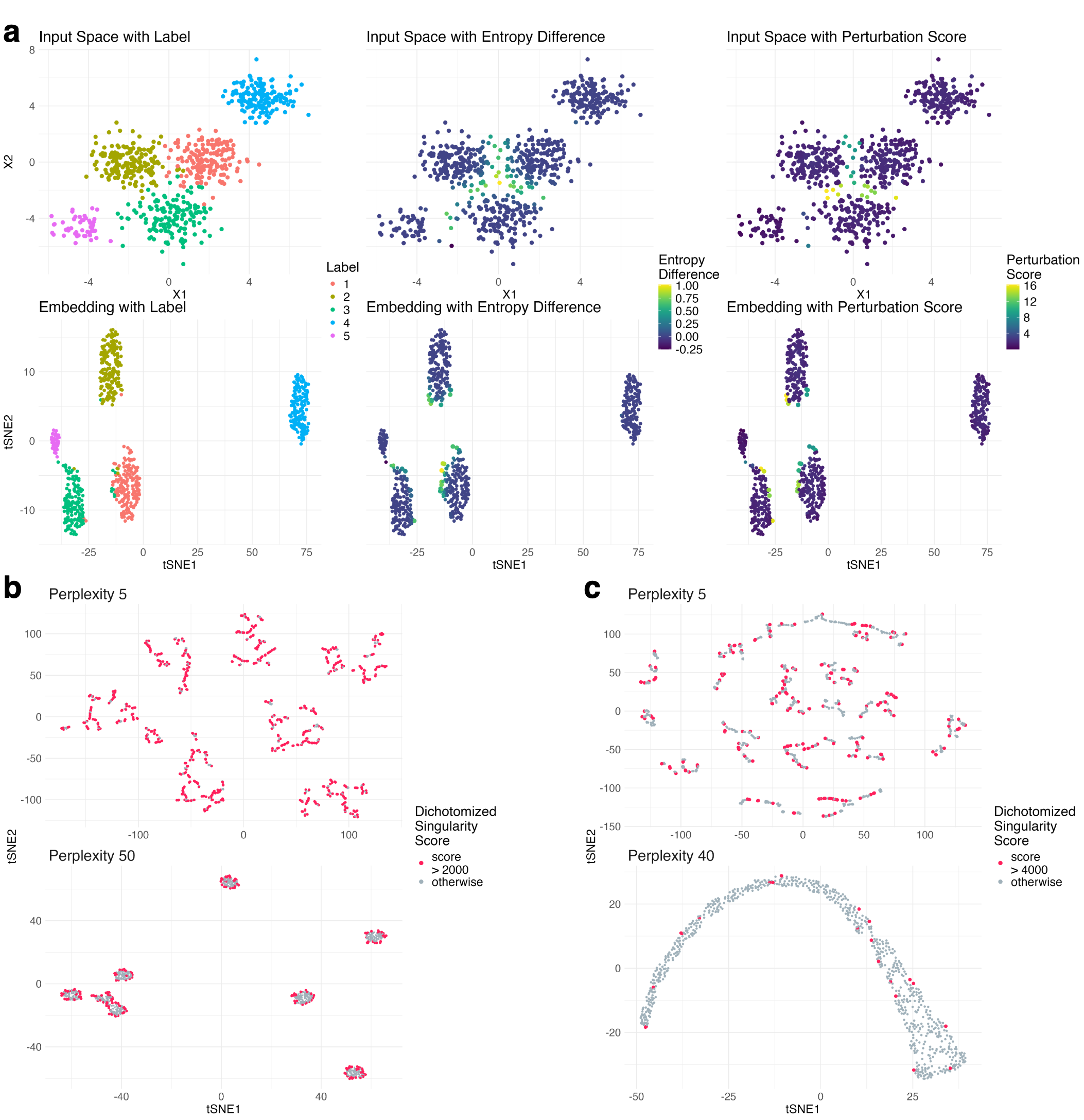}
    \caption{\textbf{Simulation studies demonstrate the effectiveness of proposed scores.} \textbf{a} Perturbation scores identify unreliable embedding points that have reduced uncertainty. Input points from 5-component Gaussian mixture data form separated clusters in the embedding space. t-SNE reduces perceived uncertainty for input points in the overlapping region (left), as captured by the label-dependent measurements, namely the entropy difference (middle). Our perturbation scores can identify the same unreliable embedding points without label information (right). \textbf{b-c} Singularity scores reveal spurious sub-clusters on Gaussian mixture data (\textbf{b}) and Swiss roll data (\textbf{c}). At a low perplexity, t-SNE creates many spurious sub-clusters. Embedding points receiving high singular scores at random locations are an indication of such spurious structures.}
    \label{fig: simulatedstudy}
\end{figure}


We implement our proposed point-wise scores for t-SNE as an example. We evaluate our diagnostic scores on two types of simulated datasets (Methods): (i) 2D Gaussian mixture data with 5 centers (unequal mixture probabilities, $n=700$) and 8 centers (equal probabilities, $n=800$), and (ii) Swiss roll data, where $n=800$ points are sampled from a 3D Swiss-roll manifold.


We apply perturbation scores to the 5-component Gaussian mixture data, where t-SNE creates misleadingly distinct cluster boundaries (Fig.~\ref{fig: simulatedstudy}a left). Without label information, our scores identify unreliable points with deceptively low uncertainty (Fig.~\ref{fig: simulatedstudy}a right). Meanwhile, the entropy differences use the ground-truth labels to calculate the reduced class entropies (Methods) in the embedding space, thus providing an objective evaluation of the degree of confidence (Fig.~\ref{fig: simulatedstudy}a middle). Our perturbation scores are closely aligned with the entropy differences.


Next, we apply singularity scores to the 8-component Gaussian mixture and Swiss roll data under two perplexity settings (Fig.~\ref{fig: simulatedstudy}b-c). Each embedding is colored by ground-truth labels, singularity scores, and dichotomized singularity scores (binary thresholding). The embeddings differ visually: a low perplexity creates spurious sub-clusters, while a high perplexity preserves cluster and manifold structures. Additionally, the distributions of dichotomized scores vary: a low perplexity results in more high scores at randomly scattered locations, whereas a high perplexity yields fewer high-scoring points.


Moreover, we quantitatively assess the clustering quality for the 8-component Gaussian mixture data using three indices: DB index \cite{DBIndex}, within-cluster distance ratio (Methods), and Wilks' $\Lambda$ \cite{WilksLambda}. All three indices (small values are better) indicate that t-SNE visualizations with less severe FI discontinuity, i.e., lower singularity scores, achieve better clustering quality, with the DB index dropping from 0.5982 to 0.3038, the within-cluster distance ratio from 0.0480 to 0.0024, and Wilks' $\Lambda$ from 0.0028 to $9.0\times 10^{-6}$. To further study the change in clustering quality, we generate 6 simulated datasets with varying cluster structures and dimensions. Across all datasets, tuning perplexity using singularity scores consistently improves clustering quality, reducing the DB index by approximately 50\%, the within-cluster distance ratio by 65\%-91\%, and Wilks' $\Lambda$ by 57\%-99\% (Supplementary Table~\ref{tab: simulatedindex}).

\subsection*{Use case 1: detecting out-of-distribution image data} \label{use1}
\label{sec:OOD Detection}

One common practical issue for statistical methods or machine learning algorithms is the \textit{distribution shift}, where the training dataset and test dataset have different distributions, often because they are collected at different sources \cite{OODsurvey2024,wilds2021, OODhospital2018}. These test data are called out-of-distribution (OOD) data.


In this case study, we identify one rarely recognized pitfall of t-SNE visualization: OOD data may become harder to discern in t-SNE embeddings because they tend to be absorbed into other clusters. Our perturbation score is able to identify the misplaced OOD embedding points.

\begin{figure}[htbp]
    \centering
    \includegraphics[scale = 0.9]{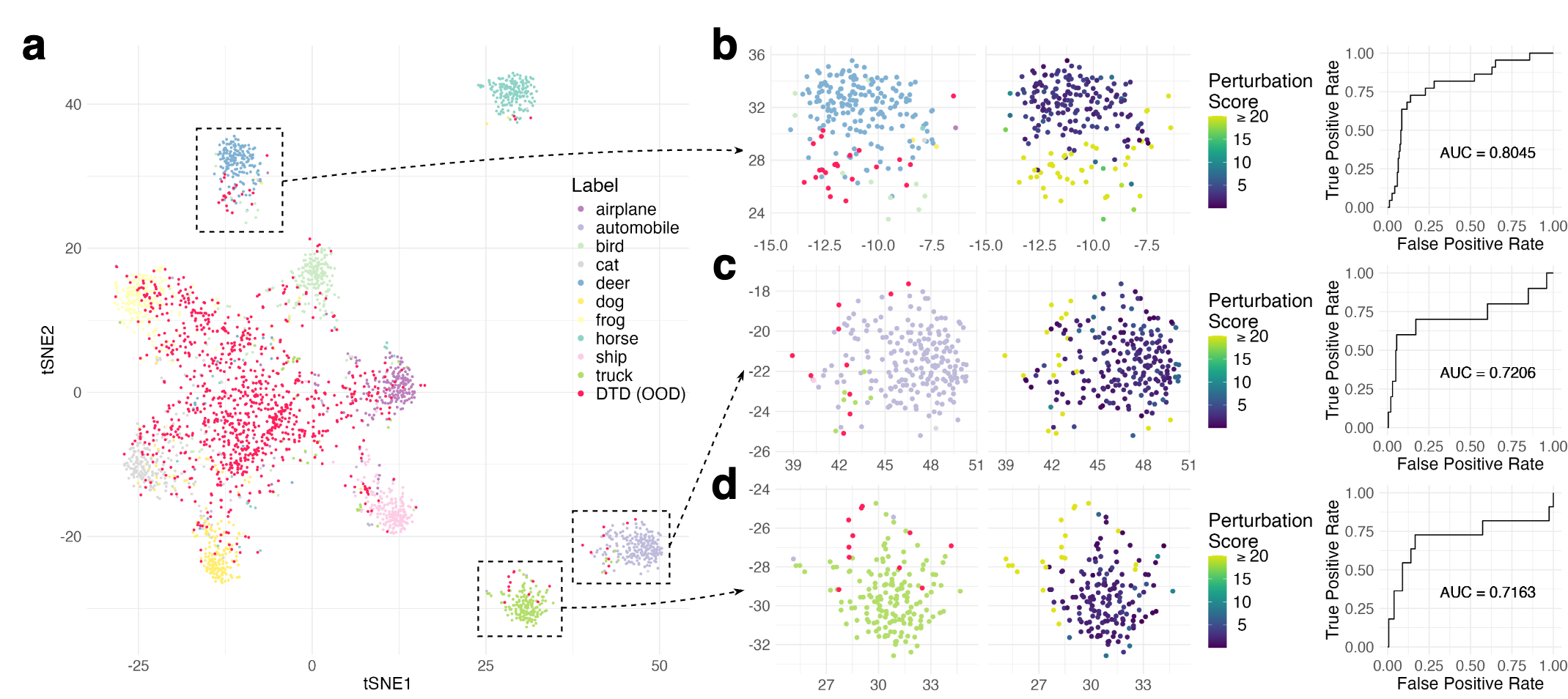}
    \caption{\textbf{Perturbation scores detect out-of-distribution (OOD) image data.} \textbf{a} 
    We use a pretrained ResNet-18 model to extract features of CIFAR-10 images and, as out-of-distribution data, of DTD texture images. Then we visualize the features using t-SNE with perplexity 100. A fraction of OOD embedding points are absorbed into clusters that represent CIFAR-10 image categories such as deer, truck, and automobile.
    \textbf{b-d} Perturbation scores can effectively identify misplaced out-of-distribution data points. The ROC curves show the proportion of OOD points correctly identified by the perturbation scores. 
    }
    \label{fig:CIFAR10ood}
\end{figure}

We use a standard ResNet-18 model \cite{resnet} trained on the CIFAR-10 dataset \cite{KrizhevskyCIFAR102009} to extract features of its test dataset and an OOD dataset known as DTD (describable textures dataset) \cite{CimpoiDTD2014}. Ideally, visualization of the features of test images and OOD images would reveal the distribution shift. However, the t-SNE embedding shows that a fraction of OOD features are absorbed into compact, well-defined CIFAR-10 clusters (Fig.~\ref{fig:CIFAR10ood}a). Without the label information, one may mistakenly assume that the misplaced OOD embedding points belong to the regular and well-separated classes in CIFAR-10. We find that the embedding misplacement results from OI discontinuity. Our inspection of the original feature space shows that the misplaced OOD data points appear to have mixed membership, resembling both CIFAR-10 and OOD data---thus their cluster membership is, in fact, less certain than what t-SNE suggests.

Our perturbation scores can successfully identify most of these misplaced OOD embedding points (Fig.~\ref{fig:CIFAR10ood}b-d). The areas under the ROC curves (AUROC) are on average 0.75 for the three selected clusters. Compared with other methods aiming for OOD detection, our perturbation score demonstrates superior performance, with kernel PCA \cite{kernelPCA} achieving an average AUROC of 0.698 and the one-class support vector machine \cite{OCSVM} achieving an average AUROC of 0.410 (Supplementary Fig.~\ref{fig: appendix: compare with kpca and ocsvm}, Methods). Additionally, we use the prediction probabilities given by the neural network to calculate the entropy of each point, and find that the entropies significantly correlate with the perturbation scores; specifically, the correlations are 0.49, 0.58, and 0.64 for the selected clusters. These findings suggest that perturbation scores are effective in detecting OOD data and can help safeguard against misinterpretation of t-SNE visualizations.



\subsection*{Use case 2: enhancing interpretation of single-cell data} \label{use2}





Our second example concerns the application of singularity scores in single-cell data. In this case study, we investigate how incorrect choices of perplexity induce spurious sub-clusters. We also provide a guide of choosing perplexity based on singularity scores, thereby reducing such spurious sub-clusters.



\begin{figure}[htbp]
    \centering
    \includegraphics[scale = 0.9]{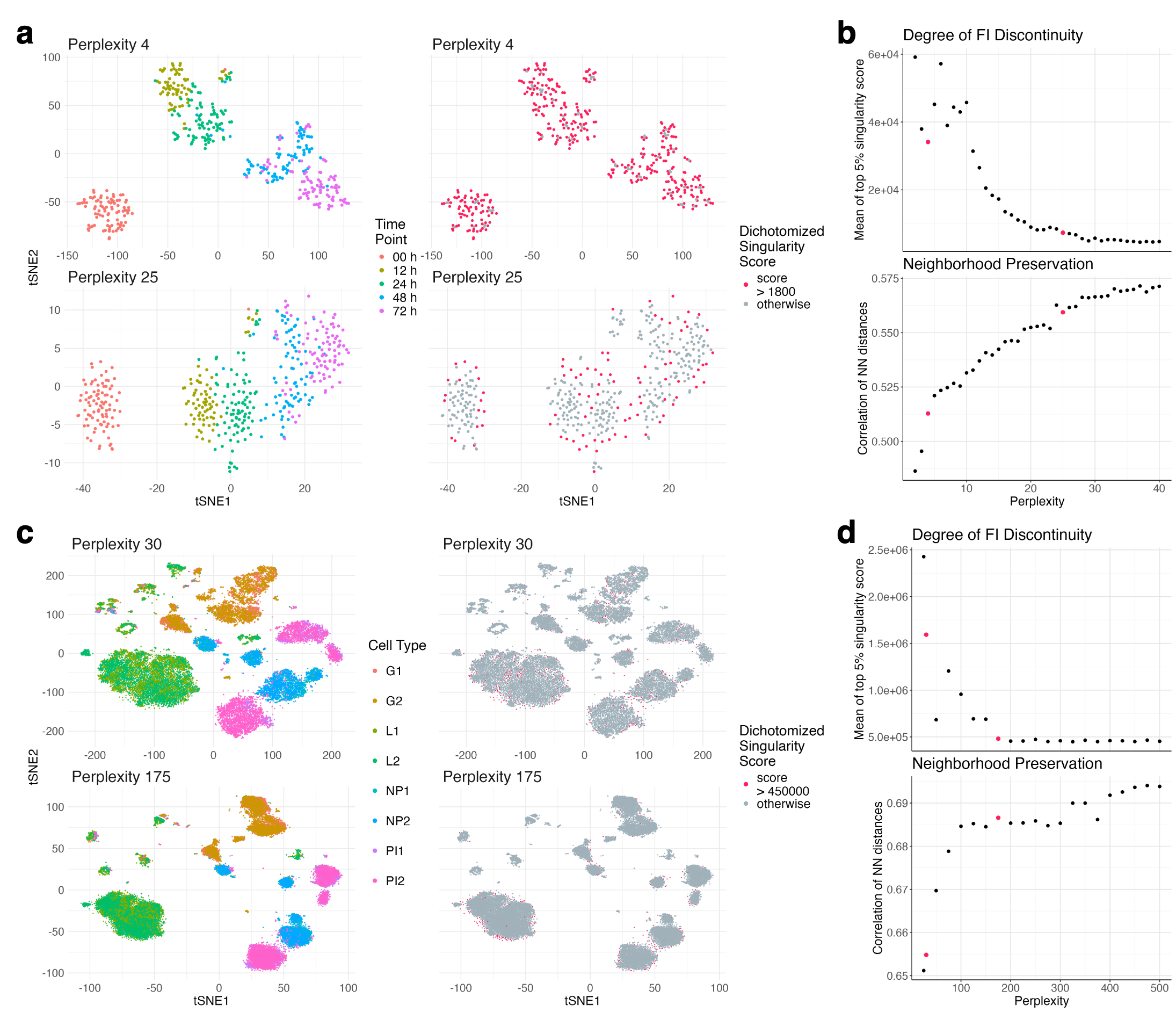}
    \caption{\textbf{Singularity scores inform the selection of the perplexity parameter.} 
    Comparative t-SNE embeddings and the corresponding singularity scores at two different perplexities in mouse embryonic cell differentiation data (\textbf{a}) and in mouse mammary epithelial cell data (\textbf{c}). The perplexity as a tuning parameter has a large impact on t-SNE visualization qualitatively. At a small perplexity, there are many spurious sub-clusters. Embeddings with high singular scores appear in random locations, which indicates the presence of such spurious structures and severe FI discontinuity. Plots of the degree of FI discontinuity and neighborhood preservation versus perplexity are shown for mouse embryonic cell differentiation data (\textbf{b}) and for mouse mammary epithelial cell data (\textbf{d}). We recommend choosing a perplexity no smaller than the elbow point, as this ensures that randomly positioned points with high singularity scores largely disappear, remaining only at cluster peripheries. Consequently, the neighborhoods of most points are embedded more faithfully, resulting in better neighborhood preservation score.
    }
    \label{fig:mousebrain}
\end{figure}

The first dataset we examined is single-cell RNA-seq data from 421 mouse embryonic stem cells (mESCs) collected at 5 sampling time points during differentiation \cite{HayashiData}. The second dataset is another single-cell RNA-seq data from 25,806 mouse mammary epithelial cells across 4 developmental stages \cite{Bach2017MammaryData}. We also include our analysis on a mid-sized mouse brain chromatin accessibility data in the Supplementary File.
Through analysis of the datasets, we find that a small perplexity tends to create spurious sub-clusters (Fig.~\ref{fig:mousebrain}a, c, Supplementary Fig.~\ref{fig: appendix: use case 2 Mouse Brain}a). Our singularity scores can provide informative insights into the spurious clusters even without the ground-truth labels, as summarized below.
\begin{enumerate}
    \item (Distribution difference) Embedding points with large singularity scores tend to appear in random and scattered locations if the perplexity is too small. In contrast, under an appropriate perplexity, embedding points with large singular scores are mostly in the periphery of clusters.
    \item (Elbow point) As the perplexity increases, the magnitude of large singular scores (calculated as the average of the top $5\%$) rapidly decreases until the perplexity reaches a threshold.
\end{enumerate}
The distribution of large singular values indicates spurious sub-clusters, reflecting the irregular LOO loss landscape (Supplementary Fig.~\ref{fig: appendix: use case 2}a, c). We extensively validated the distribution difference between small and large perplexities through statistical tests, including Spearman's correlation test between singularity scores and cluster center distances, F-tests, and permutation tests for local regression models (singularity scores regressed against locations). At low perplexities, Spearman’s correlation tests showed non-significant results for all five mESCs clusters and five mammary epithelial cell clusters (average $p$-values: 0.36 and 0.29, Supplementary Table~\ref{tab: mousebraintest}). Increasing perplexities to the singularity score elbow points (Fig.~\ref{fig:mousebrain}b, d) yielded significant correlations in four of five mESCs clusters and five of eight mammary epithelial cell classes. Similarly, F-tests and permutation tests showed $p$-values dropping from $\sim0.3$ to $<0.001$ (mESCs) and $<10^{-13}$ (mammary epithelial cells), confirming the dependence of singularity scores on location at higher perplexities. This transition aligns with LOO loss geometry: low perplexities create scattered local minima, forming spurious sub-clusters (Supplementary Fig.~\ref{fig: appendix: use case 2}a, c), whereas higher perplexities smooth the loss landscape (Supplementary Fig.~\ref{fig: appendix: use case 2}b, d), reducing artifacts.

We also observe that the degree of FI discontinuity, as indicated by the magnitude of the singularity scores, decreases rapidly until the perplexity reaches the elbow point (Fig.~\ref{fig:mousebrain}b, d). Beyond the elbow point, the spurious sub-clusters largely disappear, aligning with the improvement of neighborhood preservation (Fig.~\ref{fig:mousebrain}b, d), as measured by the nearest-neighbor distance correlation between the input and embedding spaces (Methods). However, we would not suggest increasing perplexity excessively, as it may merge clusters \cite{wattenberg2016how}, result in the loss of certain microscopic structures \cite{kobak2019art}, and often lead to longer computational running time \cite{BelkinaAutoOptParaTSNE2019}. Therefore, we suggest choosing a perplexity around the elbow point.

\subsection*{Computational costs}
\paragraph{Perturbation score.} 

Theoretically, the computational complexity for solving the LOO loss optimization takes $O(n)$ flops, instead of $O(n^2)$ flops of the original loss which involves every pairwise interaction term. Practically, our R package has the following running time.
\begin{itemize}
    \item For exact perturbation scores, it takes 35.2 seconds to compute the score per point for the CIFAR-10 images in Fig.~\ref{fig: overview} on a MacBook Air (Apple M2 chip).
    \item Leveraging pre-computed quantities, we also provide an approximation method to reduce the running time per point to 7.1 seconds, while preserving high accuracy relative to the exact score (Supplementary Fig.~\ref{fig: appendix: approx}).
    \item In addition to the approximation, we introduce a pre-screening step to increase the computational efficiency by 14X for the same dataset. This pre-screening step identifies a subset of embedding points most likely to yield high scores, and thus significantly reduces computational cost while still providing a comparable assessment of OI discontinuity locations (Supplementary Fig.~\ref{fig: appendix: prescreen}). Combining the approximation method and the pre-screening step results in an average of 0.47 seconds.
\end{itemize}

\paragraph{Singularity score.} Theoretically, the computational complexity for calculating the singularity scores for the entire dataset is $O(n^2)$ flops, primarily due to matrix operations when calculating Hessian matrices. Practically, the running time for computing singularity scores for CIFAR-10 is 15.9 seconds for all 5,000 points on a MacBook Air (Apple M2 chip).

\subsection*{Comparison with other assessment metrics} \label{compare.sec}
There are multiple recent papers on assessing and improving the reliability of neighbor embedding methods. None of these papers view the observed artifacts as an intrinsic map discontinuity, and as a result, cannot reliably identify topological changes in their proposed diagnosis. For illustration, we compare our method with EMBEDR \cite{EMBEDR}, scDEED \cite{scDEED2023}, and DynamicViz \cite{dynamicviz}.
\begin{itemize}
    \item \textbf{EMBEDR} identifies dubious embedding points by using statistical significance estimates as point-wise reliability scores. This process begins by computing point-wise KL divergences between the kernels in the input and embedding spaces, followed by a permutation test to determine whether the neighborhood preservation is significantly better than random chance. Lower $p$-values from the test indicate higher embedding reliability. EMBEDR selects the perplexity by minimizing the median $p$-values.
    \item \textbf{scDEED} calculates point-wise $p$-values by conducting a similar permutation test on the correlations of nearest-neighbor distances. Similarly, lower $p$-values indicate higher embedding reliability. ScDEED provides two approaches for parameter selection based on dubious embedding points: the first locates the elbow point and the second selects the perplexity to minimize the number of dubious points.
    \item \textbf{DynamicViz} employs a bootstrap approach to assess the stability of embeddings. Point-wise variance scores are constructed based on resampling, defined as the average variance of distances to the neighbors. Embedding points with lower variance scores are considered more reliable. DynamicViz selects the perplexity by minimizing the median variance score.
\end{itemize}



\begin{figure}[ht]
    \centering
    \includegraphics[scale = 0.9]{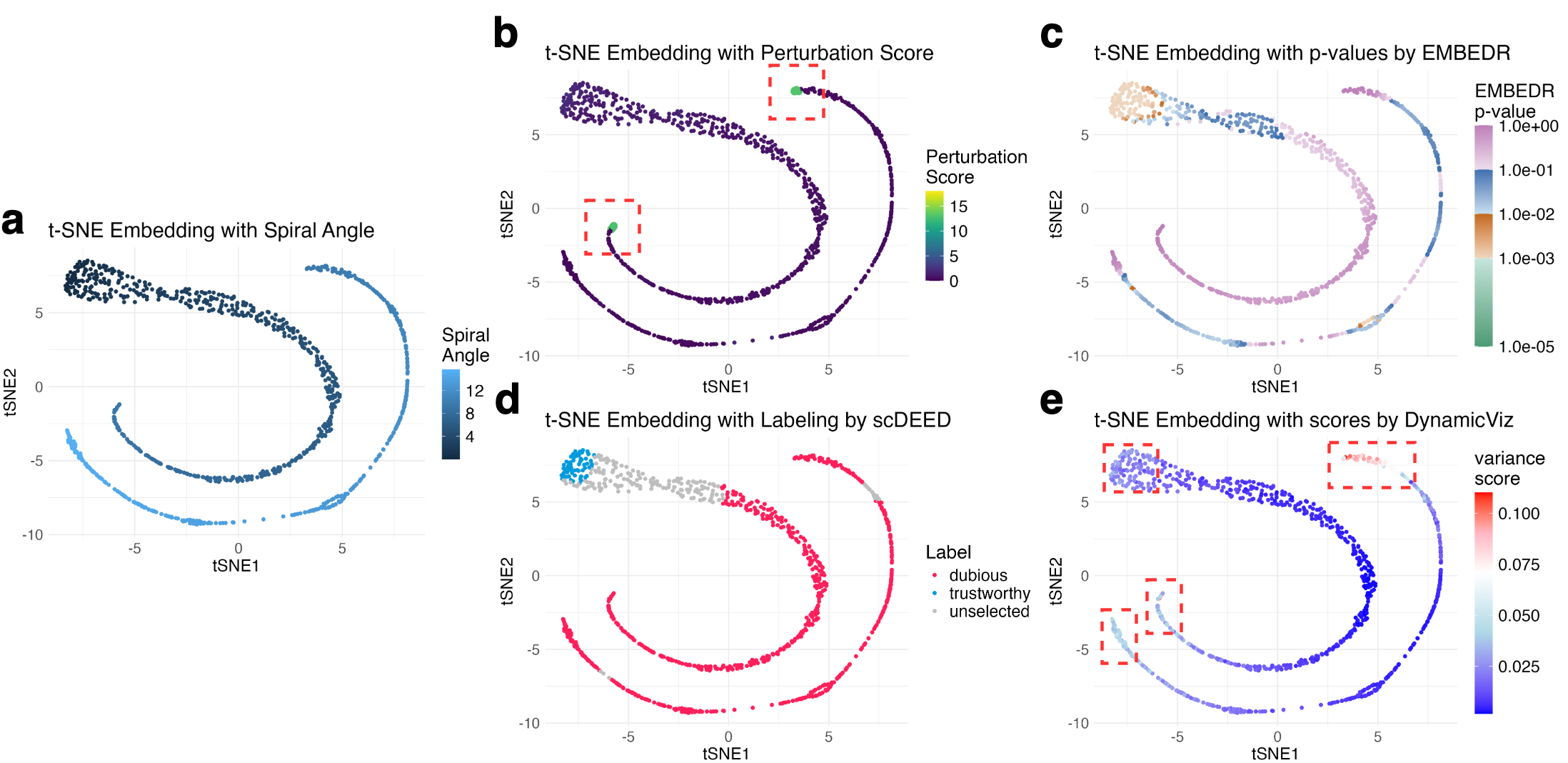}
    \caption{\textbf{Comparing perturbation scores with three diagnostic scores for the t-SNE embedding on the simulated Swiss roll dataset}. \textbf{a} The t-SNE embedding of $n=1000$ simulated points from the Swiss roll manifold under perplexity $150$. The colors correspond to the ground-truth spiral angles of the points. t-SNE algorithm erroneously breaks the smooth manifold into two disconnected parts, which indicates OI discontinuity.
    \textbf{b} Perturbation scores clearly mark the unreliable embedding points where disconnection (discontinuity) occurs.
    \textbf{c} EMBEDR suggests that most embedding points are unreliable (lower $p$-values are more reliable), but it does not identify the discontinuity location.
    \textbf{d} ScDEED evaluates most embedding points as dubious, but similar to EMBEDR, it does not identify the discontinuity location.
    \textbf{e} DynamicViz marks both the discontinuity location and the areas at both ends of the Swiss roll as unstable, making it difficult to distinguish the actual discontinuity locations. Furthermore, while it can roughly identify the discontinuity location, it still fails to pinpoint the exact points where the split occurs.
    }
    \label{fig: methods-compare}
\end{figure}

 \paragraph{Detecting distortion of global structure.} Compared with existing methods, our perturbation scores have the following advantages. First, perturbation scores are better at locating the topological changes of global structures by pinpointing the exact points. By design, they capture embedding points close to the intrinsic discontinuity of the embedding map. In a simulated Swiss roll dataset, t-SNE erroneously splits the smooth manifold into two disconnected pieces (Fig.~\ref{fig: methods-compare}a), which is a severe visualization artifact caused by OI discontinuity. Our perturbation scores accurately highlight unreliable points exactly at the disconnection location (Fig.~\ref{fig: methods-compare}b). In contrast, EMBEDR and scDEED label most points as unreliable, failing to pinpoint the discontinuity (Fig.~\ref{fig: methods-compare}c, d), as they emphasize neighborhood preservation rather than topological changes. DynamicViz identifies the general region but lacks precision (Fig.~\ref{fig: methods-compare}e).


Second, perturbation scores are also robust to low-density regions. In the simulated Gaussian mixture dataset (Supplementary Fig.~\ref{fig: methods-compare-2}a), DynamicViz fails to accurately characterize discontinuity locations in areas with a lower point density, as these areas are prone to insufficient sampling (Supplementary Fig.~\ref{fig: methods-compare-2}c). In contrast, our perturbation scores are more robust to the low-density regions (Supplementary Fig.~\ref{fig: methods-compare-2}b).

\paragraph{Aiding hyperparameter selection.} Compared with existing methods, our singularity score consistently selects a perplexity that is neither too small nor too large, thus reducing sub-clusters yet still producing fine-grained structures. We illustrate the advantage of consistency using three datasets (Supplementary Table~\ref{tab: comparesscore}).

For the mouse embryonic cell differentiation data (Fig.~\ref{fig:mousebrain}a), scDEED recommends two approaches for perplexity selection; the first is based on the elbow point and yields $3$, and the second is based on minimizing the number of dubious points and does not produce a unique value (Supplementary Fig.~\ref{fig: methods-compare-singularityscore-hayashi}a). EMBEDR fails to suggest a valid hyperparameter because we encountered errors potentially due to a small dataset size. DynamicViz and singular scores select moderate perplexity (20 and 25), reducing spurious sub-clusters compared to perplexity of 3 (Supplementary Fig.~\ref{fig: methods-compare-singularityscore-hayashi}b) and achieving the higher neighborhood preservation score (0.5594 (singularity score, highest), 0.4955 (scDEED), 0.5524 (DynamicViz)).

For the mouse brain chromatin accessibility data (Supplementary Fig.~\ref{fig: appendix: use case 2 Mouse Brain}a), scDEED selects 10 (elbow point) and 145 (minimizing dubious points). EMBEDR chooses perplexity of 145, showing a tendency of favoring larger perplexity that is also observed by \cite{scDEED2023}. DynamicViz selects perplexity of 10. Our singularity score selects perplexity of 95 (Supplementary Fig.~\ref{fig: methods-compare-singularityscore-mousebrain}a). By visual inspection, perplexity of 10 is inappropriately small because visualization exhibits numerous spurious sub-clusters. In contrast, perplexities of 95 and 145 avoid spurious sub-clusters while maintaining fine-grained structures (Supplementary Fig.~\ref{fig: methods-compare-singularityscore-mousebrain}b). Quantitatively, the perplexities suggested by singular scores, scDEED and EMBEDR lead to similar neighborhood preservation scores (0.4108, 0.4223, 0.4223). 

In the mouse mammary epithelial cell dataset, similar phenomena are observed: our singularity score selects a balanced perplexity while scDEED and EMBEDR select perplexities that are either too small or too large, and DynamicViz lacks scalability for large datasets due to its bootstrap-based approach, which requires repeated execution of visualization algorithms (Supplementary Fig.~\ref{fig: methods-compare-singularityscore-bachmammary}). Overall, the singularity score offers robust guardrail perplexities that significantly reduce spurious sub-clusters while producing informative visualization.



\subsection*{Theoretical insights: landscape of LOO loss}\label{sec:theory}

By analyzing the LOO loss function in Equation~\ref{def:loo} under a simple setting, we will 
show that OI discontinuity is caused by a hyperbolic saddle point in the LOO loss function, thereby theoretically justifying Fig.~\ref{fig: discontinuity examples}b.

Suppose that $n$ input points $\vx_1,\ldots,\vx_{n}$ are generated from a data mixture with two well-separated and balanced groups, where the first group is represented by the index set $\gI_+ \subset \{1,2,\ldots,n\}$ with $|\gI_+| = n/2$ and the second group represented by $\gI_- = \{1,2,\ldots,n\} \setminus \gI_+$. Without loss of generality, we assume that the mean vectors of $(\vy_i)_{i \in \gI_+}$ and $(\vy_i)_{i \in \gI_-}$ are $\vtheta$ and $-\vtheta$ respectively since embeddings are invariant to global shifts and rotations. Equivalently, we write
\begin{equation*}
\vy_i = \begin{cases}
    \vtheta + \vdelta_i & i \in \gI_+ \\
    -\vtheta + \vdelta_i & i \in \gI_-
\end{cases}
\end{equation*}
where $\sum_{i\in \gI_+} \vdelta_i= \sum_{i\in \gI_-} \vdelta_i = \mathbf{0}$. To simplify the loss function, we make an asymptotic assumption: consider (implicitly) a sequence of problems where input data have increasing distances between the two groups, so we expect an increasing separation of clusters in the embedding space:
\begin{equation*}
    \| \vtheta \| \to \infty, \qquad \max_{i \le n} \| \vdelta_i \| = O(1)\, .
\end{equation*}
Now consider adding an input point (`mixed' point) to a location close to the midpoint of the two groups. We assume that its similarity to the other inputs is 
\begin{equation*}
    v_{i,n+1} = \begin{cases}
        p_0 + \veps + o(\veps) \\ p_0 - \veps + o(\veps)
    \end{cases}
\end{equation*}
for $1\le i \le n$, where $p_0>0$ and $\veps$ is a small perturbation parameter. This assumption is reasonable because the similarity of the added point $\vx := \vx^\veps$ has roughly equal similarities to existing inputs up to a small perturbation. We make the asymptotic assumption $\| \vtheta \|^{-1} \asymp \veps$, namely $\veps\| \vtheta \| = O(1)$ and $[\veps\| \vtheta \|]^{-1} = O(1)$.

\begin{figure}[ht]
    \centering 
    \includegraphics[scale = 0.9]{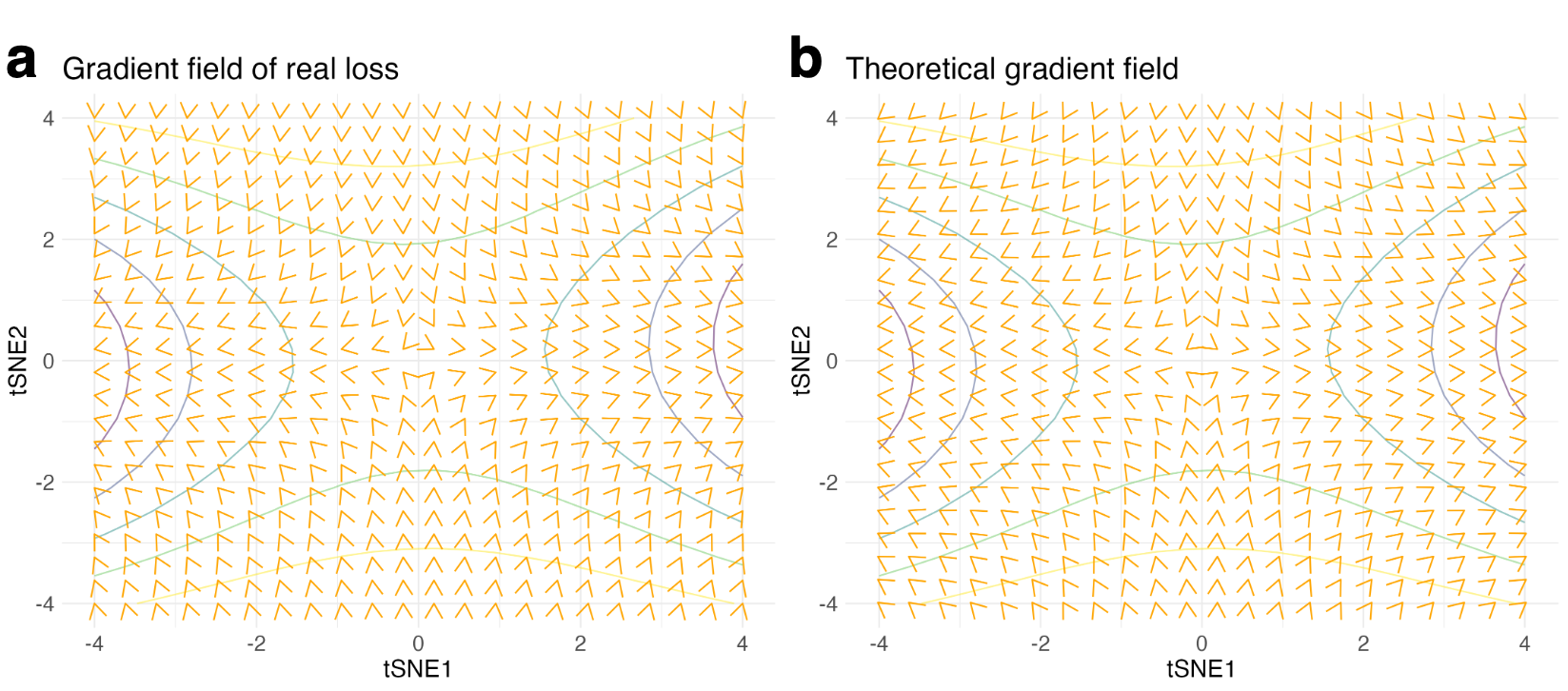}
    \caption{\textbf{Negative gradient fields of the real/theoretical LOO loss.} \textbf{a} We draw the negative gradient fields (force fields) $-\nabla_{\vy} L(\vy; \vx^\veps)$ based on the LOO loss under the same setting as in Fig.~\ref{fig: discontinuity examples}b. \textbf{b} We draw a similar field plot based on the hyperbolic term $\frac{\vy_{\parallelsum}-\vy_\bot}{\|\vtheta\|^2}$ from Theorem~\ref{thm:1}, where we take $\vtheta = (\vc_1 - \vc_2) / 2$ and $\vc_1, \vc_2$ are the centers of two clusters in the embedding. In addition, we add loss contours to both plots, which show hyperbolic paraboloids around the origin. 
    We observe excellent alignment between the negative gradient field of the LOO loss and that of the theoretical analysis. Both field plots show a pull force towards the x-axis and a push force away from the y-axis.
    }
    \label{fig: appendix: 2_10_gradient_field}
\end{figure}

\begin{thm}\label{thm:1}
Consider the LOO loss function for t-SNE given in Equation~\ref{def:tsne-w} and~\ref{def:loo}. Under the assumptions stated above, the negative gradient of the loss is 
\begin{equation*}
    -\nabla_{\vy} L(\vy; \vx^\veps) = (1 + o(1)) \Big( \underbrace{\frac{\vy_{\parallelsum} - \vy_{\bot}}{\| \vtheta\|^2}}_{\text{hyperbolic term}} + \underbrace{\frac{\veps \vtheta}{\| \vtheta\|^2}}_{\text{perturbation term}} \Big)
\end{equation*}
where $\vy_{\parallelsum} =  \vtheta\vtheta^\top \vy / \| \vtheta  \|^2$ is projection of $\vy$ in the direction of $\vtheta$, and $\vy_{\bot} = \vy - \vy_{\parallelsum}$.
\end{thm}
This result explains how the hyperbolic geometry creates OI discontinuity. 
\begin{itemize}
    \item The hyperbolic term indicates the unstable saddle point of the loss at $\vy = \mathbf{0}$. Indeed, it is exactly the tangent vector of a hyperbola, so in the embedding force (negative gradient) field there is a pull force towards the x-axis and a push force away from the y-axis (Fig.~\ref{fig: appendix: 2_10_gradient_field}).
    \item The perturbation term reflects the effects of input point $\vx^\veps$. It tilts the negative gradients slightly in the direction of $\vtheta$ if $\veps>0$ or $-\vtheta$ if $\veps<0$, which causes the algorithm to jump between widely separated local minima of $L(\vy; \vx)$ under small perturbations.
\end{itemize}

\section*{Discussion}



We developed a framework to interpret distortions in neighbor embedding methods as map discontinuities by leveraging the LOO strategy. Based on our LOO-map, we introduce two diagnostic scores to identify OI and FI discontinuities. While being generally effective, our method may not capture all distortion patterns, as factors like initialization, iterative algorithms, and other hyperparameters can introduce different types of distortions. We also recognize the absence of a formal mathematical framework for rigorously characterizing the LOO-map.

In future research, we aim to explore links between classical parametric and implicit embedding maps to fully address topological issues and improve interpretability. We also aim to enhance the scalability of our methods through efficient optimization, sparsity, tree-based approximations, and parallel computation.



\clearpage

\newpage
\appendix
\section*{Methods}

\subsection*{Verify Leave-one-out assumption empirically} \label{sec: method: LOO} \label{sec: supp: LOO}

Our LOO approach assumes that adding (or deleting/modifying) a single input
point does not change the embeddings of other points on average significantly. To verify the LOO assumption, we conduct the following experiment.

Let $\mX = [\vx_1,\ldots,\vx_n]^\top$ be the input data matrix, and $\mY = [\vy_1,\ldots,\vy_n]^\top$ be the matrix of embedding points. We then add one point $\vx$ to $\mX$ to have the new input data $\mX_+ = [\vx_1,\ldots,\vx_n,\vx]^\top$. We then run the t-SNE algorithm to obtain the embedding of $\mX_+$ as $[\tilde\vy_1,\ldots,\tilde\vy_n,\tilde\vy]^\top$. Denoted $[\tilde\vy_1,\ldots,\tilde\vy_n]$ as $\widetilde{\mY}$. To verify LOO empirically, we keep track of the difference between $\mY$ and $\widetilde{\mY}$:
\begin{equation*}
    \epsilon_n = \frac{1}{\|\mY\|_F}\|\mY-\widetilde \mY\|_F
\end{equation*}
and expect $\epsilon_n$ to be small. 

We initialize the t-SNE algorithm in the second run by the embedding points we obtain from the first run: when calculating the embedding of $\mX_+$, we use $\mY$ as the initialization for the first $n$ points. This initialization scheme aims to address two issues: (i) the loss function in a neighbor embedding method is invariant to a global rotation and a global shift of all embedding points, so it is reasonable to choose embedding points with an appropriate initialization. (ii) There are potentially multiple local minima of the loss function due to non-convexity. We verify the LOO assumption at a given local minimum (namely $\mY$) obtained from the first run.


The experiment is conducted with different sample sizes $n$ and with different types of datasets (simulated cluster data, simulated manifold data, real single-cell data, deep learning feature data). The comprehensive results showing the values of $\epsilon_n$ under different settings are presented in Supplementary Table~\ref{tab: LOO Epsilon}. We observe that the approximation errors $\epsilon_n$ are small and generally decreasing in $n$, which supports our LOO assumption.


\subsection*{Perturbation score} \label{sec: method: perturbation score}
\label{sec: appendix: diagnosis metrics}
For implementation convenience, our calculation of the perturbation score and the singularity score is based on modifying an input point instead of adding a new input point. According to the LOO assumption, the difference is negligible. 

Given an input data matrix $\mX = [\vx_1,\ldots,\vx_n]^\top$ and its embedding matrix $\mY = [\vy_1,\ldots,\vy_n]^\top$, we view $\vy_i$ as the mapping of $\vx_i$ by the partial LOO-map $\vf_i$:
\begin{equation}
\begin{aligned}
    &\vf_i(\vx) = \argmin_{\vy \in \R^2} L_i(\vy;\vx),\quad \text{where}\\&L_i(\vy;\vx)=\sum_{k\neq i} \mathcal{L}\Big(w(\vy_k, \vy); v_{i,k}(\bar \mX)\Big) 
+ Z(\bar \mY),
\end{aligned}
\label{eqn: partial loss LOO map}
\end{equation}
where $\bar \mX = [\vx_1,\ldots,\vx_{i-1}, \vx, \vx_{i+1},\ldots,\vx_n]^\top$ differs from $\mX$ only at the $i$-th input point, and $\bar \mY = [\vy_1,\ldots,\vy_{i-1}, \vy, \vy_{i+1},\ldots,\vy_n]^\top$ has frozen embedding points except for the $i$-th point which is the decision variable in the optimization problem. This partial LOO-map $\vf_i$ is based on perturbing (or modifying) a single input point rather than adding a new point, thus maintaining $n$ points in total. According to the LOO assumption, $\vf_i \approx \vf$, so we calculate the perturbation score for the $i$-th point based on $\vf_i$.

To assess the susceptibility of $\vy_i$ under moderate perturbations in $\vx_i$, we apply a perturbation of length $\lambda$ in the direction of $\ve$ to $\vx_i$ and measure the resulting change in the embedding map determined by the partial LOO-map $\vf_i$.
In our implementation, we search the perturbation directions among the first 3 principal directions of the data $\{\ve_1, \ve_2, \ve_3\}$ and their opposites  $\{-\ve_1, -\ve_2, -\ve_3\}$, and the perturbation length $\lambda$ is specified by the user. In this way, we can define the perturbation score of the $i$-th data point as 
\begin{equation}\label{eq:lambda}
    \max_{\ve\in\{\pm\ve_1, \pm\ve_2, \pm\ve_3\}}\|\vf_i(\vx_i + \lambda\ve) - \vy_i\|_2.
\end{equation}

In general, perturbation scores are not sensitive to perturbation lengths. Supplementary Fig.~\ref{fig: appendix: sensitivity_pscore} illustrates the perturbation scores of the CIFAR10 deep learning feature data for three perturbation lengths ($\lambda \in \{1, 2, 3\}$). Points with high perturbation scores remain consistent across different perturbation lengths. In practice, we recommend that users run perturbation scores on a subset of data points and test with a few different perturbation lengths. Conceptually, the perturbation score detects points that fall within a radius of $\lambda$ around the location of the OI discontinuity.
 \label{sec: supp: choose perturbation length}

Moreover, we provide two approximation algorithms to accelerate the calculation of the perturbation score for t-SNE along with a strategy for users to pre-screen points for which the perturbation score should be computed. 

\label{sec: supp: Implementation of pscore}

\paragraph{Approximation method 1.} For high-dimensional input data, often PCA as a pre-processing step is implemented before calculating the similarity scores. 
As similarity scores are recalculated for each perturbation we consider, PCA is repeated numerous times, leading to a significant increase in computation. Since PCA is robust to perturbing a single input point, we reuse the pre-processed input points after one PCA calculation based on the original input data. This approximation avoids multiple calculations of PCA.
We find that this approximation is sufficiently accurate, as the differences between perturbation scores by approximation method 1 and the exact perturbation scores are empirically negligible (Supplementary Fig.~\ref{fig: appendix: approx}a).

\paragraph{Approximation method 2.} Besides reducing PCA computations, we can further accelerate the calculation of perturbation scores by approximating the similarity scores. 

Given the input data matrix $\mX = [\vx_1,\ldots,\vx_n]^\top$ and perplexity $\mathcal{P}$, the computation of (exact) similarity scores $(v_{i,j}(\mX))_{i<j}$ in the t-SNE algorithm follows the steps below.
\begin{enumerate}
    \item Calculate the pairwise distance $d_{ij} = \|\vx_i - \vx_j\|_2$ for $i,j = 1,\ldots,n$.
    \item Find $\sigma_i$, $i=1,\ldots,n$ that satisfies
    \begin{equation}
        -\sum_{j\neq i}\frac{\exp(-d_{ij}^2/2\sigma_i^2)}{\sum_{k\neq i}\exp(-d_{ik}^2/2\sigma_i^2)}\log_2\Big(\frac{\exp(-d_{ij}^2/2\sigma_i^2)}{\sum_{k\neq i}\exp(-d_{ik}^2/2\sigma_i^2)}\Big) = \log_2(\mathcal{P}).
        \label{eqn: PMatrix solve sigma}
    \end{equation}
    \item Calculate $p_{j|i} = \frac{\exp(-d_{ij}^2/2\sigma_i^2)}{\sum_{k\neq i}\exp(-d_{ik}^2/2\sigma_i^2)}$, $i,j = 1,\ldots,n$. And 
    \begin{equation*}
        v_{i,j}(\mX) = \frac{p_{j|i} + p_{i|j}}{2n}.
    \end{equation*}
\end{enumerate}
The main computational bottleneck is at step 2, where we conduct a binary search algorithm for $n$ times to solve $(\sigma_i)_{1\leq i\leq n}$.  

To provide an approximation method, we note that when perturbing the $k$-th point, for $i\neq k$, Equation~\ref{eqn: PMatrix solve sigma} still approximately holds for the original standard deviation $\sigma_i$ since only one of the terms has been changed. Therefore, we can set $\tilde \sigma_i \approx \sigma_i$ for $i\neq k$ as an approximation to $(\tilde\sigma_i)_{1\leq i \leq n}$, the standard deviations after perturbation. In this way, we only need to conduct the binary search once to solve $\tilde \sigma_k$, which significantly speeds up the calculation of the similarity scores after perturbation. 

In terms of computational performance, approximation method 2 leads to a reduction of running time by nearly $80\%$ for a dataset of size $5000$.
We also find that approximation method 2 is highly accurate. As shown in Supplementary Fig.~\ref{fig: appendix: approx}b, perturbation scores based on approximation method 2 are approximately equal to the exact perturbation scores for most of the points.

\paragraph{Pre-screening of points.} To further speed up the computation, we use the heuristic that embedding points receiving high perturbation scores are often found at the peripheries of clusters. This heuristic motivates us to calculate the perturbation scores only for the peripheral points in the embedding space, as these points are most likely to be unreliable. We find that applying this pre-screening step tends to find most of the unreliable points (Supplementary Fig.~\ref{fig: appendix: prescreen}) with significantly increased computational speed.


We use the function \texttt{dbscan} in the R package dbscan (version 1.2-0) to identify embeddings on the periphery of clusters. 

\subsection*{Singularity score}\label{sec:append-sing}

Given an input data matrix $\mX = [\vx_1,\ldots,\vx_n]^\top$ and its embedding matrix $\mY = [\vy_1,\ldots,\vy_n]^\top$, we describe our derivation of singularity scores. 
If we add an infinitesimal perturbation $\epsilon\ve$ to $\vx_i$, then by the Taylor expansion of the partial LOO-map $\vf_i$, the resulting change in the $i$-th embedding point is expressed as 
\begin{equation}
         \vf_i(\vx_i+\epsilon\ve)- \vy_{i} = -\epsilon\mH_i^{-1}\sum_{k: k \neq i}\frac{\partial^2 \mathcal{L}(w(\vy_i, \vy_k); v_{i,k}(\mX))}{\partial\vy_i\partial \vx_k^\top} \ve + o(\epsilon),
        \label{eqn: singularity score hessian}
\end{equation}
where $\mH_i$ denotes the Hessian matrix of the partial LOO loss $L_i(\vy;\vx_i)$ with respect to $\vy$ at $\vy = \vy_i$. Notably, when $\epsilon=0$ (no perturbation), we have $\vf_i(\vx_i) = \vy_i$. Denote the total loss as
\begin{equation*}
    \mathfrak{L}(\vy_1,\ldots,\vy_n;\mX)= \sum_{1\le i<j\le n} \mathcal{L}(w(\vy_i, \vy_j); v_{i,j}(\mX)) + Z(\mY).
\end{equation*}
Then, $\mH_i$ can be written as 
\begin{equation*}
    \mH_i = \left.\frac{\partial^2 L_i(\vy;\vx_i)}{\partial \vy\partial \vy^\top}\right|_{\vy = \vy_i}  = \frac{\partial^2\mathfrak{L}(\vy_1,\ldots,\vy_n;\mX)}{\partial \vy_i\partial \vy_i^\top},
\end{equation*}
i.e., $\mH_i$ is also equal to the Hessian matrix of the total loss $\mathfrak{L}$ with respect to the $i$-th variable taking value at $\vy_i$.

Importantly, $\mH_i$ is independent of the perturbation direction $\ve$. The more singular $\mH_i$ is, the more sensitive the embedding point of $\vx_i$ becomes to infinitesimal perturbations. Thus, we define the singularity score of the $i$-th data point as the inverse of the smallest eigenvalue of the Hessian matrix of $\mathfrak{L}$, that is $\lambda_{\min}^{-1}(\mH_i)$. Supplementary Methods~\ref{sec: appendix: derivation of sscore} provides detailed derivations of Equation~\ref{eqn: singularity score hessian} and Supplementary Methods~\ref{sec: method: derivation of sscore} provides expressions of singularity scores for t-SNE, UMAP and LargeVis.

\subsection*{Scoring Metrics and Statistical Tests}

\paragraph{Entropy of class probabilities.}
For a classification task, a statistical or machine learning algorithm outputs predicted class probabilities for a test data point. For example, in neural networks, the probabilities are typically obtained through a softmax operation in the final layer. Often, the model predicts a class with the largest probability among all classes. The entropy of the probabilities can quantify how confident the model is in its prediction.

For a classification task of $k$ classes, if we denote the outputs class probabilities for one data point $\vx$ as $\vp = (p_1,\ldots,p_k)$, then we define the entropy as $E(\vp) = -\sum_{j=1}^k p_j \log(p_j)$. This quantity is widely used for measuring class uncertainty. 

\paragraph{Entropy difference.}
\label{sec: entropy difference}
We will describe an uncertainty measurement given access to the labels of input points. For a dataset $(\vx_i)_{i\le n}$ with clustering structures, we posit the following $k$-component Gaussian mixture model (GMM) from which each $\vx_i$ is sampled.
Consider a uniform prior on the $k$ clusters, i.e., $p(A_j) = \frac{1}{k}$, $j=1,2,\ldots,k$. Given cluster membership $A_j$, we define the conditional probability density function
\begin{equation*}
    p(\vx|A_j) = g(\vx|\vmu_j,\mSigma_j),
\end{equation*}
where $\vmu_j$, $\mSigma_j$ are the mean and covariance matrix in the $j$-th component, and $g(\vx|\vmu_j,\mSigma_i)$, $j=1,2,\ldots,k$ are the Gaussian density functions with mean $\vmu_j$ and covariance matrix $\mSigma_j$. 
We then have the posterior probability of $A_j$ given an observation $\vx$ as
\begin{equation}\label{eq:posterior}
    p(A_j|\vx) = \frac{p(\vx|A_j)}{\sum_{j=1}^kp(\vx|A_j)}.
\end{equation}

In the analysis of neighbor embedding methods, we will use the posterior probabilities as an uncertainty measurement. Given the ground-truth labels of the data points, we can fit two GMMs, one in the input space and the other in the embedding space, yielding estimated parameters $(\vmu_j, \mSigma_j)_{j \le k}$ for each fitted GMM. Then we can calculate the posterior probabilities of each data point belonging to the $k$ components by Equation~\ref{eq:posterior} with fitted parameters, in both the input space and the embedding space. For any data point, denote the posterior probabilities in input space as $\vp = (p_1,p_2,\ldots,p_k)$ and in embedding space as $\vq = (q_1,q_2,\ldots,q_k)$. Finally, we define the entropy difference for each point as the difference between the entropy of $\vp$ and the entropy of $\vq$, i.e., $E(\vp) - E(\vq) = -\sum_{j=1}^k p_j \log(p_j)+\sum_{j=1}^k q_j \log(q_j)$. 

The entropy difference measures the amount of decreased uncertainty of cluster membership. A positive entropy difference means $E(\vq) < E(\vp)$, so the associated data point appears to be less ambiguous in cluster membership after embedding. Vice versa, a negative entropy difference means increased uncertainty after embedding.

Since calculating entropy differences is based on ground-truth labels and fitting a clear statistical model, we believe that entropy differences are a relatively objective evaluation of visual uncertainty. If a diagnostic score without label information is aligned with the entropy difference, then the diagnostic score is likely to be reliable.


\paragraph{Evaluation score of neighborhood preservation.} \label{sec: supp: Metric to evaluate neighborhood preservation}
We calculate point-wise neighborhood preservation scores to evaluate how well the local structures are preserved by an embedding algorithm. Given the input matrix $\mX$ and the embedding matrix $\mY$, to calculate the neighborhood preservation score for the $i$-th point, we first identify its $k$-nearest neighbors in the input space, with their indices denoted as $\mathcal{N}_i = \{i_1, i_2, \ldots, i_k\}$. Then, we compute the distances from the $i$-th point to its neighbors in both the input and embedding spaces:
\begin{equation*}
    \begin{aligned}
        \vd_i^{\text{input}} &= [d(\vx_i,\vx_{i_1}),\ldots,d(\vx_i,\vx_{i_k})]^\top\\
        \vd_i^{\text{embedding}}&= [d(\vy_i,\vy_{i_1}),\ldots,d(\vy_i,\vy_{i_k})]^\top.
    \end{aligned}
\end{equation*}
The neighborhood preservation score for the $i$-th point is defined as the correlation between $\vd_i^{\text{input}}$ and $\vd_i^{\text{embedding}}$. A higher correlation indicates better preservation of the neighborhood structure.

We use the median neighborhood preservation score across all points in the dataset to assess the overall neighborhood preservation of the embedding. For hyperparameters, we choose $k=[n/5]$ and use the Euclidean distance as the metric $d$ in implementation.

\label{sec: supp: indices to evaluate clustering}

\paragraph{Davies-Bouldin Index.}
\label{sec: dbindex}


We calculate the DB index \cite{DBIndex} using the R function \texttt{index.DB} in the R package clusterSim (version 0.51-3) with $p=q=2$, i.e., using the Euclidean distance.

\paragraph{Within-cluster distance ratio.}
\label{sec: wcdr}
Consider $m$ clusters and in each cluster $i$, there are $n_i$ data points, denoted as $\{\vx_{ij}\}_{1\leq j\leq n_i}$. The centroid for each cluster is denoted as $\vx_{i\cdot} = \frac{1}{n_i}\sum_{j=1}^{n_i}\vx_{ij}$ and the mean of all data points is denoted as $\vx_{\cdot\cdot} = \frac{1}{n}\sum_{i=1}^m\sum_{j=1}^{n_i}\vx_{ij}$. 

Denote the total sum of squares (TSS) and the within-cluster sum of squares (WSS) by
\begin{equation*}
    \text{TSS} = \sum_{i=1}^m\sum_{j=1}^{n_i}\|\vx_{ij} - \vx_{\cdot\cdot}\|_2^2,\quad\text{WSS} = \sum_{i=1}^m\sum_{j=1}^{n_i}\|\vx_{ij} - \vx_{i\cdot}\|_2^2.
\end{equation*}
The within-cluster distance ratio is defined as $\text{WCDR} = \frac{\text{WSS}}{\text{TSS}}$. 
A smaller within-cluster distance ratio \text{WCDR} indicates a more pronounced clustering effect.

\paragraph{Wilks' $\Lambda$.}
We compute Wilks' $\Lambda$ statistic \cite{WilksLambda} by performing a multivariate analysis of variance using the \texttt{manova} function from the R package stats (version 4.2.1), followed by a statistical test. 

\paragraph{Statistical tests for distribution difference of singularity scores.} \label{sec: supp: tests for sc data}
We have claimed that embedding points with large singularity scores tend to appear in random locations under small perplexities but appear in the periphery of clusters under large perplexities. To quantitatively verify such distinction, we conduct several statistical tests and find that the results of the tests support our claim about the distribution difference (see Supplementary Table~\ref{tab: mousebraintest}). We provide the details of the tests as follows.

\paragraph{Tests for Spearman's rank correlation.} Given the embedding $\mY = [\vy_1,\ldots,\vy_n]^\top$ and the cluster label of each point as well as their singularity scores $\vs = [s_1,\ldots,s_n]^\top$, we can first calculate the distance of each point to its cluster center. The distance vector is denoted as $\vd=[d_1,\ldots,d_n]^\top$. We then conduct the Spearman's rank correlation test \cite{SpearmanRho} on the singularity scores $\vs$ and the distances to cluster center $\vd$. The tests show that there is no significant correlation under low perplexity but a significant correlation under larger perplexity (see $p$-values in Supplementary Table~\ref{tab: mousebraintest}).

We use the function \texttt{cor.test} in the R package stat (version 4.2.1) to perform Spearman's rank correlation tests.

\paragraph{Tests for the local regression model.} To test for distribution differences, we first fit a local regression model \cite{ClevelandLOESS1992} using the singularity scores as the response variables and the coordinates of embedding points as predictors. Next, we fit a null model with the singularity scores as the response and only the intercept as the predictor. An F-test is then conducted to determine whether the magnitude of the singularity scores is associated with the locations of the embedding points.

We also perform permutation tests by shuffling the singularity scores and fitting a local regression model for each shuffle to approximate a null distribution for the residual sum of squares. Empirical $p$-values are then computed to assess whether the singularity scores are distributed randomly. Lower $p$-values suggest rejecting the null hypothesis of random distribution.

We use the \texttt{loess} function from the R package stat (version 4.2.1) to fit the local regression models.

\subsection*{Benchmark Methods for OOD Detection}
\paragraph{Kernel PCA.}
We implemented the state-of-the-art kernel PCA method for out-of-distribution detection \cite{kernelPCA} to benchmark against the perturbation score. Since our perturbation score does not require separate training and testing steps and was directly applied to the dataset, kernel PCA was trained on the dataset and then evaluated on the same dataset to ensure a fair comparison. Additionally, to maintain consistency with the default PCA preprocessing step in the t-SNE algorithm, we applied PCA before training, retaining the first 50 principal components.

\paragraph{One-class SVM.}
We implemented the one-class SVM \cite{OCSVM} using the \texttt{OneClassSVM} function from the Python package \texttt{scikit-learn}, employing a polynomial kernel for optimal performance. Since our perturbation score does not require separate training and testing steps and was directly applied to the dataset, one-class SVM was also trained on the dataset and then evaluated on the same dataset to ensure a fair comparison. To align with the preprocessing step in t-SNE, we first applied PCA, reducing the data to its top 50 principal components before training.

\subsection*{Datasets}

\paragraph{Gaussian mixture data.}
\label{sec: supp: GMM}
A Gaussian mixture model with $k$ components is a linear combination of $k$-component Gaussian densities. The probability density function of the random variable $\vx$ generated by Gaussian mixture model \cite{ReynoldsGMM2009} is
\begin{equation*}
    p(\vx) = \sum_{i=1}^k\pi_i g(\vx|\vmu_i,\mSigma_i),
\end{equation*}
where $\vmu_i$, $\mSigma_i$ are the mean and covariance matrix in the $i$-th component, the scalars $\pi_i$, $i=1,2,\ldots,k$ are the mixture weights satisfying $\sum_{i=1}^k\pi_i = 1$, and $g(\vx|\vmu_i,\mSigma_i)$, $i=1,2,\ldots,n$ are the probability density functions of the Gaussian distribution family with mean $\vmu_i$ and covariance matrix $\mSigma_i$.

We randomly generated Gaussian mixture datasets with various numbers of components and mixture weights using the function \texttt{rGMM} in the R package MGMM (version 1.0.1.1).

\paragraph{Swiss roll data.}
\label{sec: supp: swissroll}
The Swiss roll data is a classical manifold data. Usually, the dataset consists of three-dimensional i.i.d. data points, denoted as $(x,y,z)^\top\in\mathbb{R}^3$, where
\begin{equation*}
    x = t\cos(t),\,y=t\sin(t),z=z.
\end{equation*}
Here, $t$ is the parameter controlling the spiral angle and is uniformly distributed in a chosen range $[a,b]$. And $z$ is the height parameter and is also uniformly distributed in the chosen span of heights $[c,d]$.

We randomly generated Swiss roll datasets and used the function \texttt{Rtsne} in the R package Rtsne (version 0.17) to obtain the t-SNE embeddings of the datasets. We computed the perturbation scores with perturbation length $1$ in Fig.~\ref{fig: methods-compare}b.

\paragraph{Deep learning feature data.}
We used the pretrained ResNet-18 model to perform a forward pass on the CIFAR-10 dataset to extract features of dimension 512. We also performed the forward pass using the same pre-trained model on the Describable Textures Dataset (DTD) dataset \cite{CimpoiDTD2014} as our out-of-distribution data in Fig.~\ref{fig:CIFAR10ood}. We also randomly subsampled both datasets to reduce computational load. Specifically, in Fig.~\ref{fig: overview}, we sampled $5000$ images from the CIFAR-10 test dataset as our deep learning feature data and obtained the t-SNE embedding under perplexity $125$. We then computed the perturbation scores with perturbation length $2$. In Fig.~\ref{fig:CIFAR10ood}, we sampled $2000$ CIFAR-10 images and $1000$ DTD images, combining them into a dataset that includes OOD data points. We obtained the t-SNE embedding under perplexity $100$ and computed the perturbation scores with perturbation length $2$. 

\paragraph{Mouse brain single-cell ATAC-seq data.}
The ATAC-seq dataset was created to capture the gene activity of mouse brain cells. The dataset has been preprocessed by Luecken et al. \cite{luecken}. We applied the R functions \texttt{CreateSeuratObject}, \texttt{FindVariableFeatures} and \texttt{NormalizeData} in R package Seurat to identify 1000 most variable genes for 3618 cells. The dataset was subsampled when being used to verify the LOO assumption.

\paragraph{Mouse embryonic stem cell differentiation data.}
The single-cell RNA-seq dataset was constructed to investigate the dynamics of gene expression of mouse embryonic stem cells (mESCs) undergoing differentiation \cite{HayashiData}. The dataset was preprocessed, normalized, and scaled by following the standard procedures by R package Seurat using functions \texttt{CreateSeuratObject}, \texttt{NormalizeData} and \texttt{ScaleData}. We also used R function \texttt{FindVariableFeatures} to identify the 2000 most variable genes for all 421 cells.

\paragraph{Human pancreatic tissue single-cell RNA-seq data.}
The single-cell RNA-seq data generated from human pancreatic tissues \cite{panc8} provides a comprehensive view of gene expression across 8 different cell types in pancreatic tissue. The dataset was preprocessed, normalized, and scaled by following the standard procedures described above. We also used R function \texttt{FindVariableFeatures} to identify the 2000 most variable genes for all 2364 cells. The dataset was subsampled when being used to verify the LOO assumption.

\paragraph{Single-cell RNA-seq data of PBMCs with treatment of interferon-beta.} 
This single-cell RNA-seq dataset profiles gene expression in peripheral blood mononuclear cells (PBMCs) following interferon-$\beta$ (IFNB) treatment, capturing cellular responses to immune stimulation \cite{ifnb}.  The dataset was preprocessed, normalized, and scaled by following the standard procedures described above. We used R function \texttt{FindVariableFeatures} to identify the 2000 most variable genes for all 6,548 cells. The dataset was subsampled when being used to verify the LOO assumption.

\paragraph{Mouse mammary epithelial single-cell data.}
This dataset contains the gene expression profile of mammary epithelial cells across from two mice at four developmental stages: nulliparous, mid-gestation, lactation, and post-involution \cite{Bach2017MammaryData}. The dataset was preprocessed, normalized, and scaled by following the standard procedures described above. We used R function \texttt{FindVariableFeatures} to identify the 2000 most variable genes for all 25,806 cells.

\subsection*{Implementation of t-SNE}
We used the function \texttt{Rtsne} in the R package Rtsne (version 0.17) to perform the t-SNE algorithm. We choose \texttt{theta = 0} to perform exact t-SNE. We also adjusted the code in Rtsne to access the similarity scores $(v_{i,j}(\mX))_{i<j}$. The adjusted function \texttt{Rtsne} can be found in \url{https://github.com/zhexuandliu/MapContinuity-NE-Reliability}.

\section*{Data availability}
CIFAR-10 raw data is available from \cite{KrizhevskyCIFAR102009} [\hyperlink{https://www.cs.toronto.edu/~kriz/cifar.html}{https://www.cs.toronto.edu/~kriz/cifar.html}]. Describable Textures Dataset is available from \cite{CimpoiDTD2014} [\hyperlink{https://www.robots.ox.ac.uk/~vgg/data/dtd/}{https://www.robots.ox.ac.uk/~vgg/data/dtd/}]. The pretrained ResNet-18 model is available at [\hyperlink{https://huggingface.co/edadaltocg/resnet18_cifar10/}{https://huggingface.co/edadaltocg/resnet18\_cifar10}]. Mouse brain single-cell ATAC-seq data can be downloaded from Figshare [\hyperlink{https://figshare.com/ndownloader/files/25721789}{https://figshare.com/ndownloader/files/25721789}]. The ATAC-seq datasets have been preprocessed by Luecken et al. (\cite{luecken}) to characterize gene activities. Mouse embryonic stem cell differentiation data is available in Gene Expression Omnibus with accession code [\hyperlink{https://www.ncbi.nlm.nih.gov/geo/query/acc.cgi?acc=GSE98664}{GSE98664}]. The single-cell RNA-seq dataset generated from PBMCs treated with interferon-$\beta$ is available from the R package Seurat (version 5.0.3) under the name \texttt{ifnb}. The single-cell RNA-seq data of human pancreatic tissues is available from the \texttt{smartseq2} dataset in the R package Seurat (version 5.0.3) under the name \texttt{panc8}. Mouse mammary epithelial single-cell data is available from the R package scRNAseq (version 2.20.0) under the name \texttt{BachMammaryData}. Source data are also provided.

\section*{Code availability}
The code for calculating the two diagnostic scores (as an R package), and the code for reproducing the simulation and analysis of this paper are available at \url{https://github.com/zhexuandliu/MapContinuity-NE-Reliability}.


\bibliographystyle{naturemag}
\bibliography{refs}

\section*{Acknowledgements}
Y.Z. is supported by NSF-DMS grant 2412052 and by the Office of the Vice Chancellor for Research and Graduate Education at the UW Madison with funding from the Wisconsin Alumni Research Foundation. Z.L. and Y.Z. would like to thank Yixuan Li for suggesting out-of-distribution detection, and thank Sebastien Roch, Zexuan Sun, Xinyu Li and Jingyang Lyu for helpful discussions. R.M. would like to thank Jonas Fischer, Dmitry Kobak, Stefan Steinerberger and Bin Yu for helpful discussions on t-SNE and UMAP.

\section*{Author contributions}
Y.Z. and R.M. conceived the study. Z.L. designed and implemented the method with input from Y.Z. and R.M. Z.L. contributed to the numerical analysis and software implementation. Y.Z. and R.M. designed and developed the theoretical results for the study. Z.L. prepared a draft of the manuscript. Y.Z. and R.M. edited the manuscript.

\section*{Competing interests}
The Authors declare no competing interests.

\newpage
\setcounter{table}{0} 
\setcounter{figure}{0}
\renewcommand{\thetable}{S\arabic{table}}
\renewcommand{\thefigure}{S\arabic{figure}}
\setcounter{equation}{0} 
\renewcommand{\theequation}{S\arabic{equation}}
\section{Supplement to “Assessing and improving reliability of neighbor embedding methods: a map-continuity perspective”}

\subsection{Derivation of singularity score} \label{sec: appendix: derivation of sscore}
Given an input data matrix $\mX = [\vx_1,\ldots,\vx_n]^\top$ and its embedding matrix $\mY = [\vy_1,\ldots,\vy_n]^\top$, without loss of generality, we will derive singularity score for the $n$-th embedding point. Singularity score measures the sensitivity of the embedding point under infinitesimal perturbation. Therefore, we add an infinitesimal perturbation of length $\epsilon$ and direction as a vector $\ve$ of unit length to $\vx_n$ to analyze the sensitivity of its embedding $\vy_n$.

We note that $\vy_n$ is the minimizer of the partial LOO loss involving the $n$-th embedding point: 
\begin{equation*}
    \vy_n = \argmin_{\vy \in \R^2} \sum_{1\le i \le n-1} \mathcal{L}(w(\vy_i, \vy); v_{i,n}(\mX)) 
+ Z( [\vy_1,\ldots,\vy_{n-1},\vy]^\top)
\end{equation*}
and denote the perturbed dataset as $\widetilde{\mX}$ and $\tilde{\vy}_n = \vf_n(\vx_n+\epsilon\ve)$, which is the mapping of perturbed $\vx_n$ by LOO-map, we have
\begin{equation*}
    \tilde{\vy}_n = \argmin_{\vy \in \R^2} \sum_{1\le i \le n-1} \mathcal{L}(w(\vy_i, \vy); v_{i,n}(\widetilde{\mX})) 
+ Z([\vy_1,\ldots,\vy_{n-1},\vy]^\top).
\end{equation*}
By first order condition, we have

\begin{align}
    \left.\Bigg[\sum_{1\le i \le n-1}\Big(\frac{\partial \mathcal{L}(w(\vy_i, \vy); v_{i,n}(\mX))}{\partial\vy}\Big) + \frac{\partial Z([\vy_1,\ldots,\vy_{n-1},\vy]^\top) }{\partial\vy}\Bigg]\right|_{\vy = \vy_{n}}=\mathbf{0}, \label{eqn: sscore derivation 1st cond 1}\\
    \left.\Bigg[\sum_{1\le i \le n-1}\Big(\frac{\partial \mathcal{L}(w(\vy_i, \vy); v_{i,n}(\widetilde{\mX}))}{\partial\vy}\Big) + \frac{\partial Z([\vy_1,\ldots,\vy_{n-1},\vy]^\top) }{\partial\vy}\Bigg]\right|_{\vy = \tilde{\vy}_{n}}=\mathbf{0}.\label{eqn: sscore derivation 1st cond 2}
\end{align}
By doing Taylor's expansion to $\left.\frac{\partial \mathcal{L}(w(\vy_i, \vy); v_{i,n}(\widetilde{\mX}))}{\partial\vy}\right|_{\vy = \tilde{\vy}_{n}}$ on $\tilde{\vy}_{n}$ and $v_{i,n}(\widetilde{\mX})$ for $i=1,\ldots, n-1$, we obtain
\begin{equation*}
    \begin{aligned}
        &\left.\frac{\partial \mathcal{L}(w(\vy_i, \vy); v_{i,n}(\widetilde{\mX}))}{\partial\vy}\right|_{\vy = \tilde{\vy}_{n}}\\ =& \left.\frac{\partial \mathcal{L}(w(\vy_i, \vy); v_{i,n}(\mX))}{\partial\vy}\right|_{\vy = \vy_{n}} + \left.\frac{\partial^2 \mathcal{L}(w(\vy_i, \vy); v_{i,n}(\mX))}{\partial\vy\partial\vy^\top}\right|_{\vy = \vy_{n}}(\tilde{\vy}_n - \vy_{n})\\
        &+\left.\frac{\partial^2 \mathcal{L}(w(\vy_i, \vy); v)}{\partial\vy\partial v}\right|_{\vy = \vy_{n},v = v_{i,n}(\mX)}(v_{i,n}(\widetilde{\mX}) - v_{i,n}(\mX))\\
        &+o(\|\tilde{\vy}_n - \vy_{n}\| + \|v_{i,n}(\widetilde{\mX}) - v_{i,n}(\mX)\|).
    \end{aligned}
\end{equation*}
Then by doing Taylor's expansion to $\left.\frac{\partial Z([\vy_1,\ldots,\vy_{n-1},\vy]^\top) }{\partial\vy}\right|_{\vy = \tilde{\vy}_{n}}$  on $\tilde{\vy}_{n}$, we have
\begin{equation*}
    \begin{aligned}
        &\left.\frac{\partial Z([\vy_1,\ldots,\vy_{n-1},\vy]^\top) }{\partial\vy}\right|_{\vy = \tilde{\vy}_{n}}\\ =& \left.\frac{\partial Z([\vy_1,\ldots,\vy_{n-1},\vy]^\top) }{\partial\vy}\right|_{\vy = \vy_{n}} + \left.\frac{\partial Z([\vy_1,\ldots,\vy_{n-1},\vy]^\top) }{\partial\vy\partial\vy^\top}\right|_{\vy = \vy_{n}}(\tilde{\vy}_n - \vy_{n})+o(\|\tilde{\vy}_n - \vy_{n}\|).
    \end{aligned}
\end{equation*}
Note that the embedding points and similarity scores are all functions of the input $\mX$. We can do Taylor’s expansion to both $\tilde{\vy}_n$ and $v_{i,n}(\widetilde{\mX})$:
\begin{equation*}
    \begin{aligned}
        \tilde{\vy}_n - \vy_{n} &= \epsilon \Big(\frac{\partial \vy_{n}}{\partial \vx_n}\Big)^\top \ve + o(\epsilon),\\
        v_{i,n}(\widetilde{\mX}) - v_{i,n}(\mX) &= \epsilon \Big(\frac{\partial v_{i,n}(\mX)}{\partial \vx_n}\Big)^\top \ve + o(\epsilon).
    \end{aligned}
\end{equation*}
Plug in, we have
\begin{equation}
    \begin{aligned}
        &\left.\frac{\partial \mathcal{L}(w(\vy_i, \vy); v_{i,n}(\widetilde{\mX}))}{\partial\vy}\right|_{\vy = \tilde{\vy}_{n}}\\ =& \left.\frac{\partial \mathcal{L}(w(\vy_i, \vy); v_{i,n}(\mX))}{\partial\vy}\right|_{\vy = \vy_{n}} + \epsilon\left.\frac{\partial^2 \mathcal{L}(w(\vy_i, \vy); v_{i,n}(\mX))}{\partial\vy\partial\vy^\top}\right|_{\vy = \vy_{n}}\Big(\frac{\partial \vy_{n}}{\partial \vx_n}\Big)^\top \ve\\
        &+\epsilon\left.\frac{\partial^2 \mathcal{L}(w(\vy_i, \vy); v)}{\partial\vy\partial v}\right|_{\vy = \vy_{n},v = v_{i,n}(\mX)}\Big(\frac{\partial v_{i,n}(\mX)}{\partial \vx_n}\Big)^\top \ve+o(\epsilon)
    \end{aligned}
    \label{eqn: sscore deri taylor L}
\end{equation}
and 
\begin{equation}
    \begin{aligned}
        &\left.\frac{\partial Z([\vy_1,\ldots,\vy_{n-1},\vy]^\top) }{\partial\vy}\right|_{\vy = \tilde{\vy}_{n}}\\ =& \left.\frac{\partial Z([\vy_1,\ldots,\vy_{n-1},\vy]^\top) }{\partial\vy}\right|_{\vy = \vy_{n}} + \epsilon\left.\frac{\partial Z([\vy_1,\ldots,\vy_{n-1},\vy]^\top) }{\partial\vy\partial\vy^\top}\right|_{\vy = \vy_{n}}\Big(\frac{\partial \vy_{n}}{\partial \vx_n}\Big)^\top \ve+o(\epsilon).
    \end{aligned}
    \label{eqn: sscore deri taylor Z}
\end{equation}
By summing up Eqn.~\ref{eqn: sscore deri taylor L} for $i=1,\ldots,n-1$ and Eqn.~\ref{eqn: sscore deri taylor Z}, we have
\begin{equation*}
    \begin{aligned}
        &\left.\Bigg[\sum_{1\le i \le n-1}\Big(\frac{\partial \mathcal{L}(w(\vy_i, \vy); v_{i,n}(\widetilde{\mX}))}{\partial\vy}\Big) + \frac{\partial Z([\vy_1,\ldots,\vy_{n-1},\vy]^\top) }{\partial\vy}\Bigg]\right|_{\vy = \tilde{\vy}_{n}}\\
        =& \left.\Bigg[\sum_{1\le i \le n-1}\Big(\frac{\partial \mathcal{L}(w(\vy_i, \vy); v_{i,n}(\mX))}{\partial\vy}\Big) + \frac{\partial Z([\vy_1,\ldots,\vy_{n-1},\vy]^\top) }{\partial\vy}\Bigg]\right|_{\vy = \vy_{n}}\\
        &+ \epsilon\left.\frac{\partial^2[\sum_{1\le i \le n-1}\mathcal{L}(w(\vy_i, \vy); v_{i,n}(\mX)) + Z([\vy_1,\ldots,\vy_{n-1},\vy]^\top)]}{\partial \vy\partial \vy^\top}\right|_{\vy = \vy_{n}}\Big(\frac{\partial \vy_{n}}{\partial \vx_n}\Big)^\top \ve \\
        &+ \epsilon\sum_{1\le i \le n-1}\Big[\left.\frac{\partial^2 \mathcal{L}(w(\vy_i, \vy); v)}{\partial\vy\partial v}\right|_{\vy = \vy_{n},v = v_{i,n}(\mX)}\Big(\frac{\partial v_{i,n}(\mX)}{\partial \vx_n}\Big)^\top \Big]\ve+o(\epsilon).
    \end{aligned}
\end{equation*}
Also note that 
\begin{equation*}
    \frac{\partial^2[\sum_{1\le i \le n-1}\mathcal{L}(w(\vy_i, \vy); v_{i,n}(\mX)) + Z([\vy_1,\ldots,\vy_{n-1},\vy]^\top)]}{\partial \vy\partial \vy^\top} = \frac{\partial^2\mathfrak{L}(\vy_1,\ldots,\vy_{n-1},\vy;\mX)}{\partial \vy\partial \vy^\top}
\end{equation*}
where $\mathfrak{L}$ denotes the total loss:
\begin{equation*}
\mathfrak{L}(\vy_1,\ldots,\vy_n;\mX) = \sum_{1\le i<j\le n} \mathcal{L}(w(\vy_i, \vy_j); v_{i,j}(\mX)) + Z([\vy_1,\ldots,\vy_{n}]^\top).
\end{equation*}
Plug in the first order condition Eqn.~\ref{eqn: sscore derivation 1st cond 1} and Eqn.~\ref{eqn: sscore derivation 1st cond 2}, we have the change of $\vy_n$ after the infinitesimal perturbation as
\begin{equation*}
    \begin{aligned}
        \tilde{\vy}_n - \vy_{n} &= \epsilon \Big(\frac{\partial \vy_{n}}{\partial \vx_n}\Big)^\top \ve + o(\epsilon)\\
        & = -\epsilon\mH_n^{-1}\sum_{1\le i \le n-1}\Big[\left.\frac{\partial^2 \mathcal{L}(w(\vy_i, \vy); v)}{\partial\vy\partial v}\right|_{\vy = \vy_{n},v = v_{i,n}(\mX)}\Big(\frac{\partial v_{i,n}(\mX)}{\partial \vx_n}\Big)^\top\Big] \ve + o(\epsilon)
    \end{aligned}
\end{equation*}
where $\mH_n$ denote the Hessian matrix of the loss function with respect to $\vy_n$:
\begin{equation*}
    \mH_n = \left.\frac{\partial^2\mathfrak{L}(\vy_1,\ldots,\vy_{n-1},\vy;\mX)}{\partial \vy\partial \vy^\top}\right|_{\vy = \vy_{n}}.
\end{equation*}

Importantly, $\mH_n$ is independent of the perturbation direction $\ve$. A singular Hessian matrix results in the most extreme local discontinuity. The more singular $\mH_n$ is, the more sensitive the embedding of $\vx_n$ becomes to infinitesimal perturbations. This is why we define the singularity score for the $n$-th data point as $\lambda_{\min}^{-1}(\mH_n)$. The higher the singularity score, the more singular $\mH_n$ is, and the greater the sensitivity of the embedding of $\vx_n$ to infinitesimal perturbations.

\subsection{Singularity scores for t-SNE, UMAP, LargeVis}
We have the detailed singularity scores for t-SNE, UMAP, LargeVis as follows.

\label{sec: method: derivation of sscore}
\paragraph{Singularity score for t-SNE.} The total loss for t-SNE is
\begin{align*}
&\mathfrak{L}(\vy_1,\ldots,\vy_n;\mX) = \sum_{1\le i<j\le n} \mathcal{L}(w(\vy_i, \vy_j); v_{i,j}(\mX)) + Z([\vy_1,\ldots,\vy_{n}]^\top),\qquad \text{where}\\
&\mathcal{L}(w(\vy_i, \vy_j); v_{i,j}(\mX)) = -2v_{i,j}(\mX)\log \big(w(\vy_i, \vy_j)\big),\   w(\vy_i, \vy_j) = (1+\|\vy_i-\vy_j\|_2^2)^{-1},\\ &Z([\vy_1,\ldots,\vy_{n}]^\top) = \log\Big(\sum_{k,l:k\neq l}(1+\|\vy_k-\vy_l\|_2^2)^{-1}\Big).
\end{align*}
Then, the singularity score for the $i$-th t-SNE embedding point is $\lambda_{\min}^{-1}(\mH_{i})$, where
\begin{equation*}
\begin{aligned}
    &\mH_{i} = \frac{\partial \mathfrak{L}}{\partial \vy_i\partial\vy_i^\top}=-\sum_{j:j\neq i}\mH_{ij},\qquad\text{where}\\
    &\mH_{ij} =\frac{\partial \mathfrak{L}}{\partial \vy_i\partial\vy_j^\top} =  -4v_{i,j}(\mX)w(\vy_i, \vy_j)\mI_2 + 8v_{i,j}(\mX)w^2(\vy_i, \vy_j)(\vy_i-\vy_j)(\vy_i-\vy_j)^\top\\
    &\quad -16\big(\sum_{k,l:k\neq l}w(\vy_k, \vy_l)\big)^{-2}\Big(\sum_{l:l\neq j}\big(w^2(\vy_j,\vy_l)(\vy_j-\vy_l)\big)\Big)\Big(\sum_{l:l\neq i}\big(w^2(\vy_i,\vy_l)(\vy_i-\vy_l)\big)\Big)^\top\\
    &\quad +4\big(\sum_{k,l:k\neq l}w(\vy_k, \vy_l)\big)^{-1}w^2(\vy_i,\vy_j)\mI_2 -16 \big(\sum_{k,l:k\neq l}w(\vy_k, \vy_l)\big)^{-1}w^3(\vy_i,\vy_j)(\vy_i-\vy_j)(\vy_i-\vy_j)^\top.
\end{aligned}
\end{equation*}

\paragraph{Singularity score for UMAP.} The total loss for UMAP is
\begin{align*}
&\mathfrak{L}(\vy_1,\ldots,\vy_n;\mX) = \sum_{1\le i<j\le n} \mathcal{L}(w(\vy_i, \vy_j); v_{i,j}(\mX)) + Z([\vy_1,\ldots,\vy_{n}]^\top),\qquad \text{where}\\
&\mathcal{L}(w(\vy_i, \vy_j); v_{i,j}(\mX)) = -v_{i,j}(\mX)\log \big(w(\vy_i, \vy_j)\big) - (1-v_{i,j}(\mX))\log\big(1-w(\vy_i, \vy_j)\big),\\   &w(\vy_i, \vy_j) = (1+a\|\vy_i-\vy_j\|_2^{2b})^{-1},\quad Z([\vy_1,\ldots,\vy_{n}]^\top) = 0,
\end{align*}
in which $a$ and $b$ are the hyperparameters chosen by user.

Then, the singularity score for the $i$-th UMAP embedding point is $\lambda_{\min}^{-1}(\mH_{i})$, where
\begin{equation*}
\begin{aligned}
    \mH_{i} =& \sum_{k:k\neq i}2abv_{i,k}(\mX)w(\vy_i, \vy_k)\|\vy_i-\vy_k\|_2^{2(b-1)}\mI_2\\
    &- \sum_{k:k\neq i}4a^2b^2v_{i,k}(\mX)w^2(\vy_i, \vy_k)\|\vy_i-\vy_k\|_2^{4(b-1)}(\vy_i-\vy_k)(\vy_i-\vy_k)^\top\\
    &+\sum_{k:k\neq i}4ab(b-1)v_{i,k}(\mX)w(\vy_i, \vy_k)\|\vy_i-\vy_k\|_2^{2 (b-2)}(\vy_i-\vy_k)(\vy_i-\vy_k)^\top\\
    &-\sum_{k:k\neq i}2ab(1-v_{i,k}(\mX))w^2(\vy_i, \vy_k)(1-w(\vy_i, \vy_k))^{-1}\|\vy_i-\vy_k\|_2^{2(b-1)}\mI_2\\
    & + \sum_{k:k\neq i}\frac{a^2b^2w^4(\vy_i, \vy_k)(8w^{-1}(\vy_i, \vy_k)-4)}{(1-w(\vy_i, \vy_k))^{2}}(1-v_{i,k}(\mX))\|\vy_i-\vy_k\|_2^{4(b-1)}(\vy_i-\vy_k)(\vy_i-\vy_k)^\top\\
    &-\sum_{k:k\neq i}\frac{4ab(b-1)w^2(\vy_i, \vy_k)}{1-w(\vy_i, \vy_k)}(1-v_{i,k}(\mX))\|\vy_i-\vy_k\|_2^{2 (b-2)}(\vy_i-\vy_k)(\vy_i-\vy_k)^\top.
\end{aligned}
\end{equation*}

\paragraph{Singularity score for LargeVis.} The total loss for LargeVis is
\begin{align*}
&\mathfrak{L}(\vy_1,\ldots,\vy_n;\mX) = \sum_{1\le i<j\le n} \mathcal{L}(w(\vy_i, \vy_j); v_{i,j}(\mX)) + Z([\vy_1,\ldots,\vy_{n}]^\top),\qquad \text{where}\\
&\mathcal{L}(w(\vy_i, \vy_j); v_{i,j}(\mX)) = \mathbbm{1}_{\{(i,j)\in E\}}v_{i,j}(\mX)\log \big(w(\vy_i, \vy_j)\big) + \gamma\mathbbm{1}_{\{(i,j)\notin E\}}\log \big(1-w(\vy_i, \vy_j)\big),\\   &w(\vy_i, \vy_j) = f(\|\vy_i-\vy_j\|_2),\quad f(x) = (1+x^{2})^{-1},\quad Z([\vy_1,\ldots,\vy_{n}]^\top) = 0,
\end{align*}
in which $E$ is the set of edges in the pre-constructed neighbor graph and $\gamma$ is an unified weight assigned to the negative edges.

Then, the singularity score for the $i$-th LargeVis embedding point is $\lambda_{\min}^{-1}(\mH_{i})$, where 
\begin{equation*}
    \begin{aligned}
        \mH_i =& -\sum_{k:k\neq i}\big(\mathbbm{1}_{\{(i,k)\in E\}}v_{i,k}(\mX) + \gamma\mathbbm{1}_{\{(i,k)\notin E\}}\big)\big(2w(\vy_i, \vy_k)\mI_2-4w^2(\vy_i, \vy_k)(\vy_i-\vy_k)(\vy_i-\vy_k)^\top\big)\\
        &+\sum_{k:k\neq i}\gamma\mathbbm{1}_{\{(i,k)\notin E\}}\Big(\frac{2\mI_2}{\|\vy_i-\vy_k\|_2^2} - \frac{4(\vy_i-\vy_k)(\vy_i-\vy_k)^\top}{\|\vy_i-\vy_k\|_2^4}\Big).
    \end{aligned}
\end{equation*}

\subsection{Theoretical Results} \label{sec:append-proof}
\subsubsection{Interpolation Property of the LOO-map}
For a dataset $\widetilde{\mX} = [\vx_1,\ldots,\vx_n,\vx_{n+1}]^\top$ where $\vx_{n+1} = \vx_{n}$ and their embedding points $[\tilde\vy_1,\ldots,\tilde\vy_n,\tilde\vy_{n+1}]^\top$, note that the similarity scores are equal for $\vx_n$ and $\vx_{n+1}$, i.e., $v_{i,n}(\widetilde{\mX})=v_{i,n+1}(\widetilde{\mX})$, $\forall\, i$. Thus,
\begin{equation*}
    \frac{\partial \mathfrak{L}(\tilde\vy_1,\ldots,\tilde\vy_{n},\tilde\vy_{n+1};\widetilde{\mX})}{\partial \tilde\vy_n} = \frac{\partial \mathfrak{L}(\tilde\vy_1,\ldots,\tilde\vy_{n},\tilde\vy_{n+1};\widetilde{\mX})}{\partial \tilde\vy_{n+1}}.
\end{equation*}
Using standard gradient descent or momentum methods, as employed in the t-SNE algorithm, with equal initializations for $\vy_{n}$ and $\vy_{n+1}$ (e.g., PCA initialization), we observe that for any iteration step $t$, $\tilde\vy_n^{(t)}=\tilde\vy_{n+1}^{(t)}$. In this way, a local minima of $\mathfrak{L}$, i.e., the embedding of $\widetilde{\mX}$ can be obtained with $\tilde\vy_n=\tilde\vy_{n+1}$. 

The LOO-map is defined as
\begin{equation*}
\begin{aligned}
&\vf(\vx) = \argmin_{\vy}L(\vy; \vx),\quad\text{where}\\
&L(\vy; \vx) = \sum_{1\le i \le n} \mathcal{L}\Big(w(\vy_i, \vy); v_{i,n+1}\big(\begin{bmatrix}\mX \\  \vx \end{bmatrix}\big)\Big) 
+ Z\Big(\begin{bmatrix}\mY \\  \vy\end{bmatrix}\Big).
\end{aligned}
\end{equation*}
The LOO assumption yields that $\vy_i \approx \tilde{\vy}_{i}$ for $i=1,\ldots,n$. Therefore, we have
\begin{equation*}
    L(\vy; \vx) \approx L_{n+1}(\vy; \vx)=\sum_{1\le i \le n} \mathcal{L}\Big(w(\tilde\vy_i, \vy); v_{i,n+1}\big(\begin{bmatrix}\mX \\  \vx \end{bmatrix}\big)\Big) 
+ Z\Big(\begin{bmatrix}\widetilde\mY \\  \vy\end{bmatrix}\Big),
\end{equation*}
where $\widetilde\mY = [\tilde\vy_1,\ldots,\tilde\vy_n]^\top$.
Note that $\tilde{\vy}_{n}=\tilde{\vy}_{n+1} = \argmin_\vy L_{n+1}(\vy; \vx)\approx \argmin_\vy L(\vy; \vx)$. Therefore, we have $\vf(\vx_n)\approx \tilde{\vy}_{n}$. Using similar reasoning for $\vx_{n+1} = \vx_i$, $i = 1, \ldots, n-1$, we have that
\begin{equation*}
    \vf(\vx_i) \approx \vy_i,\quad i = 1, \ldots, n,
\end{equation*}
which implies the approximate interpolation property.

When implementing the two diagnostic scores, the calculation is based on modifying an input point instead of adding a new input point. In detail, given an input data matrix $\mX = [\vx_1,\ldots,\vx_n]^\top$ and its embedding matrix $\mY = [\vy_1,\ldots,\vy_n]^\top$, we view $\vy_i$ as the mapping of $\vx_i$ by the partial LOO-map $\vf_i$:
\begin{equation}
\begin{aligned}
    &\vf_i(\vx) = \argmin_{\vy \in \R^2} L_i(\vy;\vx),\quad \text{where}\\&L_i(\vy;\vx)=\sum_{k\neq i} \mathcal{L}\Big(w(\vy_k, \vy); v_{i,k}(\bar \mX)\Big) 
+ Z(\bar \mY).
\end{aligned}
\end{equation}
where $\bar \mX = [\vx_1,\ldots,\vx_{i-1}, \vx, \vx_{i+1},\ldots,\vx_n]^\top$ differs from $\mX$ only at the $i$-th input point, and $\bar \mY = [\vy_1,\ldots,\vy_{i-1}, \vy, \vy_{i+1},\ldots,\vy_n]^\top$ has frozen embedding points except for the $i$-th point which is the decision variable in the optimization problem. Since $\mY$ is the minimizer of the total loss $\mathfrak{L}$, each $\vy_i$ is the minimizer of the partial loss $L_i(\vy;\vx_i)$, i.e., $f_i(\vx_i) = \vy_i$, which exhibits the exact interpolation property.

\subsubsection{Hyperbolic Structure in the LOO Loss: Proof of Theorem 1} 
In this subsection, we will prove Theorem~\ref{thm:1}. From basic calculation, we have
\begin{align*}
    & -\nabla_{\vy} L(\vy; \vx^\veps) = \vF_a + \vF_r, \qquad \qquad \text{where} \\
    &\vF_a = 4\sum_{i=1}^{n} \frac{v_{i,n+1}}{1 + \| \vy_i - \vy\|^2} (\vy_i - \vy), \\
    &\vF_r = -\frac{4}{Z} \sum_{i=1}^{n} \frac{1}{\big(1 + \| \vy_i - \vy\|^2\big)^2} (\vy_i - \vy)
\end{align*}
Let us simplify $\vF_a$ and $\vF_r$ using the asymptotics we assumed in the theorem. First we observe that
\begin{align*}
    \frac{1}{1 + \| \pm \vtheta - \vy + \vdelta_i\|^2} &= \frac{1}{1 + \| \vtheta\|^2 + 2 \langle \pm \vtheta, -\vy + \vdelta_i \rangle + \| -\vy + \vdelta_i\|^2} \\
    &= \frac{1}{\| \vtheta\|^2}\left[1 - 2 \Big\langle \pm \frac{\vtheta}{\| \vtheta\|^2}, -\vy + \vdelta_i \Big\rangle + O\Big(\frac{1}{\| \vtheta\|^2} \Big)\right]
\end{align*}
If $i \in \gI_+$, then
\begin{align*}
    \frac{\vy_i - \vy}{1 + \| \vtheta - \vy + \vdelta_i\|^2} &= \frac{\vtheta+\vdelta_i - \vy}{\| \vtheta \|^2} \left[1 - 2\Big\langle \frac{\vtheta}{\| \vtheta\|^2}, -\vy + \vdelta_i \Big\rangle + O\Big(\frac{1}{\| \vtheta \|^2} \Big)  \right] \\
    &= \frac{\vtheta}{\| \vtheta\|^2} + \frac{\vdelta_i - \vy}{\| \vtheta \|^2} - \frac{2 \vtheta}{\| \vtheta \|^2} \Big\langle \frac{\vtheta}{\| \vtheta\|^2}, -\vy + \vdelta_i\Big\rangle + O\Big(\frac{1}{\| \vtheta\|^3}\Big);
\end{align*}
if $i \in \gI_-$, then
\begin{align*}
    \frac{\vy_i - \vy}{1 + \| -\vtheta - \vy + \vdelta_i\|^2} &= \frac{-\vtheta+\vdelta_i - \vy}{\| \vtheta \|^2} \left[1 - 2\Big\langle -\frac{\vtheta}{\| \vtheta\|^2}, -\vy + \vdelta_i \Big\rangle + O\Big(\frac{1}{\| \vtheta \|^2} \Big)  \right]\\
    & = -\frac{\vtheta}{\| \vtheta\|^2} + \frac{\vdelta_i - \vy}{\| \vtheta \|^2} - \frac{2 \vtheta}{\| \vtheta \|^2} \Big\langle \frac{\vtheta}{\| \vtheta\|^2}, -\vy + \vdelta_i\Big\rangle + O\Big(\frac{1}{\| \vtheta\|^3}\Big)\, .
\end{align*}
Since $\sum_i \vdelta_i = \mathbf{0}$, we have
\begin{equation*}
    \frac{1}{n} \sum_{i=1}^{n} \frac{\vy_i - \vy}{1 + \| \vy_i - \vy\|^2} = -\frac{\vy}{\| \vtheta\|^2} + \frac{2}{\| \vtheta \|^2} \vy_{\parallelsum} + O\Big( \frac{1}{\| \vtheta \|^3} \Big) \, .
\end{equation*}
Similarly,
\begin{equation*}
    \frac{1}{n} \sum_{i \in \gI_+} \frac{\vy_i - \vy}{1 + \| \vy_i - \vy\|^2} - \frac{1}{n} \sum_{i \in \gI_-} \frac{\vy_i - \vy}{1 + \| \vy_i - \vy\|^2} = \frac{2\vtheta}{\| \vtheta\|^2} + O\Big( \frac{1}{\| \vtheta \|^3} \Big) \, .
\end{equation*}
Therefore,
\begin{equation}\label{eq:Fa}
    \vF_a = \frac{4np_0(2 \vy_{\parallelsum} - \vy)}{\| \vtheta \|^2} + \frac{8n\epsilon \vtheta}{\| \vtheta \|^2} + O\Big(\frac{1}{\|\vtheta\|^3} \Big). 
\end{equation}
To handle $\vF_r$, we notice that 
\begin{equation*}
    \frac{\vy_i - \vy}{\big(1 + \| \vy_i - \vy \|^2\big)^2} = O\Big( \frac{1}{\| \vtheta \|^3} \Big)\,.
\end{equation*}
Moreover, 
\begin{equation*}
    \frac{1}{1 + \|\vy_i - \vy_j \|^2} \le 1, \qquad \Longrightarrow \qquad Z \le  n(n+1)\,.
\end{equation*}
and also
\begin{equation*}
    Z \ge  \sum_{i,j \in \gI_+} \frac{1}{1 + \|\vy_i - \vy_j \|^2}  =  \sum_{i,j \in \gI_+} \frac{1}{1 + \|\vdelta_i - \vdelta_j \|^2} 
    ,.
\end{equation*}
Thus, $Z^{-1}$ is of constant order, so 
\begin{equation}\label{eq:Fr}
    \vF_r = O\Big( \frac{1}{\| \vtheta \|^3} \Big)\,.
\end{equation}
Combining Eqn.~\ref{eq:Fa} and~\ref{eq:Fr}, we reach our conclusion.

\subsection{Supplementary Tables and Figures}
\begin{table}[h]
\footnotesize
\centering
\begin{tabular}{ccccc}
\hline
Dataset                     &Points & Dimensions & Domain  & Usage      \\ \hline
CIFAR-10 \cite{KrizhevskyCIFAR102009}                          &  10,000  &   512               & Deep learning& Perturbation score, verification of LOO assumption \\
DTD \cite{CimpoiDTD2014}         &  5,640      &   512            & Deep learning& Perturbation score (OOD detection)\\
Embryo \cite{HayashiData}&  421       &  2,000              & Single-cell &                 Singularity score\\
Brain \cite{luecken}          &  3,618    &  1,000               & Single-cell&   Singularity score, verification of LOO assumption               \\
IFNB \cite{ifnb}                          &  6,548      &  2,000             & Single-cell &   Verification of LOO assumption              \\
Panc8 \cite{panc8}                  &  2,364        &   2,000          & Single-cell  &    Verification of LOO assumption      \\      
Mammary \cite{Bach2017MammaryData}     &  25,806        &   2,000          & Single-cell  &   Singularity score            \\ 
\hline
\end{tabular}
\caption{\textbf{Datasets analyzed in this work.} This paper analyzes seven real-world datasets: CIFAR-10, Describable Textures Dataset (DTD), the mouse embryonic stem cell differentiation data (Embryo), the single-cell RNA-seq dataset generated from PBMCs treated with interferon-$\beta$ (IFNB),  the mouse brain single-cell ATAC-seq data (Brain), the single-cell RNA-seq dataset generated from human pancreatic tissues (Panc8), the single-cell dataset of mammary epithelial cells (Mammary). The table lists the number of points, dimensions, domains, and usage of the datasets.}  
\label{tab: datasetoverview}
\end{table}

\vspace*{\fill}
\renewcommand{\arraystretch}{1.5}
\begin{table}[ht]
\centering
\small
\begin{tabular}{|c|c|c|c|c|c|}
\hline
Dataset   & Number of points & Perplexity 5 & Perplexity 25 & Perplexity 50 & Perplexity 75 \\ \hline
          & $n=1000$& $0.080\ (0.0017)$&$0.068\ (0.0035)$&$0.054\ (0.0056)$&$0.048\ (0.0124)$\\ \cline{2-6} 
2-GMM      & $n=3000$&$0.071\ (0.0008)$&$0.044\ (0.0018)$&$0.036\ (0.0022)$&$0.032\ (0.0010)$\\ \cline{2-6} 
          & $n=5000$&$0.062\ (0.0006)$&$0.034\ (0.0007)$&$0.033\ (0.0019)$&$0.032\ (0.0017)$\\ \hline
          
          & $n=1000$& $0.081\ (0.0019)$&$0.074\ (0.0110)$&$0.003\ (0.0018)$&$0.002\ (0.0005)$\\ \cline{2-6} 
Swissroll & $n=3000$& $0.072\ (0.0006)$&$0.043\ (0.0041)$&$0.0047\ (0.0062)$&$0.038\ (0.0068)$\\ \cline{2-6} 
          & $n=5000$&$0.063\ (0.0007)$&$0.033\ (0.0014)$&$0.031\ (0.0034)$&$0.037\ (0.0061)$\\ \hline
          
          & $n=1000$& $0.082\ (0.0032)$&$0.021\ (0.0046)$&$0.005\ (0.0030)$&$0.004\ (0.0026)$\\ \cline{2-6} 
Brain     & $n=2000$& $0.069\ (0.0041)$&$0.038\ (0.0042)$&$0.005\ (0.0019)$&$0.003\ (0.0012)$\\ \cline{2-6} 
          & $n=3000$&$0.063\ (0.0016)$&$0.041\ (0.0027)$&$0.016\ (0.0030)$&$0.003\ (0.0012)$\\ \hline
          
          & $n=500$& $0.094\ (0.0024)$&$0.046\ (0.0099)$&$0.030\ (0.0357)$&$0.025\ (0.0273)$\\ \cline{2-6} 
Panc8     & $n=1200$& $0.083\ (0.0012)$&$0.064\ (0.0016)$&$0.045\ (0.0037)$&$0.028\ (0.0066)$\\ \cline{2-6} 
          & $n=2000$&$0.079\ (0.0013)$&$0.060\ (0.0013)$&$0.049\ (0.0028)$&$0.039\ (0.0017)$\\ \hline

          & $n=1000$& $0.085\ (0.0036)$&$0.069\ (0.0022)$&$0.059\ (0.0052)$&$0.052\ (0.0107)$\\ \cline{2-6} 
IFNB      & $n=3000$& $0.064\ (0.0015)$&$0.049\ (0.0019)$&$0.048\ (0.0033)$&$0.046\ (0.0029)$\\ \cline{2-6} 
          & $n=5000$&$0.059\ (0.0010)$&$0.044\ (0.0010)$&$0.043\ (0.0022)$&$0.040\ (0.0015)$\\ \hline
          
          & $n=1000$& $0.086\ (0.0019)$&$0.042\ (0.0081)$&$0.017\ (0.0231)$&$0.006\ (0.0066)$\\ \cline{2-6} 
CIFAR10   & $n=3000$& $0.072\ (0.0009)$&$0.044\ (0.0013)$&$0.029\ (0.0037)$&$0.023\ (0.0113)$\\ \cline{2-6} 
          & $n=5000$&$0.065\ (0.0004)$&$0.039\ (0.0006)$&$0.029\ (0.0009)$&$0.025\ (0.0031)$\\ \hline
\end{tabular}
\caption{\textbf{Averaged (and std of) $\boldsymbol{\epsilon_n}$ across multiple trials under different datasets and perplexities.} We measure the approximation error $\epsilon_n$ across 20 independent trials and report the mean (and standard error) of $\epsilon_n$ for each setting. We observe from the table that all $\epsilon_n$'s are small and noticeably, $\epsilon_n$ is generally decreasing in $n$, which supports the LOO assumption.}
\label{tab: LOO Epsilon}
\end{table}
\vspace*{\fill}

\renewcommand{\arraystretch}{1.5}
\begin{table}[htbp]
\centering
\small
\begin{tabular}{|c|c|c|c|c|}
\hline
Data                       & Perplexity                  & DB Index &WCDR &Wilks' $\Lambda$ \\ \hline
\multirow{2}{*}{2-GMM 2d}  & Perplexity $5$              & $0.7470$ ($0.0401$)             & $0.3625$ ($0.0242$)                      &$0.2244$ ($0.0170$)\\ \cline{2-5} 
                           &\cellcolor[HTML]{EFEFEF} Perplexity $50$ &\cellcolor[HTML]{EFEFEF} $0.3821$ ($0.0653$)            &\cellcolor[HTML]{EFEFEF} $0.1287$ ($0.0353$)                      &\cellcolor[HTML]{EFEFEF}$0.0960$ ($0.0234$)\\ \hline
\multirow{2}{*}{2-GMM 10d} & Perplexity $5$              & $0.4580$  ($0.0082$)            & $0.1760$  ($0.0050$)                     &$0.0803$  ($0.0066$)\\ \cline{2-5} 
                           &\cellcolor[HTML]{EFEFEF}Perplexity $50$ &\cellcolor[HTML]{EFEFEF} $0.1232$ ($0.0034$)             &\cellcolor[HTML]{EFEFEF} $0.0152$ ($0.0008$)                      &\cellcolor[HTML]{EFEFEF}$0.0077$ ($0.0006$) \\ \hline
\multirow{2}{*}{2-GMM 50d} & Perplexity $5$              & $0.3943$  ($0.0090$)            & $0.1372$  ($0.0053$)                     &$0.0697$  ($0.0045$)\\ \cline{2-5} 
                           &\cellcolor[HTML]{EFEFEF} Perplexity $45$ &\cellcolor[HTML]{EFEFEF} $0.1046$ ($0.0058$)             &\cellcolor[HTML]{EFEFEF} $0.0110$ ($0.0011$)                      &\cellcolor[HTML]{EFEFEF}$0.0056$ ($0.0006$) \\ \hline
\multirow{2}{*}{5-GMM 2d}  & Perplexity $5$              & $0.5226$  ($0.0286$)            & $0.0776$  ($0.0067$)                     &$0.0061$  ($0.0012$)\\ \cline{2-5} 
                           &\cellcolor[HTML]{EFEFEF} Perplexity $65$ &\cellcolor[HTML]{EFEFEF} $0.2211$ ($0.0230$)             &\cellcolor[HTML]{EFEFEF} $0.0076$ ($0.0002$)                      &\cellcolor[HTML]{EFEFEF}$6.00\times10^{-6}$ ($2.26\times10^{-6}$) \\ \hline
\multirow{2}{*}{5-GMM 10d} & Perplexity $5$              & $0.3550$  ($0.0228$)            & $0.0331$  ($0.0025$)                     &$0.0014$  ($0.0003$)\\ \cline{2-5} 
                           &\cellcolor[HTML]{EFEFEF}Perplexity $40$ &\cellcolor[HTML]{EFEFEF} $0.2549$ ($0.0225$)             &\cellcolor[HTML]{EFEFEF} $0.0060$ ($0.0016$)                      &\cellcolor[HTML]{EFEFEF}$0.0001$ ($6.16\times10^{-5}$) \\ \hline
\multirow{2}{*}{5-GMM 50d} & Perplexity $5$              & $0.2813$  ($0.0066$)            & $0.0244$  ($0.0008$)                     &$0.0006$  ($0.0001$)\\ \cline{2-5} 
                           &\cellcolor[HTML]{EFEFEF}Perplexity $35$ &\cellcolor[HTML]{EFEFEF} $0.1627$ ($0.00091$)             &\cellcolor[HTML]{EFEFEF} $0.0048$ ($0.0005$)                      &\cellcolor[HTML]{EFEFEF}$3.02\times10^{-5}$ ($4.85\times10^{-6}$) \\ \hline
\end{tabular}
\caption{\textbf{Clustering quality improves after selecting a perplexity based on singularity scores.} Across the various data distributions and dimensions listed in the table, increasing the perplexity up to the elbow point leads to a decrease in all three quantitative metrics, indicating an improvement in clustering quality.}
\label{tab: simulatedindex}
\end{table}

\renewcommand{\arraystretch}{1.2}

\begin{table}[htbp]
\footnotesize
\centering
\begin{tabular}{cccccc}
\hline
\multicolumn{1}{l|}{}                                                                                                  & \multicolumn{1}{c|}{}        & \multicolumn{2}{c|}{Perplexity 4}                                       & \multicolumn{2}{c}{Perplexity 25}                            \\ \hline
\multicolumn{1}{c|}{\multirow{6}{*}{\begin{tabular}[c]{@{}c@{}}Test for\\ Spearman's\\ Rank Correlation\end{tabular}}} & \multicolumn{1}{c|}{Class}   & \multicolumn{1}{c|}{$p$-value}  & \multicolumn{1}{c|}{Corrected $p$-value*} & \multicolumn{1}{c|}{$p$-value}            & Corrected $p$-value* \\ \cline{2-6} 
\multicolumn{1}{c|}{}                                                                                                  & \multicolumn{1}{c|}{1}       & \multicolumn{1}{c|}{$0.0757$} & \multicolumn{1}{c|}{$0.3026$}           & \multicolumn{1}{c|}{$0.0000$}           & $0.0000$           \\
\multicolumn{1}{c|}{}                                                                                                  & \multicolumn{1}{c|}{2}       & \multicolumn{1}{c|}{$0.1298$} & \multicolumn{1}{c|}{$0.3895$}           & \multicolumn{1}{c|}{$6.3\times10^{-5}$} & $1.9\times10^{-4}$ \\
\multicolumn{1}{c|}{}                                                                                                  & \multicolumn{1}{c|}{3}       & \multicolumn{1}{c|}{$0.0389$} & \multicolumn{1}{c|}{$0.1945$}           & \multicolumn{1}{c|}{$2.7\times10^{-5}$} & $1.1\times10^{-4}$ \\
\multicolumn{1}{c|}{}                                                                                                  & \multicolumn{1}{c|}{4}       & \multicolumn{1}{c|}{$0.2493$} & \multicolumn{1}{c|}{$0.4533$}           & \multicolumn{1}{c|}{$0.7408$}           & $0.7408$           \\
\multicolumn{1}{c|}{}                                                                                                  & \multicolumn{1}{c|}{5}       & \multicolumn{1}{c|}{$0.2266$} & \multicolumn{1}{c|}{$0.4533$}           & \multicolumn{1}{c|}{$4.0\times10^{-3}$} & $8.0\times10^{-3}$           \\ \hline
\multicolumn{1}{c|}{\begin{tabular}[c]{@{}c@{}}F-test for\\ Local Regression Model\end{tabular}}                       & \multicolumn{1}{l|}{$p$-value} & \multicolumn{2}{c|}{$0.3598$}                                           & \multicolumn{2}{c}{$9.7\times10^{-4}$}                     \\ \hline
\multicolumn{1}{c|}{\begin{tabular}[c]{@{}c@{}}Permutation Test for\\ Local Regression Model\end{tabular}}             & \multicolumn{1}{c|}{$p$-value} & \multicolumn{2}{c|}{$0.3195$}                                           & \multicolumn{2}{c}{$0.0074$}                                 \\ \hline
\multicolumn{6}{l}{* Use Holm-Bonferroni correction for multiple testing.}           
\end{tabular}
\caption{\textbf{Tests of the distribution difference of singularity scores in the single-cell RNA-seq data of mouse embryonic stem cells differentiation.} We verifiy the distribution difference using Spearman’s rank correlation tests between singularity scores and distances to cluster centers, F-tests and per- mutation tests for a local regression model (singularity scores regressed against locations). The results of all three tests confirm the distribution difference of singularity scores between small and large perplexities.}  
\label{tab: hayashitest}
\end{table}

\begin{table}[htbp]
\footnotesize
\centering
\begin{tabular}{cccccc}
\hline
\multicolumn{1}{c|}{}                                                                                                  & \multicolumn{1}{c|}{}                   & \multicolumn{2}{c|}{Perplexity 5}                                       & \multicolumn{2}{c}{Perplexity 95}                            \\ \hline
\multicolumn{1}{c|}{\multirow{7}{*}{\begin{tabular}[c]{@{}c@{}}Test for\\ Spearman's\\ Rank Correlation\end{tabular}}} & \multicolumn{1}{c|}{Class}              & \multicolumn{1}{c|}{$p$-value}  & \multicolumn{1}{c|}{Corrected $p$-value*} & \multicolumn{1}{c|}{$p$-value}            & Corrected $p$-value* \\ \cline{2-6} 
\multicolumn{1}{c|}{}                                                                                                  & \multicolumn{1}{c|}{Astrocytes}         & \multicolumn{1}{c|}{$0.2136$} & \multicolumn{1}{c|}{$1.0000$}           & \multicolumn{1}{c|}{$0.5048$}           & $0.5048$           \\
\multicolumn{1}{c|}{}                                                                                                  & \multicolumn{1}{c|}{Endothelial Cells}  & \multicolumn{1}{c|}{$0.5594$} & \multicolumn{1}{c|}{$1.0000$}           & \multicolumn{1}{c|}{$0.0651$}           & $0.1893$           \\
\multicolumn{1}{c|}{}                                                                                                  & \multicolumn{1}{c|}{Excitatory Neurons} & \multicolumn{1}{c|}{$0.7461$} & \multicolumn{1}{c|}{$1.0000$}           & \multicolumn{1}{c|}{$1.2\times10^{-5}$} & $5.9\times10^{-5}$ \\
\multicolumn{1}{c|}{}                                                                                                  & \multicolumn{1}{c|}{Inhibitory Neurons} & \multicolumn{1}{c|}{$0.0470$} & \multicolumn{1}{c|}{$0.2820$}           & \multicolumn{1}{c|}{$0.0631$}           & $0.1893$           \\
\multicolumn{1}{c|}{}                                                                                                  & \multicolumn{1}{c|}{Microglia}          & \multicolumn{1}{c|}{$0.3735$} & \multicolumn{1}{c|}{$1.0000$}           & \multicolumn{1}{c|}{$1.2\times10^{-7}$} & $7.3\times10^{-7}$ \\
\multicolumn{1}{c|}{}                                                                                                  & \multicolumn{1}{c|}{Oligodendrocytes}   & \multicolumn{1}{c|}{$0.6364$} & \multicolumn{1}{c|}{$1.0000$}           & \multicolumn{1}{c|}{$3.9\times10^{-3}$} & $0.0156$           \\ \hline
\multicolumn{1}{c|}{\begin{tabular}[c]{@{}c@{}}F-test for\\ Local Regression Model\end{tabular}}                       & \multicolumn{1}{c|}{$p$-value}            & \multicolumn{2}{c|}{$0.1013$}                                          & \multicolumn{2}{c}{$<2.2\times10^{-16}$}                     \\ \hline
\multicolumn{1}{c|}{\begin{tabular}[c]{@{}c@{}}Permutation Test for\\ Local Regression Model\end{tabular}}             & \multicolumn{1}{c|}{$p$-value}            & \multicolumn{2}{c|}{$0.0610$}                                           & \multicolumn{2}{c}{$0.0000$}                                 \\ \hline
\multicolumn{6}{l}{* Use Holm-Bonferroni correction for multiple testing.}  
\end{tabular}
\caption{\textbf{Tests of the distribution difference of singularity scores in the mouse brain single-cell ATAC-seq data.} We verifiy the distribution difference using Spearman’s rank correlation tests between singularity scores and distances to cluster centers, F-tests and per- mutation tests for a local regression model (singularity scores regressed against locations). The results of all three tests confirm the distribution difference of singularity scores between small and large perplexities.}  
\label{tab: mousebraintest}
\end{table}

\begin{table}[htbp]
\footnotesize
\centering
\begin{tabular}{cccccc}
\hline
\multicolumn{1}{c|}{}                                                                                                  & \multicolumn{1}{c|}{}                   & \multicolumn{2}{c|}{Perplexity 30}                                       & \multicolumn{2}{c}{Perplexity 175}                            \\ \hline
\multicolumn{1}{c|}{\multirow{9}{*}{\begin{tabular}[c]{@{}c@{}}Test for\\ Spearman's\\ Rank Correlation\end{tabular}}} & \multicolumn{1}{c|}{Class}              & \multicolumn{1}{c|}{$p$-value}  & \multicolumn{1}{c|}{Corrected $p$-value*} & \multicolumn{1}{c|}{$p$-value}            & Corrected $p$-value* \\ \cline{2-6} 
\multicolumn{1}{c|}{}                                                                                                  & \multicolumn{1}{c|}{NP1}         & \multicolumn{1}{c|}{$0.9962$} & \multicolumn{1}{c|}{$1.0000$}           & \multicolumn{1}{c|}{$0.3623$}           & $1.0000$           \\
\multicolumn{1}{c|}{}                                                                                                  & \multicolumn{1}{c|}{NP2}  & \multicolumn{1}{c|}{$0.0557$} & \multicolumn{1}{c|}{$0.2223$}           & \multicolumn{1}{c|}{$0.4702$}           & $1.0000$           \\
\multicolumn{1}{c|}{}                                                                                                  & \multicolumn{1}{c|}{G1}  & \multicolumn{1}{c|}{$0.6469$} & \multicolumn{1}{c|}{$1.0000$}           & \multicolumn{1}{c|}{$7.1\times10^{-14}$}           & $3.5\times10^{-13}$           \\
\multicolumn{1}{c|}{}                                                                                                  & \multicolumn{1}{c|}{G2}  & \multicolumn{1}{c|}{$0.0147$} & \multicolumn{1}{c|}{$0.0734$}           & \multicolumn{1}{c|}{$1.1\times10^{-15}$}           & $6.8\times10^{-15}$           \\
\multicolumn{1}{c|}{}                                                                                                  & \multicolumn{1}{c|}{L1} & \multicolumn{1}{c|}{$2.7\times10^{-25}$} & \multicolumn{1}{c|}{$2.2\times10^{-24}$}           & \multicolumn{1}{c|}{$3.3\times10^{-150}$} & $2.6\times10^{-149}$ \\
\multicolumn{1}{c|}{}                                                                                                  & \multicolumn{1}{c|}{L2}          & \multicolumn{1}{c|}{$1.5\times 10^{-8}$} & \multicolumn{1}{c|}{$1.1\times 10^{-7}$}           & \multicolumn{1}{c|}{$5.2\times10^{-76}$} & $3.6\times 10^{-75}$ \\
\multicolumn{1}{c|}{}                                                                                                  & \multicolumn{1}{c|}{PI1}          & \multicolumn{1}{c|}{$0.5188$} & \multicolumn{1}{c|}{$1.0000$}           & \multicolumn{1}{c|}{$0.8746$} & $1.0000$ \\ 
\multicolumn{1}{c|}{}                                                                                                  & \multicolumn{1}{c|}{PI2}   & \multicolumn{1}{c|}{$0.0022$} & \multicolumn{1}{c|}{$0.0131$}           & \multicolumn{1}{c|}{$4.6\times 10^{-11}$} & $1.8\times 10^{-10}$           \\ \hline
\multicolumn{1}{c|}{\begin{tabular}[c]{@{}c@{}}F-test for\\ Local Regression Model\end{tabular}}                       & \multicolumn{1}{c|}{$p$-value}            & \multicolumn{2}{c|}{$0.2979$}                                          & \multicolumn{2}{c}{$5.3\times10^{-14}$}                     \\ \hline
\multicolumn{1}{c|}{\begin{tabular}[c]{@{}c@{}}Permutation Test for\\ Local Regression Model\end{tabular}}             & \multicolumn{1}{c|}{$p$-value}            & \multicolumn{2}{c|}{$0.2379$}                                           & \multicolumn{2}{c}{$0.0000$}                                 \\ \hline
\multicolumn{6}{l}{* Use Holm-Bonferroni correction for multiple testing.}  
\end{tabular}
\caption{\textbf{Tests of the distribution difference of singularity scores in the mammary epithelial cell data.} We verifiy the distribution difference using Spearman’s rank correlation tests between singularity scores and distances to cluster centers, F-tests and per- mutation tests for a local regression model (singularity scores regressed against locations). The results of all three tests confirm the distribution difference of singularity scores between small and large perplexities.}  
\label{tab: mammarytest}
\end{table}

\renewcommand{\arraystretch}{1.2}

\begin{table}[htbp]
\footnotesize
\centering
\begin{subtable}[h]{\textwidth}
\centering
\begin{tabular}{cccccc}
\hline
Embryo                    & Singularity Score & EMBEDR & scDEED (elbow) & scDEED & DynamicViz \\ \hline
Perplexity                & 25                & -      & 3              & -      & 20         \\
Neighborhood Preservation & 0.5594            & -      & 0.4955         & -      & 0.5524     \\ \hline
\end{tabular}
\caption{Mouse embryonic stem cells (mESCs) differentiation data.}
\end{subtable}
\hfill
\vspace{0.5cm}
\begin{subtable}[h]{\textwidth}
\centering
\begin{tabular}{cccccc}
\hline
Brain                     & Singularity Score & EMBEDR & scDEED (elbow) & scDEED & DynamicViz \\ \hline
Perplexity                & 95                & 145    & 10             & 145    & 10         \\
Neighborhood Preservation & 0.4108            & 0.4223 & 0.3575         & 0.4223 & 0.3575     \\ \hline
\end{tabular}
\caption{Mouse brain single-cell ATAC-seq data.}
\end{subtable}
\hfill
\vspace{0.5cm}
\begin{subtable}[h]{\textwidth}
\centering
\begin{tabular}{cccccc}
\hline
Mammary                     & Singularity Score & EMBEDR & scDEED (elbow) & scDEED & DynamicViz \\ \hline
Perplexity                & 175                & 450    & 100             & 500    & -         \\
Neighborhood Preservation & 0.6866            & 0.6937 & 0.6846         & 0.69387 & -     \\ \hline
\end{tabular}
\caption{Mouse mammary epithelial cells data.}
\end{subtable}
\caption{\textbf{Perplexities chosen by singularity score, EMBEDR, scDEED, and DynamicViz and the corresponding neighborhood preservation score.} \textbf{(a)} Comparison of perplexity selection methods in mouse embryonic stem cell (mESC) differentiation data. EMBEDR fails to select a perplexity as we encountered an error when running the package, while scDEED’s second approach, which minimizes the number of dubious points, fails to yield a unique value. Our singular scores identify a perplexity that reduces sub-clusters while preserving the highest neighborhood continuity.  
\textbf{(b)} Comparison of perplexity selection methods in mouse brain single-cell ATAC-seq data. Our singular scores suggest a moderate perplexity that reduces sub-clusters, maintains high neighborhood preservation and retains fine-grained structure.  
\textbf{(c)} Comparison of perplexity selection methods in mouse mammary epithelial cells data. DynamicViz fails due to its lack of scalability for large datasets, as its bootstrap-based approach requires repeated execution of visualization algorithms. Among other methods, our singular scores suggest a moderate perplexity that ensures high neighborhood preservation while retaining fine-grained structure.
}  
\label{tab: comparesscore}
\end{table}
\clearpage

\begin{figure}[htbp]
    \centering 
    \includegraphics[scale = 0.9, trim = 0cm 0cm 0cm 0.3cm]{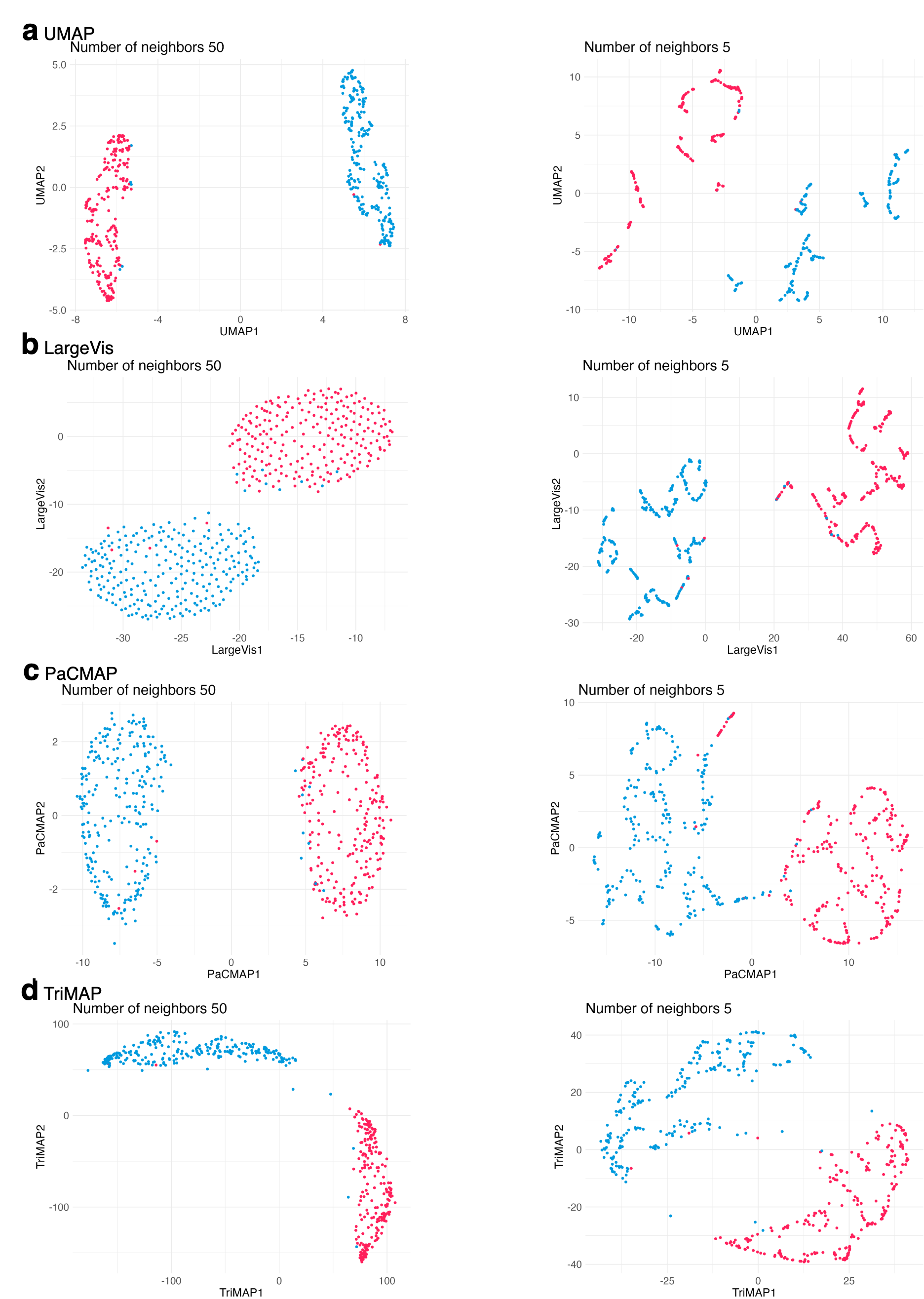}
    \caption{\textbf{OI and FI discontinuities are common among neighbor embedding methods.} We use the same setting as in Figure~\ref{fig: discontinuity examples}a, where we generate two-component Gaussian mixture data and embed the input points using various neighbor embedding methods, including UMAP (\textbf{a}), LargeVis (\textbf{b}), PaCMAP (\textbf{c}), TriMAP (\textbf{d}). The hyperparameter is the number of neighbors (similar to perplexity) chosen as $50$ (left panels) and $5$ (right panels) respectively. We observe similar OI and FI discontinuity as in the t-SNE method.
    }
    \label{fig: appendix: umapdiscontinuity}
\end{figure}

\newpage
\begin{figure}[htbp]
    \centering 
    \includegraphics[scale = 0.9, trim = 0cm 0cm 0cm 1cm]{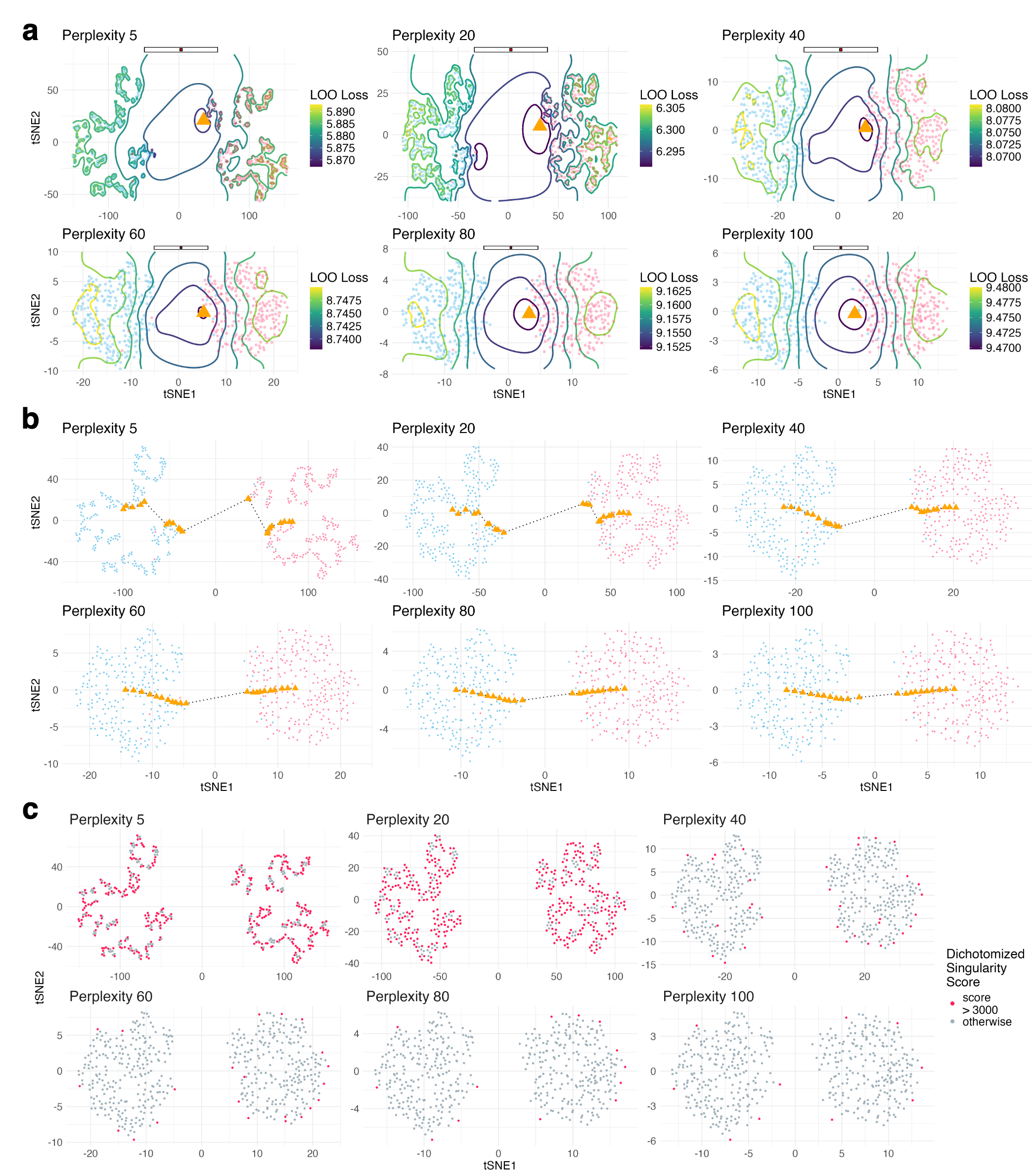}
    \caption{\textbf{Large perplexity lessens FI discontinuity.} 
    We use the same mixture data as in Figure~\ref{fig: discontinuity examples} and run the standard t-SNE algorithm 6 times at perplexity 5, 20, 40, 60, 80, 100. \textbf{a} Contour plot shows the landscape of LOO loss $L(\vy; \vx)$ for $\vx = 0.5(\vc_1 +\vc_2)$ under different perplexities. The number of local minima in the loss decreases with higher perplexity, indicating that the FI discontinuity lessens under higher perplexity. \textbf{b} Trajectories of the embedding point of $\vx$ are shown with different perplexities. Under small perplexities, numerous local minima cause an uneven trajectory of embedding points when we add $\vx$ at evenly interpolated locations; while under larger perplexities, the trajectory is more smooth. This further suggests a reduction of FI discontinuity when the perplexity increases. \textbf{c} Embeddings with dichotomized singularity scores. Embedding points with high singularity scores decrease in number when increasing the perplexity.}
    \label{fig: appendix: 2_3}
\end{figure}
\newpage
\begin{figure}[htbp]
    \centering 
    \includegraphics[scale = 0.9]{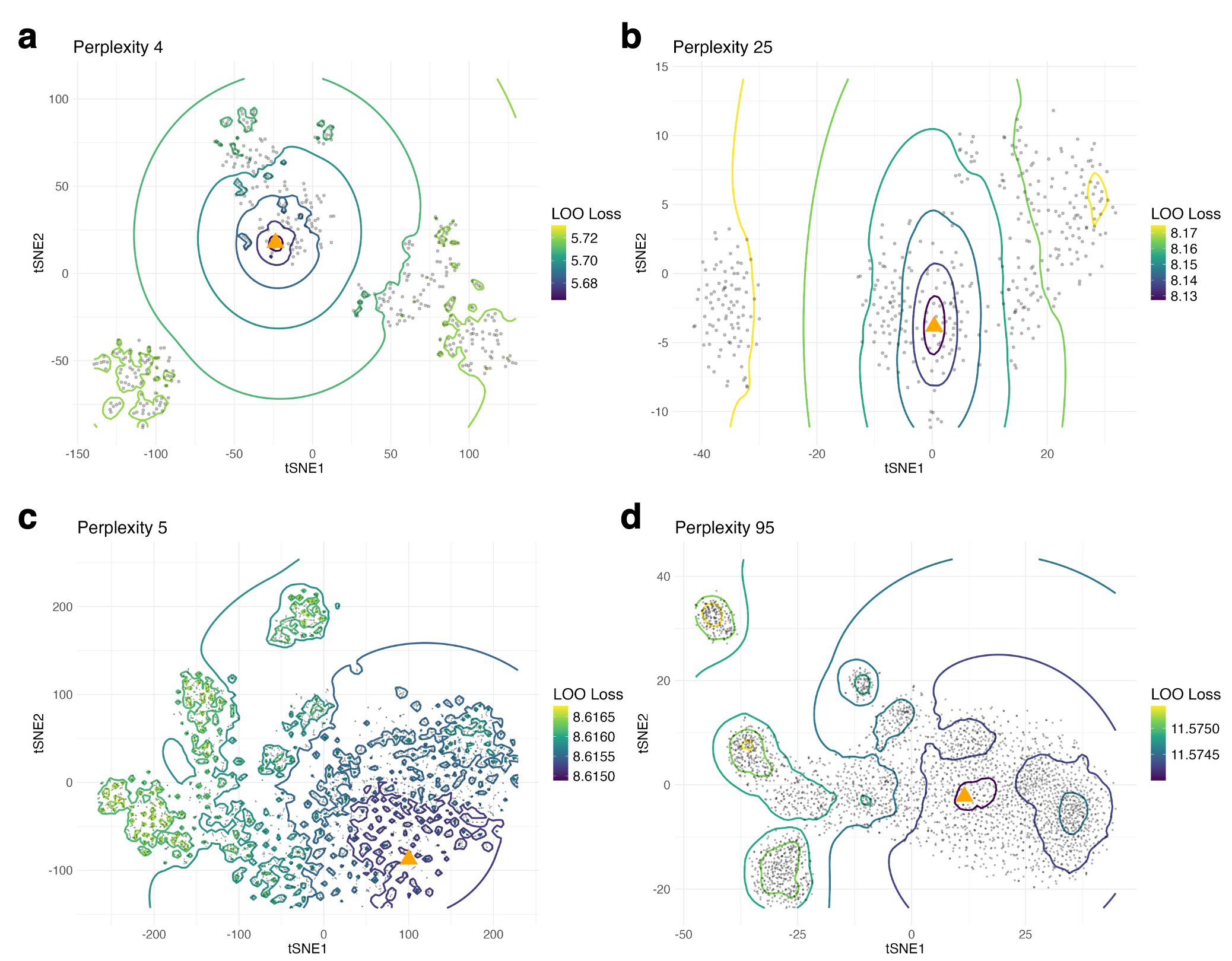}
    \caption{\textbf{Comparing loss landscapes for single-cell data with different perplexities.} We present LOO loss landscapes under small and large perplexities for mouse embryonic stem cells differentiation data (\textbf{a-b}) and mouse brain single-cell ATAC-seq data (\textbf{c-d}). 
    We randomly choose an input $\vx_i$ from the dataset and plot the landscape of the partial LOO loss $L_i(\vy;\vx_i)$.  
    We observe that more local minima emerge at random locations in the partial LOO loss landscape under a small perplexity than under a large perplexity, which supports our claim about the reduction of local minima and FI discontinuity under a large perplexity.}
    \label{fig: appendix: use case 2}
\end{figure}
\newpage
\begin{figure}[htbp]
    \centering 
    \includegraphics[scale = 0.9]{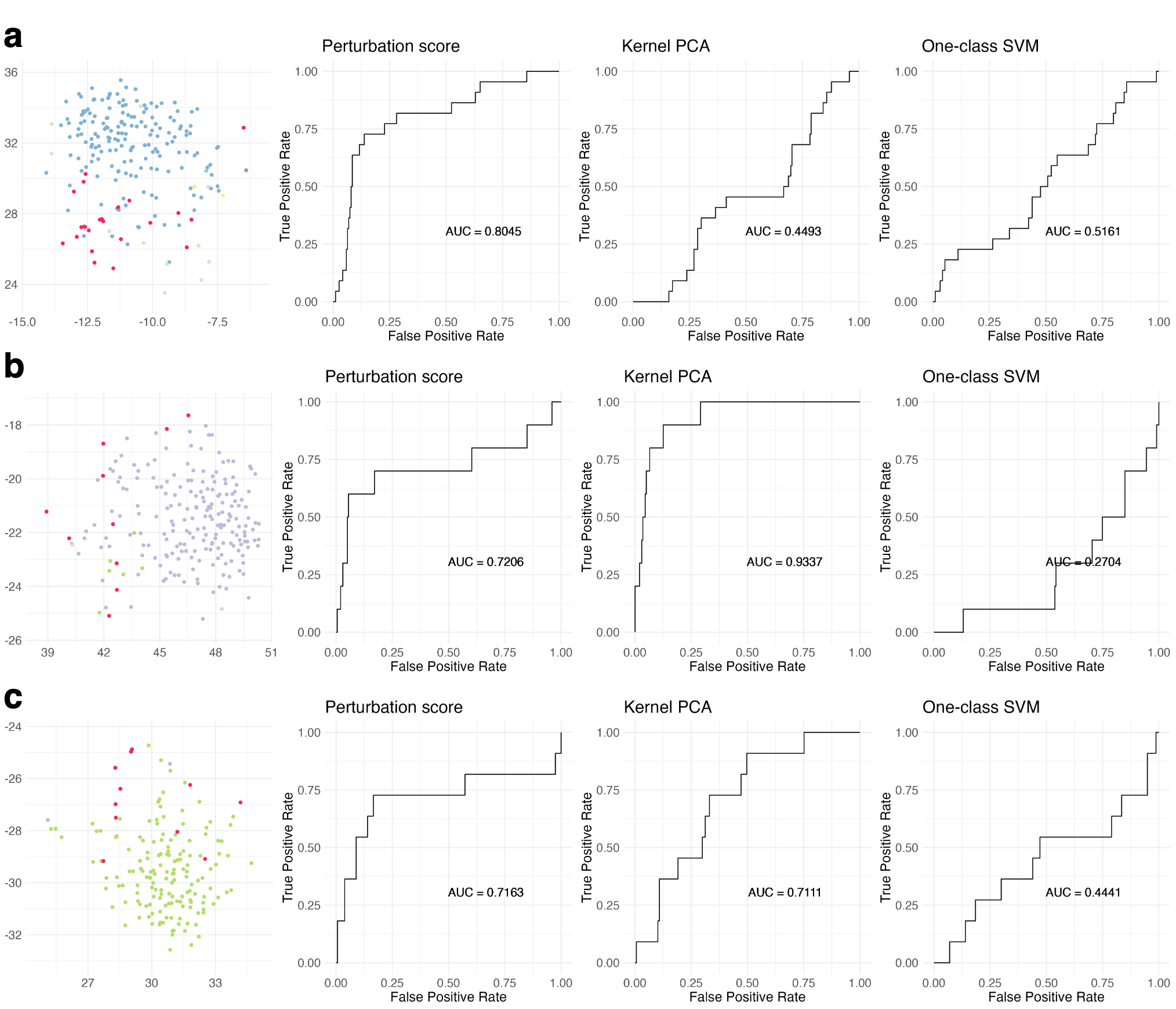}
    \caption{\textbf{Comparing perturbation score with kernel PCA and one-class SVM for OOD detection.} Using the dataset presented in Fig.~\ref{fig:CIFAR10ood}, we compare the performance of the three methods. The perturbation score outperforms the others, achieving an average AUROC of 0.747 across the three selected clusters. In contrast, kernel PCA and one-class SVM obtain averaged AUROCs of 0.698 and 0.410, respectively.}
    \label{fig: appendix: compare with kpca and ocsvm}
\end{figure}
\newpage
\begin{figure}[htbp]
    \centering 
    \includegraphics[scale = 0.9]{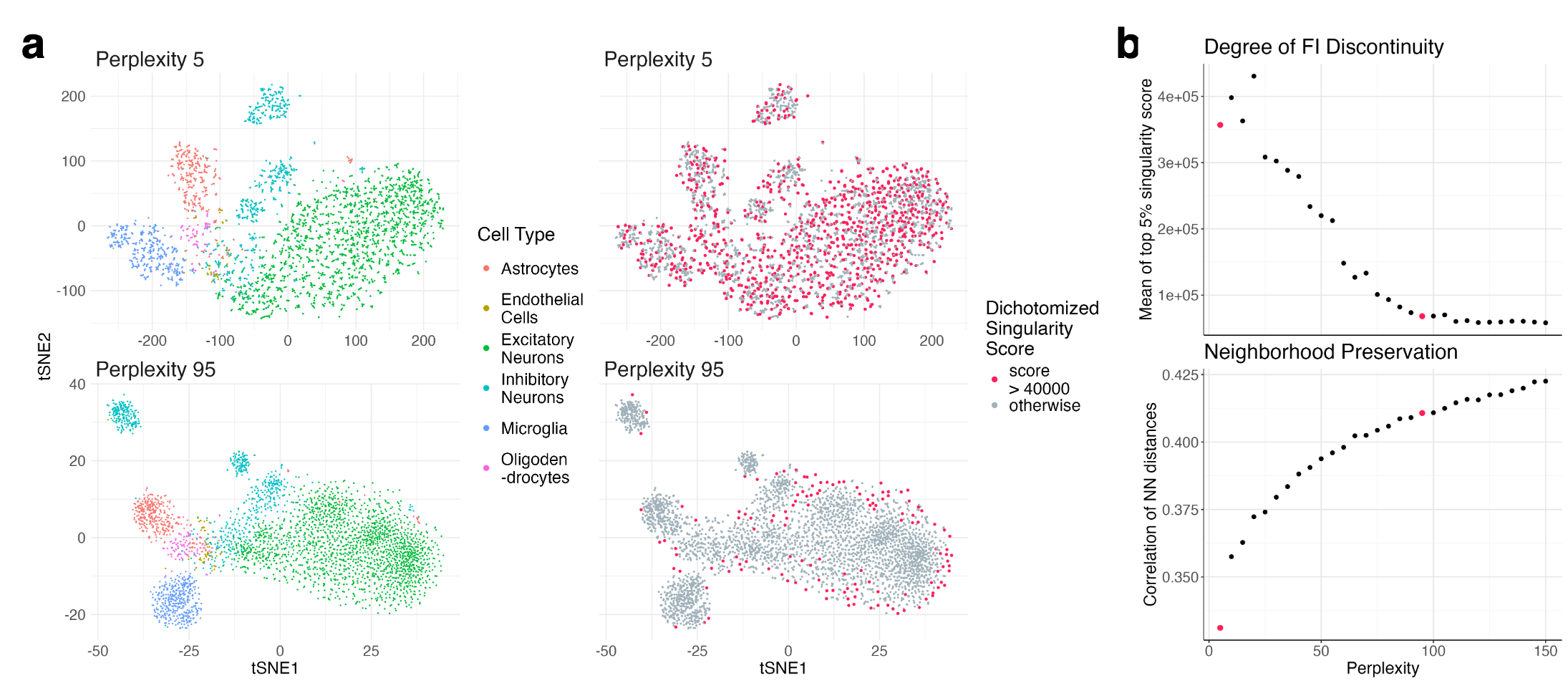}
    \caption{\textbf{Singularity scores inform selection of the perplexity parameter in the mouse brain single-cell ATAC-seq data.} In the t-SNE visualizations of the mouse brain single-cell ATAC-seq data, multiple spurious sub-clusters are introduced at perplexity 5 (default by t-SNE algorithm) (\textbf{a}). High singularity scores appear in random locations, indicating the presence of such spurious structures and severe FI discontinuity. The perplexity chosen by the singularity score creates better visualization compared to default. Qualitatively, the cluster structure is more compact and the high singularity scores disappear except in the peripheries of the clusters. Quantitatively, the neighborhoods of most points are more faithfully embedded (higher neighbor-hood preservation) (\textbf{b}).}
    \label{fig: appendix: use case 2 Mouse Brain}
\end{figure}

\newpage
\begin{figure}[htbp]
    \centering 
    \includegraphics[scale = 0.9]{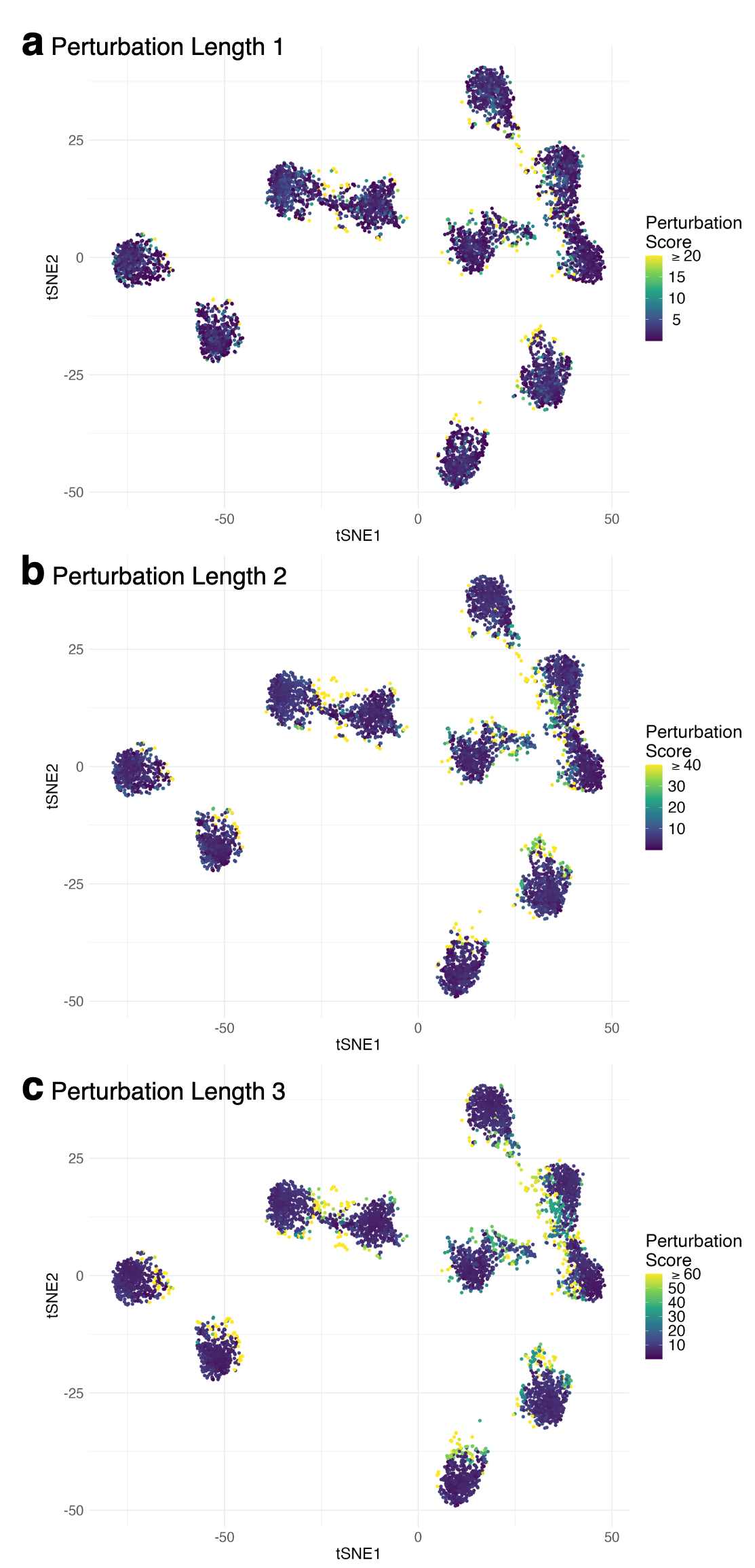}
    \caption{\textbf{Perturbation scores of CIFAR-10 image features with different choices of perturbation length.} We vary the choice of the perturbation length $\lambda \in \{1, 2, 3\}$ in the calculation of perturbation scores (see Eqn.~\ref{eq:lambda}). We find that embedding points receiving high scores are consistent across different choices of $\lambda$. This suggests that perturbation scores are not sensitive to the choice of the perturbation length.
    }
    \label{fig: appendix: sensitivity_pscore}
\end{figure}

\begin{figure}[htbp]
    \centering 
    \includegraphics[scale = 0.9]{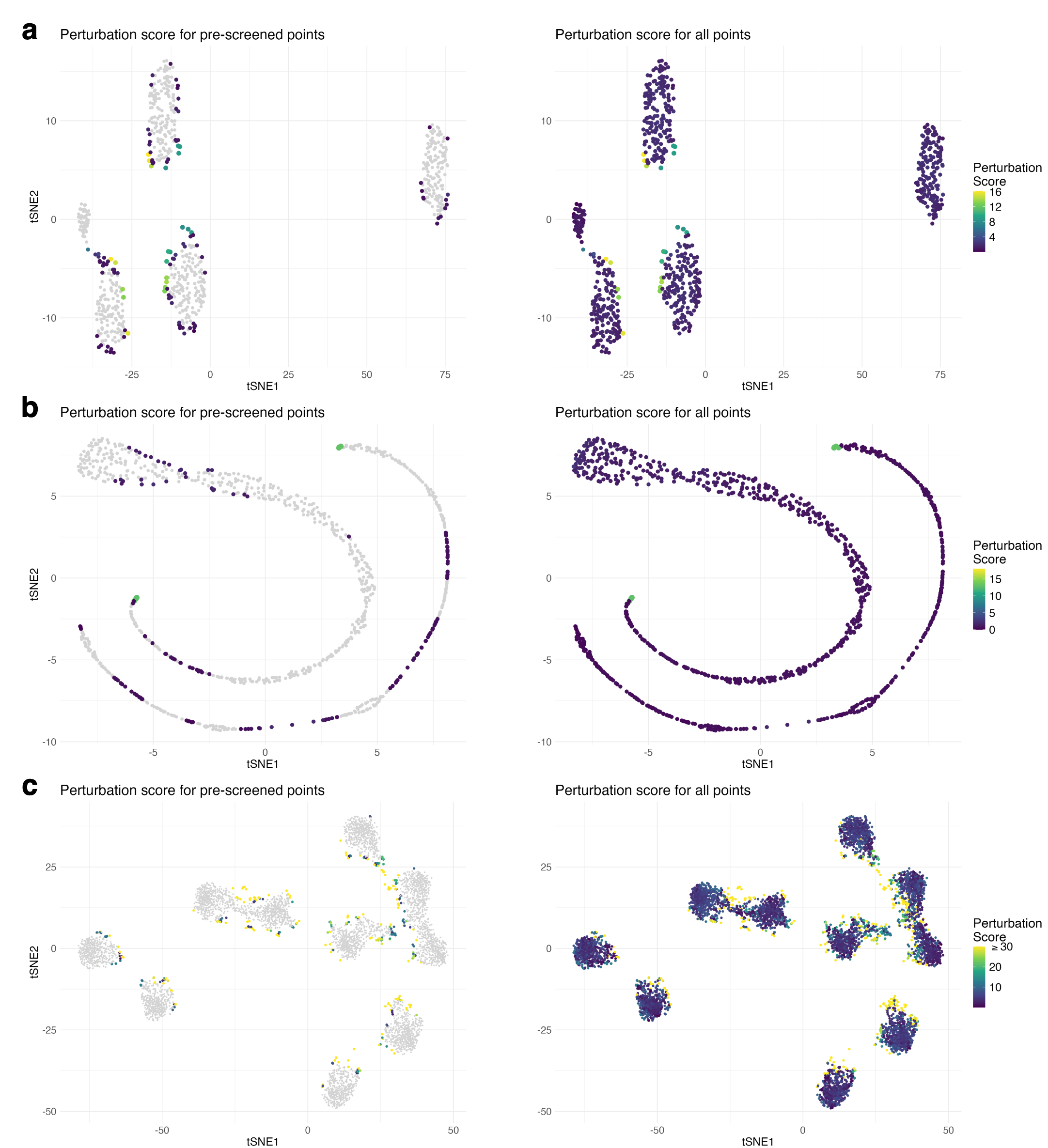}
    \caption{\textbf{Pre-screening points for calculating perturbations scores provides comparable results.} We examine the validity of the pre-screening step by comparing the perturbation scores with the pre-screening step (visualized in left panels) and those without (visualized in right panels). We present the results of perturbation scores on three datasets---5-component Gaussian mixture data (\textbf{a}), Swiss roll data (\textbf{b}), and the deep learning feature data of CIFAR-10 (\textbf{c}). 
    We find that the calculating perturbation scores with the pre-screening step still identifies most of the OI discontinuity locations, so pre-screening provides a faster and comparably reliable assessment of OI discontinuity.
    }
    \label{fig: appendix: prescreen}
\end{figure}

\begin{figure}[htbp]
    \centering 
    \includegraphics[scale = 0.9]{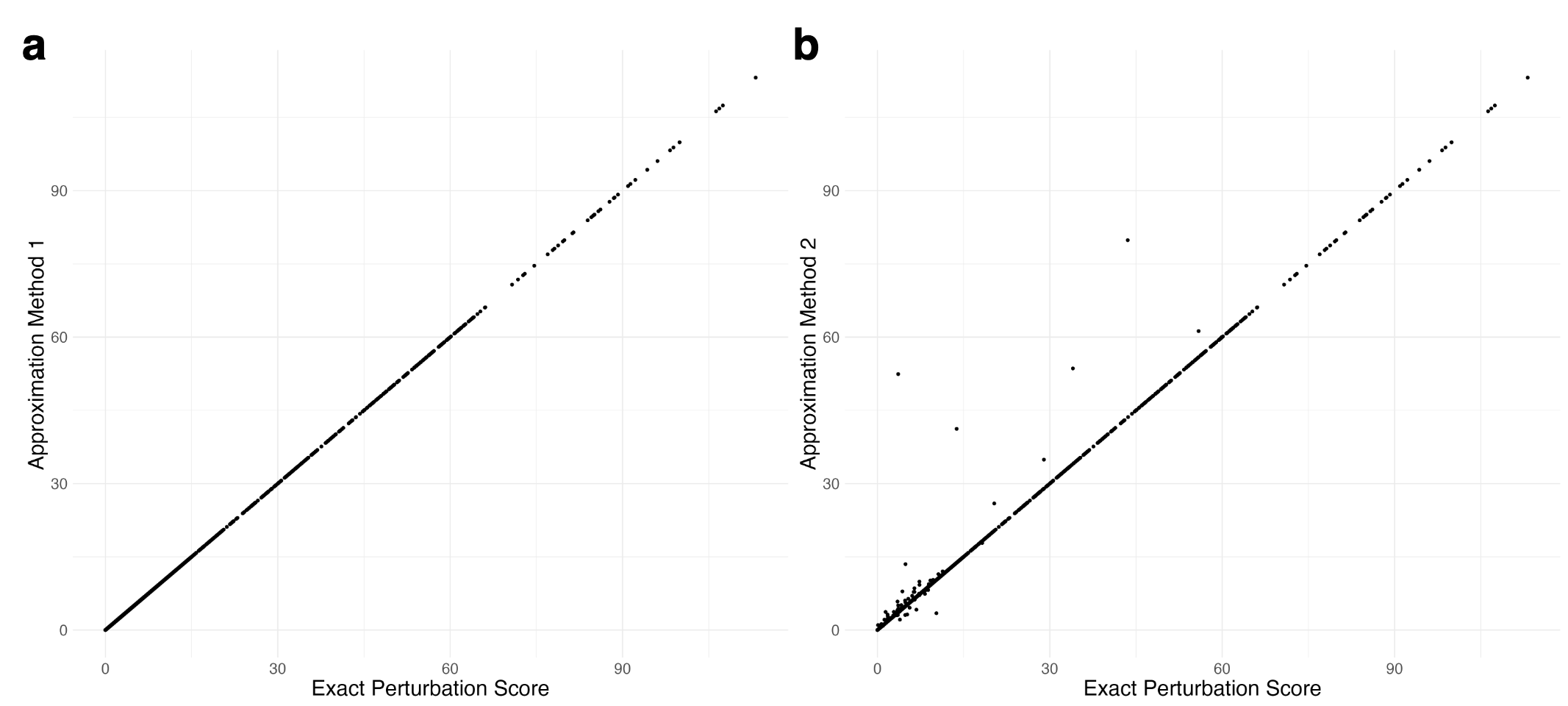}
    \caption{\textbf{The approximation method 1\&2 are accurate for calculating perturbation scores.}  We test the approximation quality on a dataset consisting of 5000 deep learning feature vectors obtained from CIFAR-10 images.
    \textbf{a} Scatter plot of exact perturbation scores v.s.~perturbation scores by the approximation method 1. \textbf{b} Scatter plot of exact perturbation scores v.s.~perturbation scores by approximation method 2. We find that the perturbation scores using both approximation methods are approximately equal to the exact perturbation scores for almost all the points.}
    \label{fig: appendix: approx}
\end{figure}

\begin{figure}[htbp]
    \centering
    \includegraphics[scale = 0.9]{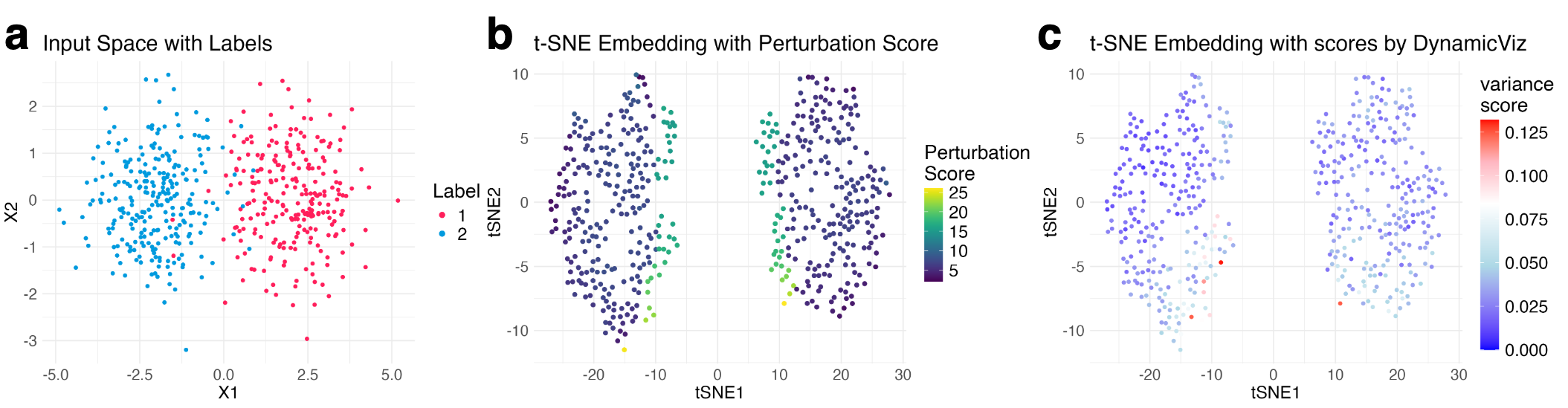}
    \caption{\textbf{Comparing perturbation scores and variance scores from DynamicViz}. \textbf{a} We use a two-dimensional two-component Gaussian mixture data as the input data.
    \textbf{b} Perturbation scores clearly mark the unreliable embedding points where the disconnection (discontinuity) occurs.
    \textbf{c} DynamicViz fails to identify most of the embedding points that are overconfidently clustered.
    }
    \label{fig: methods-compare-2}
\end{figure}

\begin{figure}[htbp]
    \centering
    \includegraphics[scale = 0.9]{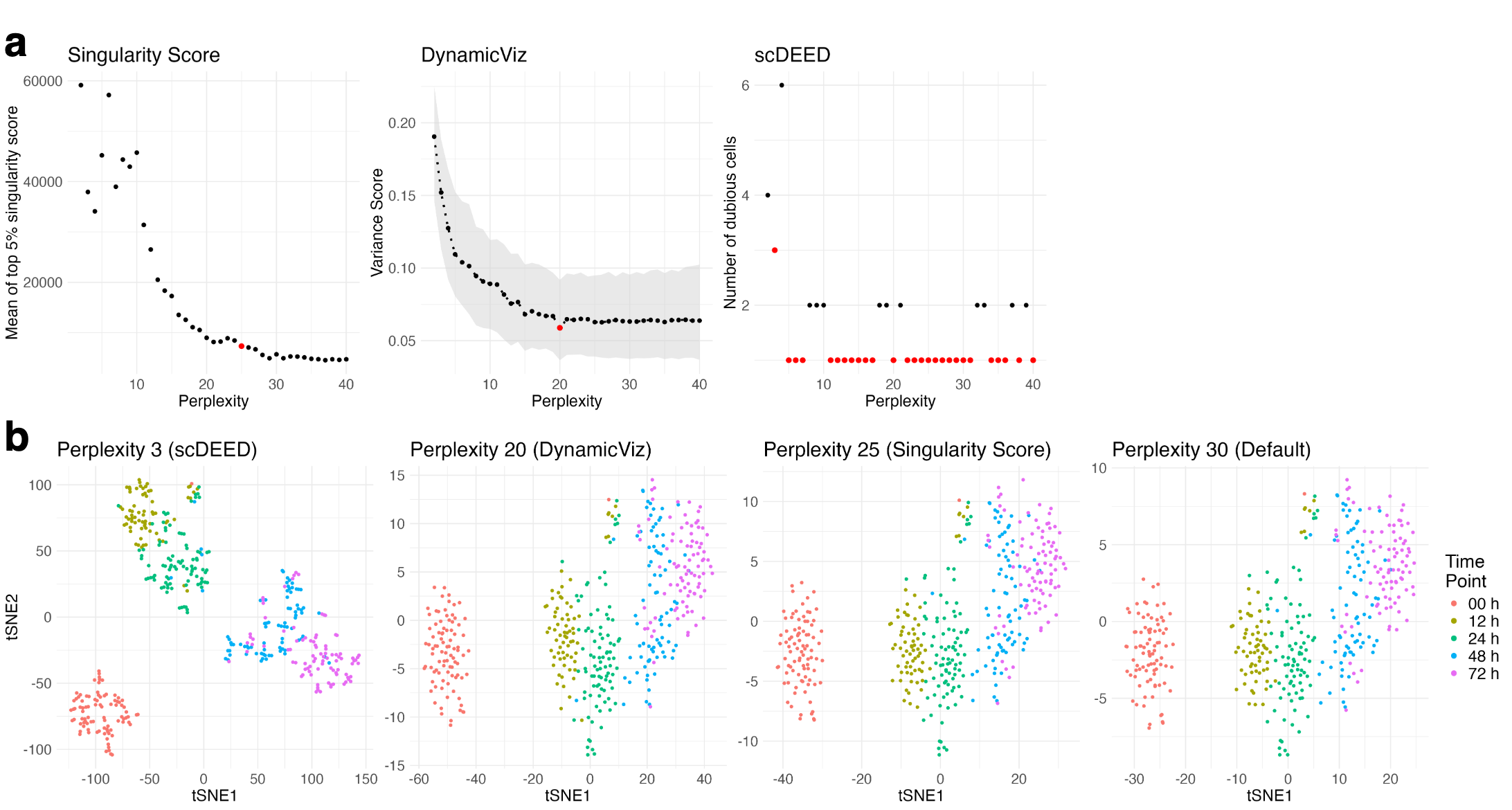}
    \caption{\textbf{Comparing perplexity selection on the mouse embryonic cell differentiation data}. \textbf{a} For mouse embryonic cell differentiation data, singularity score suggests selecting a perplexity of 25. DynamicViz chooses a perplexity of 20 by minimizing the variance score. By identifying the elbow point in the plot of dubious embedding points against perplexity, scDEED selects a perplexity of 3. However, it fails to select the optimal perplexity by minimizing the number of dubious points, as 26 out of 39 perplexity candidates reach the minimum number. EMBEDR is not applicable to the mouse embryonic cell differentiation data due to errors occurring for datasets smaller than 1,000.
    \textbf{b} The t-SNE embedding of perplexity 3 (chosen by scDEED) yields numerous spurious sub-clusters. The t-SNE embeddings of perplexity 20 (chosen by DynamicViz), 25 (chosen by singularity score), and 30 (default) are visually very similar.
    }
    \label{fig: methods-compare-singularityscore-hayashi}
\end{figure}

\begin{figure}[htbp]
    \centering
    \includegraphics[scale = 0.9]{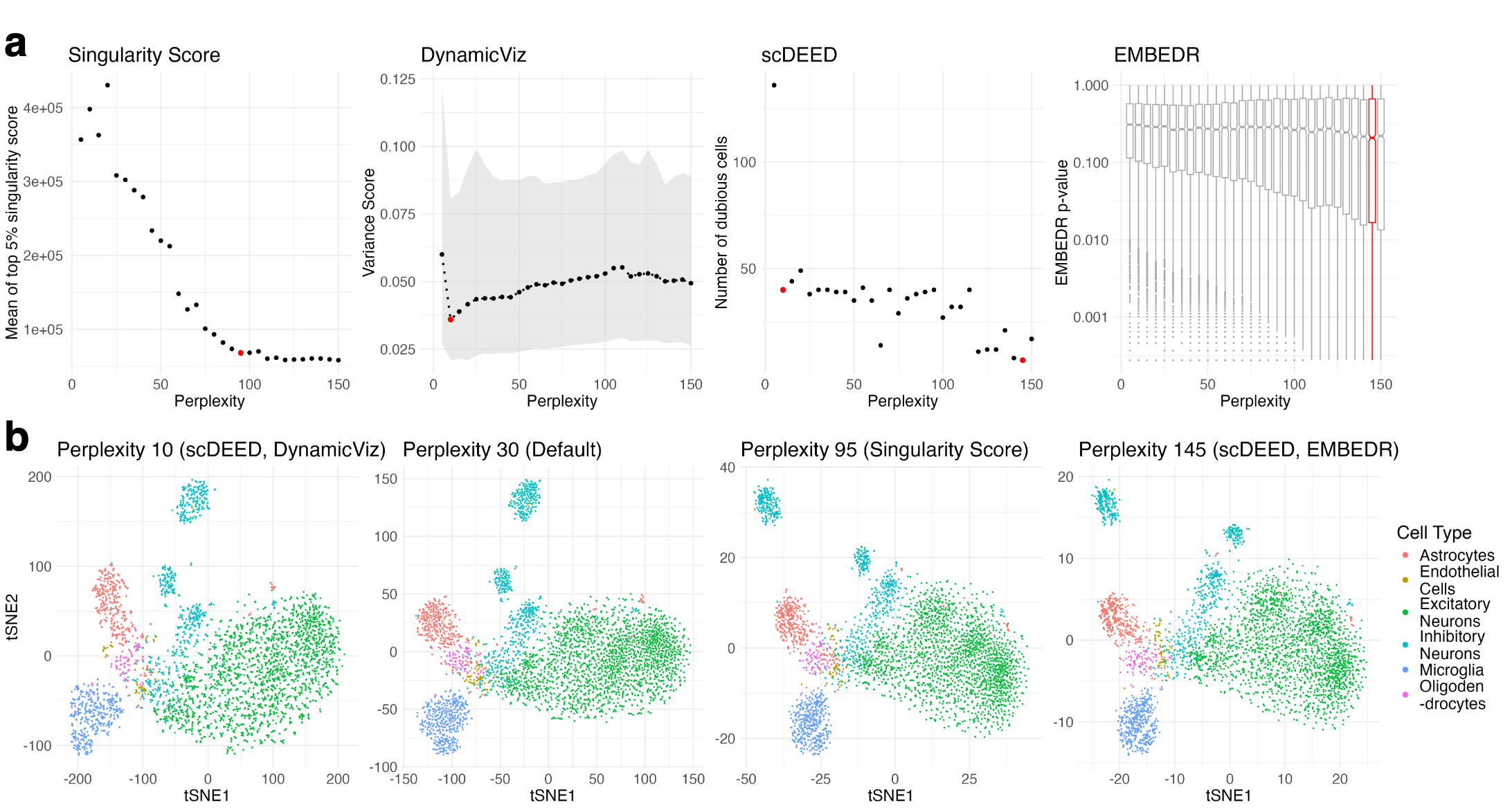}
    \caption{\textbf{Comparing perplexity selection on the mouse brain chromatin accessibility data}. 
    \textbf{a} For mouse brain chromatin accessibility data, singularity score suggests selecting a perplexity of 95. DynamicViz chooses a perplexity of 10. By identifying the elbow point in the plot of dubious embedding points against perplexity, scDEED also selects a perplexity of 10. And it selects a perplexity of 145 by minimizing the number of dubious points. EMBEDR also selects a perplexity of 145 by minimizing the median of the permutation test $p$-values, agreeing with its undesirable property of choosing the larger candidate perplexity. We used standard boxplots (center line, median; box limits, upper and lower quartiles; points, outliers).
    \textbf{b} The t-SNE embedding of perplexity 10 (chosen by DynamicViz and scDEED) yields numerous spurious sub-clusters. The t-SNE embeddings of perplexity 30 (default), 95 (chosen by singularity score), and 145 (chosen by scDEED and EMBEDR) are visually very similar. However, increasing perplexity excessively is not suggested, as it may cause clusters to merge and result in the loss of certain genuine local or microscopic structures.
    }
    \label{fig: methods-compare-singularityscore-mousebrain}
\end{figure}

\begin{figure}[htbp]
    \centering
    \includegraphics[scale = 0.9]{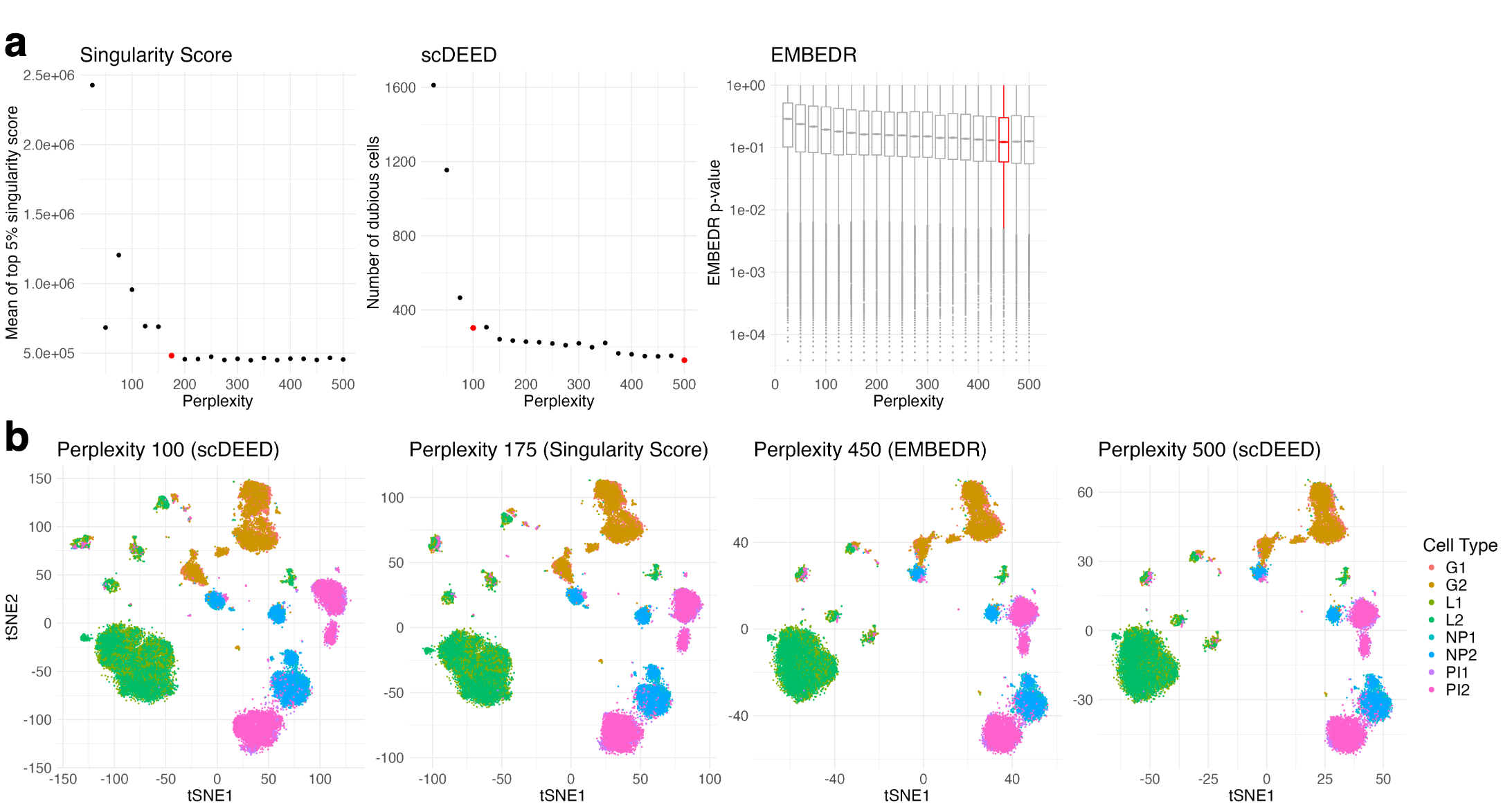}
    \caption{\textbf{Comparing perplexity selection on the mouse mammary epithelial cell dataset}. 
    \textbf{a} Singularity score suggests selecting a perplexity of 175. By identifying the elbow point in the plot of dubious embedding points against perplexity, scDEED selects a perplexity of 100. And it selects a perplexity of 500 by minimizing the number of dubious points. EMBEDR also selects a large perplexity of 450. We used standard boxplots (center line, median; box limits, upper and lower quartiles; points, outliers). DynamicViz lacks scalability for large datasets due to its bootstrap-based approach, which necessitates repeated execution of visualization algorithms.
    \textbf{b} The t-SNE embeddings of perplexity 100 (scDEED) and 175 (singularity score) are visually very similar. We can observe that clusters separated in perplexity 175 start to merge for larger perplexities of 450 and 500. Therefore, increasing perplexity excessively is not suggested.
    }
    \label{fig: methods-compare-singularityscore-bachmammary}
\end{figure}
\clearpage

\end{document}